\documentclass[prd,aps,showpacs,nofootinbib,nobibnotes,floatfix]{revtex4}
\usepackage{psfig}

\newcommand{\cqg}{Class.\ Quant.\ Grav.\ }
\newcommand{\apjsup}{Astrophys.\ J.\ Suppl.\ }
\newcommand{\jetp}{JETP\ }

\newcommand{\dd}{{\rm d}}

\newcommand{\SC}{{\scriptscriptstyle {\rm SC}}}
\newcommand{\MC}{{\scriptscriptstyle {\rm MC}}}
\newcommand{\SCAL}{{\scriptscriptstyle {\rm S}}}

\newcommand{\UPSTKS}[3]{\Upsilon^{[{#1}]}_{{#2}\,{#3}}}
\newcommand{\XITKSLM}[5]{\xi^{[{#1}]\,{#3}}_{{#2}\,{#4}\,{#5}}}
\newcommand{\XIKSLM}[4]{\xi^{{#2}}_{{#1}\,{#3}\,{#4}}}

\newcommand{\diag}{{\rm diag}}
\newcommand{\MOD}[1]{\mbox{ (mod }{#1})}

\newcommand{\BAR}{{\rm b}}
\newcommand{\OBS}{{\rm obs}}
\newcommand{\LSS}{{\rm LSS}}

\newcommand{\bb}[1]{{\bf {#1}}}
\newcommand{\EDE}{{\bb E}^2}
\newcommand{\TTR}{{\bb T}^3}
\newcommand{\ETR}{{\bb E}^3}
\newcommand{\RTR}{{\bb R}^3}

\newcommand{\CHU}{\chi}

\newcommand{\ETAL}{{\it et al.}}

\newcommand{\QED}{{\it Q.E.D.}}

\newcommand{\UUNIT}[2]{\,{\mbox{#1}^{#2}}}

\newcommand{\DIR}[1]{\delta^{{\rm D}}({#1})}
\newcommand{\KRON}[2]{\delta_{{#1}\,{#2}}}

\newcommand{\bk}{{\bf k}}
\newcommand{\bx}{{\bf x}}
\newcommand{\bT}{{\bf T}}

\newcommand{\ee}[1]{{\rm e}^{#1}}

\newcommand{\VECDEUXD}[2]{\left(\begin{array}{c} 
                                {#1} \\ 
                                {#2} \end{array} \right)}
\newcommand{\VECTROISD}[3]{\left(\begin{array}{c}
                                 {#1} \\ 
                                 {#2} \\ 
                                 {#3} \end{array} \right)}
\newcommand{\MATDEUXD}[4]{\left(\begin{array}{rrr}
                                {#1} & {#2} \\ 
	                        {#3} & {#4} \end{array} \right)}
\newcommand{\MATTROISD}[9]{\left(\begin{array}{rrr}
                                 {#1} & {#2} & {#3} \\ 
                                 {#4} & {#5} & {#6} \\ 
	                         {#7} & {#8} & {#9} \end{array} \right)}

\begin{document}
\title{Cosmic microwave background anisotropies in multi-connected
flat spaces}

\author{Alain Riazuelo}
\email{riazuelo@iap.fr}
\affiliation{Service de Physique Th{\'e}orique,
             CEA/DSM/SPhT, Unit{\'e} de recherche associ{\'e}e au
             CNRS, CEA/Saclay F--91191 Gif-sur-Yvette c{\'e}dex,
             France}

\author{Jeffrey Weeks}
\email{weeks@northnet.org}
\affiliation{15 Farmer St.,  Canton  NY 13617-1120, USA}

\author{Jean-Philippe Uzan}
\email{uzan@iap.fr}
\affiliation{Institut d'Astrophysique de Paris, GR$\varepsilon$CO,
FRE 2435-CNRS, 98bis boulevard Arago, 75014 Paris, France \\
Laboratoire de Physique Th\'eorique, CNRS-UMR 8627, Universit\'e Paris
Sud, B\^atiment 210, F--91405 Orsay c\'edex, France}

\author{Roland Lehoucq}
\email{lehoucq@cea.fr}
\affiliation{CE-Saclay, DSM/DAPNIA/Service d'Astrophysique, F--91191
Gif-sur-Yvette c\'edex, France, \\ Laboratoire Univers et Th\'eories,
CNRS-UMR 8102, Observatoire de Paris, F--92195 Meudon
c\'edex, France}

\author{Jean-Pierre Luminet}
\email{jean-pierre.luminet@obspm.fr}
\affiliation{Laboratoire Univers et Th\'eories, CNRS-UMR 8102,
Observatoire de Paris, F--92195 Meudon c\'edex, France}

\date{13 November 2003}


\begin{abstract}
This article investigates the signature of the seventeen
multi-connected flat spaces in cosmic microwave background (CMB) maps.
For each such space it recalls a fundamental domain and a set of
generating matrices, and then goes on to find an orthonormal basis for
the set of eigenmodes of the Laplace operator on that space.  The
basis eigenmodes are expressed as linear combinations of eigenmodes of
the simply connected Euclidean space.  A preceding work, which
provides a general method for implementing multi-connected topologies
in standard CMB codes, is then applied to simulate CMB maps and
angular power spectra for each space. Unlike in the 3-torus, the
results in most multi-connected flat spaces depend on the location of
the observer. This effect is discussed in detail. In particular, it is
shown that the correlated circles on a CMB map are generically not
back-to-back, so that negative search of back-to-back circles in the
WMAP data does not exclude a vast majority of flat or nearly flat
topologies.
\end{abstract}
\maketitle

\section{Introduction}

Among all multi-connected three-dimensional spaces, ``flat
spaces''\footnote{In this article, we follow the cosmological use and
we call ``flat spaces'' the eighteen types of Euclidean spaces, and
``Euclidean space'' the simply connected universal cover $\ETR$.} have
been studied the most extensively in the cosmological context. This is
due to the computational simplicity of the simplest compact flat
three-manifold, the 3-torus, which has been used extensively in
numerical simulations. The main goal of this article is to provide
tools to compute the CMB properties and produce high resolution CMB
maps for all seventeen multi-connected flat spaces~\footnote{Test maps
for these spaces are available on demand.}, following the general
method introduced in our preceding work~\cite{rulw02}.

Recent measurements show that the density parameter $\Omega_0$ is
close to unity and the observable universe is approximately flat.  CMB
data obtained by the Archeops balloon experiments~\cite{archeops} and
more recently by WMAP~\cite{wmap} place strong constraints on the
curvature. In addition, WMAP~\cite{map} and later the Planck
satellite~\cite{planck} do and will provide full sky maps of CMB
anisotropies, offering an opportunity to probe the topological
properties of our universe. This observational constraint on the
curvature radius of the universe motivates the detailed study of flat
spaces even though spherical spaces are also promising
candidates~\cite{wlu02,glluw,lwugl,luw02,prl,nat}.

At present, the status of the constraints on the topology of flat
spaces is evolving rapidly driven by the release of the WMAP
data. Previous analysis, based on the COBE data, mainly constrained
the topology of a 3-torus (see
Refs.~\cite{sokolov93,staro93,stevens93,costa95,levin98,levin99,pf,inoue00,inoue01,inoue2,roukema}
and Refs.~\cite{lachieze95,uzan99,levin02} for reviews of different
methods for searching for the topology).

The WMAP data~\cite{wmap} possess some anomalies on large angular
scales that may be explained by a topological structure. In
particular, the quadrupole is abnormally low, the octopole is very
planar and the alignment between the quadrupole and octopole is also
anomalous~\cite{teg1}. Besides many other potential
explanations~\cite{quad}, it was suggested that a toroidal universe
with a smaller dimension on the order of half the horizon scale may
explain all these anomalies~\cite{teg1} but it was latter shown, on
the basis of a finer statistical analysis, that it did
not~\cite{teg2}. Another topology was recently proposed to explain
some of this anomaly in the case of a slightly positively curved
space, namely the Poincar\'e dodecahedral space~\cite{nat}.

The first results of the search for the topology through the ``circles
in the sky'' method~\cite{cornish98} gave negative results for
back-to-back or almost back-to-back circles~\cite{teg2,cssk}. While
the first applies only to back-to-back circle with no twist, the
second study includes an arbitrary twist and conclude that ``it rules
out the possibility that we live in a universe with topology smaller
than $24 \UUNIT{Gpc}{}$''. As will be discussed in this paper,
back-to-back circles are generic only for homogeneous topologies such
as e.g. 3-tori and a subclass of lens spaces. In non-homogeneous
spaces the relative position of the circles depends on the position of
the observer in the fundamental polyhedron.

In conclusion, as demonstrated by these preliminary results, only the
toroidal spaces have been really
constrained~\cite{teg2,cssk}. Besides, a series of studies have
pointed out a departure of the WMAP data from statistical
isotropy. Copi {\em et al.}~\cite{copi} recently argued in particular
that they are inconsistent with an isotropic Gaussian distribution at
98.8\% confidence level. Previous studies pointed toward a possible
North-South asymmetry of the data~\cite{eriksen,park}. Spaces with
non-trivial topology are a class of models in which global isotropy
(and possibly global homogeneity) is broken. Simulated CMB maps of
these spaces may help to construct estimators for quantifying the
departure of the temperature distribution from isotropy, and also give
a deeper understanding of recent results.

Let us emphasize that in the case where the topological scale is
slightly larger than the size of the observable universe, no matching
circles will be observed. This might also happen for a configuration
where the circles would all lie in the direction of the galactic disk
where the signal-to-noise ratio might be too low. Contrary to the
simply connected case, the correlation matrix, $C_{\ell m}^{\ell' m'}
\equiv \left< a_{\ell m} a_{\ell' m'} \right>$, of the coefficients of
the development of the temperature fluctuations on spherical
harmonics, will not be proportional to $\delta_{\ell \ell'} \delta_{m
m'}$. The study of this correlation matrix could offer the possibility
to probe topology (slightly) beyond the horizon. Computing the
correlation matrix $C_{\ell m}^{\ell' m'}$ for different
multi-connected spaces will help design the best strategy to constrain
the deviation from the simply connected case, and gives a concrete
example of cosmological models in which the global homogeneity and
isotropy are broken.

As described in detail in our preceding work~\cite{rulw02}, what is
needed for any CMB computation are the eigenmodes of the Laplacian
\begin{equation}
\label{Helmotz1}
\Delta \UPSTKS{\Gamma}{k}{} = - k^2 \UPSTKS{\Gamma}{k}{} ,
\end{equation}
with boundary conditions compatible with the given topology.  These
eigenmodes can be developed on the basis ${\cal Y}_{k \ell m}$ of the
(spherically symmetric) eigenmodes of the universal covering space as
\begin{equation}
\label{eq:1}
\UPSTKS{\Gamma}{k}{s}
 = \sum_{\ell = 0}^\infty 
   \sum_{m = -\ell}^{\ell}
        \XITKSLM{\Gamma}{k}{s}{\ell}{m} {\cal Y}_{k \ell m} ,
\end{equation}
so that all the topological information is encoded in the coefficients
$\XITKSLM{\Gamma}{k}{s}{\ell}{m}$, where $s$ labels the various
eigenmodes sharing the same eigenvalue $- k^2$.  Ref.~\cite{rulw02}
computes these coefficients for the torus and lens spaces and
Refs.~\cite{glluw,cornish99} discuss more general cases.

To summarize, this article aims at several goals. First, it will give
the complete classification of flat spaces and the exact form of the
eigenmodes of the Laplacian for each of them. It will also provide a
set of simulated CMB maps for most of these spaces. Among other
effects, it will illustrate the effect of non-compact directions and
discuss the influence of the position of the observer in the case of
non-homogeneous spaces, which has never been discussed before.  It
also explains the structure of the observed CMB spectrum in the case
of a very anisotropic (i.e., flattened or elongated in one direction)
fundamental domain.

This article is organized as follows. We start by recalling the
properties of the eighteen flat spaces (Section~\ref{sec02}) as well
as the eigenmodes of the simply connected three-dimensional Euclidean
space $\ETR$ (Section~\ref{sec2}), and in particular how to convert
planar waves, which suit the description of topology, to spherical
waves, which are more convenient for CMB computation. Then, in
Section~\ref{SectionEigenmodesOfMulticonnected}, we explain how to
extract the modes of a given multi-connected space from the modes of
$\ETR$. This method is then applied to give the eigenmodes of the ten
compact flat spaces (Section~\ref{SectionCompactSpaces}), the five
multi-connected flat spaces with two compact directions (``chimney
spaces'', Section~\ref{SectionDoublyPeriodic}) and the two
multi-connected flat spaces with only one compact direction (``slab
spaces'', Section~\ref{SectionSinglyPeriodic}). Applying the general
formalism developed in our previous work~\cite{rulw02}, we produce CMB
maps for some of these spaces.  With three exceptions the manifolds
are not homogeneous, in the sense that a given manifold does not look
the same from all points. To discuss the implication of the observed
CMB and the genericity of the maps, we detail in
Section~\ref{SectionLocationOfObserver} the influence of the position
of the observer on the form of the eigenmodes and we study its
consequences on the observed CMB maps. We show in particular that the
matched circles are generically not back-to-back, but their relative
position depends on the topology, the precise shape of the fundamental
domain, and the position of the observer.

\section*{Notation}

The local geometry of the universe is described by a locally Euclidean
Friedmann-Lema\^{\i}tre metric
\begin{equation}
\label{fl_metric} \dd s^2
 = - c^2 \dd t^2
   + a^2 (t) \left[\dd\CHU^2 + \CHU^2\dd \omega^2 \right] ,
\end{equation}
where $a(t)$ is the scale factor, $t$ the cosmic time, $\dd
\omega^2 \equiv \dd \theta^2 + \sin^2 \theta \, \dd \varphi^2$ the
infinitesimal solid angle.

\section{The eighteen flat spaces}\label{sec02}

\begin{figure}
\centerline{\psfig{file=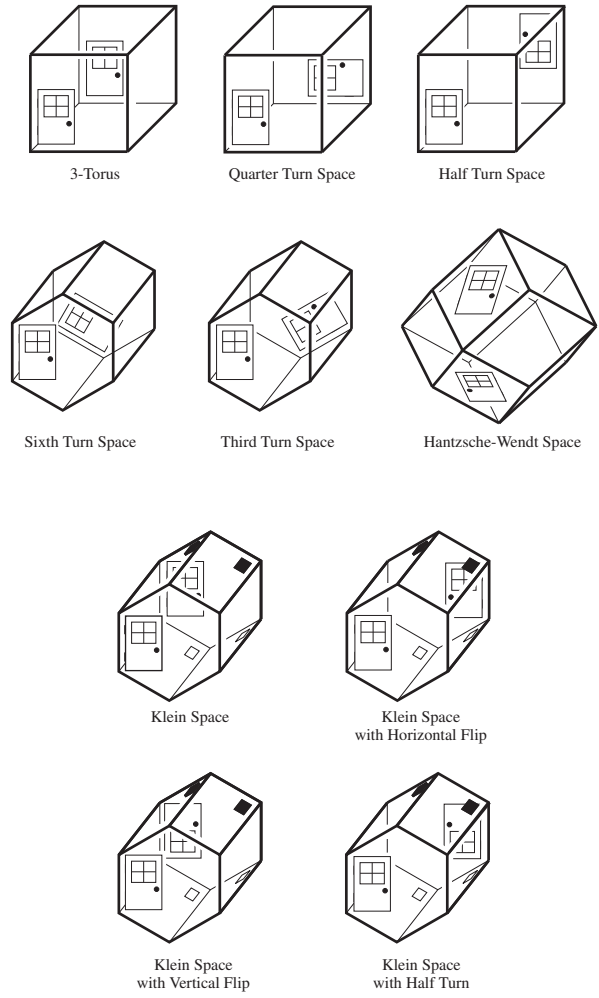,width=8cm}}
\caption{Fundamental domains for the compact flat three-manifolds.
The unmarked walls are glued straight across.  Courtesy of Adam Weeks
Marano (first published in Ref.~\cite{cipra}).}
\label{FigureCompactSpaces}
\end{figure}

\begin{figure}
\centerline{\psfig{file=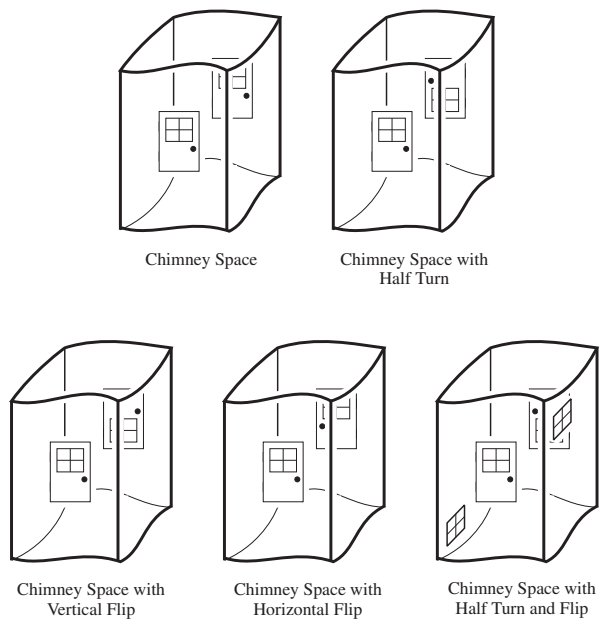,width=8cm}} 
\caption{The chimney space is made from an infinitely tall
rectangular chimney with front and back faces (resp. left and right
faces) glued straight across.  The four variations on the chimney
space space glue the front face to the back as indicated by the doors.
In all variations except the last the left and right faces are glued
straight across; in the last variation they are glued with a
top-to-bottom flip so that the windows match.  (Courtesy of Adam Weeks
Marano)}
\label{FigureChimneySpaces}
\end{figure}

\begin{figure}
\centerline{\psfig{file=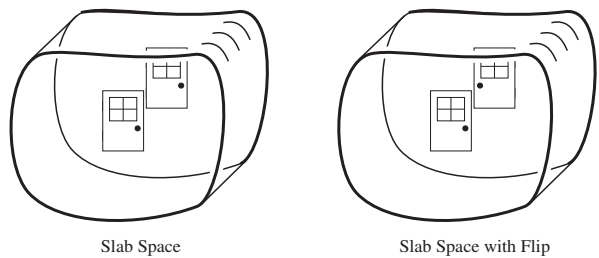,width=8cm}} 
\caption{The slab space is made from an infinitely tall and wide slab
of space with its front face glued to its back face straight across.
The variation glues the faces with a flip.  (Courtesy of Adam Weeks
Marano)}
\label{FigureSlabSpaces}
\end{figure}

Let us start by recalling the list of flat spaces. They are obtained
as the quotient $\ETR / \Gamma$ of three-dimensional Euclidean space
$\ETR$ by a group $\Gamma$ of symmetries of $\ETR$ that is discrete
and fixed point free.  The classification of such spaces has long been
known~\cite{feodoroff85,bierbach11}, motivated by the study of
crystallography and completed in 1934~\cite{novacki34}.  The ten
compact flat spaces are quotients of the 3-torus; six are orientable
and four are non-orientable.  Fig.~\ref{FigureCompactSpaces} shows
fundamental polyhedra. The non-compact spaces form two families, the
chimney space and its quotients having two compact directions
(Fig.~\ref{FigureChimneySpaces}) and the slab space and its quotient
having only one compact direction (Fig.~\ref{FigureSlabSpaces}). The
terms {\it slab space} and {\it chimney space} were coined by Adams
and Shapiro in their beautiful exposition of the flat
three-dimensional topologies~\cite{toponames}.  Table~\ref{tab1}
summarizes the properties of the whole family of flat spaces.

\begin{table}
\vskip0.25cm
\begin{tabular}{|c|c|c|c|}
\hline 
$\quad$ Symbol $\quad$ & Name & $\quad$ \# Compact Directions $\quad$
& $\quad$ Orientable $\quad$ \\
\hline\hline 
$E_{1}$ & 3-torus & 3 & Yes \\ 
$E_{2}$ & half turn space & 3 & Yes \\ 
$E_{3}$ & quarter turn space & 3 & Yes \\ 
$E_{4}$ & third turn space & 3 & Yes \\ 
$E_{5}$ & sixth turn space & 3 & Yes \\ 
$E_{6}$ & Hantzsche-Wendt space & 3 & Yes \\ 
\hline 
$E_{7}$ & Klein space & 3 & No \\ 
$E_{8}$ & Klein space with horizontal flip & 3 & No \\ 
$E_{9}$ & Klein space with vertical flip & 3 & No \\ 
$E_{10}$ & Klein space with half turn & 3 & No \\ 
\hline\hline 
$E_{11}$ & chimney space & 2 & Yes \\ 
$E_{12}$ & chimney space with half turn & 2 & Yes \\ 
\hline 
$E_{13}$ & chimney space with vertical flip & 2 & No \\
$E_{14}$ & chimney space with horizontal flip & 2 & No \\ 
$E_{15}$ & chimney space with half turn and flip & 2 & No \\
\hline\hline 
$E_{16}$ & slab space & 1 & Yes \\ 
\hline 
$E_{17}$ & slab space with flip & 1 & No \\ 
\hline\hline 
$E_{18}$ & Euclidean space & 0 & Yes \\ 
\hline
\end{tabular}
\caption{Classification of the 18 three-dimensional flat spaces.}
\label{tab1}
\end{table}

\section{Eigenmodes of $\ETR$}\label{sec2}

The eigenmodes of Euclidean space $\ETR$ admit two different bases: a
basis of planar waves and a basis of spherical waves.  The former are
more convenient when seeking eigenbases for multi-connected spaces,
while the latter are more convenient for simulating CMB maps.  This
section considers both bases and the conversion between them, as also
detailed in the particular case of the torus in Ref.~\cite{rulw02}.

\subsection{Planar waves}
\label{SectionPlanarWaves}

Each vector $\bk$ defines a planar wave
\begin{equation}
\label{PlanarWave}
\Upsilon_\bk (\bx) = \ee{i \bk \cdot \bx} .
\end{equation}
The defining vector $\bk$, called the {\it wave vector}, lives in the
dual space, so the dot product $\bk \cdot \bx$ is always
dimensionless. These modes are indeed not square integrable and are
normalized as
\begin{equation}\label{norm1}
\int_{\RTR} \Upsilon_\bk (\bx) \Upsilon_{\bk'}^* (\bx)
            \frac{\dd^3 \bx}{(2 \pi)^3}
 = \DIR{\bk - \bk'} .
\end{equation}

\subsection{Spherical waves}

Each spherical wave factors into a radial part and an angular part,
\begin{equation}
\label{SphericalWave}
{\cal Y}_{k \ell m} (\chi, \theta, \varphi)
 = \sqrt{\frac{2}{\pi}} \,
   (2 \pi)^{3 / 2} \,
    j_\ell (k \chi) \,
    Y_\ell^m (\theta, \varphi) ,
\end{equation}
where $(\chi, \theta, \varphi)$ are the usual spherical coordinates
\begin{eqnarray}
\label{SphericalCoordinates}
x & = & \chi \; \sin{\theta} \; \cos{\varphi} \nonumber \\
y & = & \chi \; \sin{\theta} \; \sin{\varphi} \nonumber \\
z & = & \chi \; \cos{\theta} .
\end{eqnarray}
The radial factor $j_\ell (k \chi)$ is the spherical Bessel function
of index $\ell$, and the angular factor $Y_\ell^m (\theta, \varphi)$
is the standard spherical harmonic.  The mode ${\cal Y}_{k \ell m}$ is
not square integrable and is normalized according to
\begin{equation}
\int_{\RTR} {\cal Y}_{k \ell m} {\cal Y}_{k' \ell' m'}^*
            \frac{\chi^2 \dd \chi \dd \cos \theta \dd \varphi}{(2 \pi)^3}
 = \frac{1}{k^2} \DIR{k - k'} \KRON{\ell}{\ell'} \KRON{m}{m'} ,
\end{equation}
which is analogous to the normalization (\ref{norm1}) and which
determines the numerical coefficient $\sqrt{2 / \pi}$.

\subsection{Conversion}

Subsequent sections will find explicit bases for the eigenmodes of
multi-connected flat three-manifolds as linear combinations of planar
waves.  The planar waves may easily be converted to spherical waves
using Eqns 5.17.3.14 and 5.17.2.9 of Ref.~\cite{vmk}:
\begin{eqnarray}
\label{Conversion}
\Upsilon_\bk (\bx)
 & = & \ee{i \bk \cdot \bx} \nonumber \\
 & = & \sum_{\ell = 0}^\infty \;
            i^\ell \; j_\ell (k \, |\bx|) \; (2\ell + 1) \; 
            P_\ell (\cos{\theta_{\bk, \bx}}) \nonumber \\
 & = & \sum_{\ell = 0}^\infty \;
            i^\ell \; j_\ell (k \, |\bx|) \;
            \left(4 \pi \sum_{m = - \ell}^{\ell}
                             Y_\ell^m ({\bf \hat x}) Y_\ell^{m*} ({\bf \hat k})
            \right) \nonumber \\
 & = & \sum_{\ell = 0}^\infty \,
            \sum_{m = - \ell}^{\ell} \;
                 i^\ell \; Y_\ell^{m*} ({\bf \hat k}) \;
                 \left[ \; 4 \pi \; j_\ell (k \, |\bx|) \;
                        Y_\ell^m ({\bf \hat x}) \; \right] \nonumber \\
 & = & \sum_{\ell = 0}^\infty \,
            \sum_{m = - \ell}^{\ell} \;
                 \left( \; i^\ell \; Y_\ell^{m*} ({\bf \hat k}) \right) 
                 {\cal Y}_{k l m} (\bx) ,
\end{eqnarray}
where $k = |\bk|$, ${\bf\hat k} \equiv \bk / |\bk|$, and ${\bf\hat x}
\equiv \bx / |\bx|$.

In particular, the conversion formula (\ref{Conversion}) lets one
easily translate a planar wave $\Upsilon_\bk$ to the framework we
developed in Ref.~\cite{rulw02}, which expresses each basis eigenmode
as a sum of spherical waves
\begin{equation}
\Upsilon_\bk
 = \Upsilon_{k, s}
 = \sum_{\ell = 0}^\infty 
        \sum_{m = - \ell}^{\ell}
             \xi^s_{k \ell m} {\cal Y}_{k \ell m} ,
\end{equation}
where $s$ indexes the different $\Upsilon_\bk$ whose wave vectors
$\bk$ share the same modulus $k$. In the Euclidean case the index may
be chosen to be $s = \hat\bk$. Comparison with (\ref{Conversion})
immediately gives the coefficients
\begin{equation}
\xi^{\hat \bk}_{k \ell m} = i^\ell \; Y_\ell^{m*} ({\bf \hat k}) .
\end{equation}

\section{Eigenmodes of Multi-Connected Spaces}
\label{SectionEigenmodesOfMulticonnected}

A multi-connected flat three-manifold is the quotient $\ETR / \Gamma$
of Euclidean space $\ETR$ under the action of a group $\Gamma$ of
isometries.  The group $\Gamma$ is called the {\it holonomy group} and
is always discrete and fixed point free. Each eigenmode $\hat\Upsilon$
of the multi-connected space $\ETR / \Gamma$ lifts to a
$\Gamma$-periodic eigenmode $\Upsilon$ of $\ETR$, that is, to an
eigenmode of $\ETR$ that is invariant under the action of the holonomy
group $\Gamma$.  Common practice blurs the distinction between
eigenmodes of $\ETR / \Gamma$ and $\Gamma$-periodic eigenmodes of
$\ETR$, and we follow that practice here.  Thus the task of finding
the eigenmodes of the multi-connected space $\ETR / \Gamma$ becomes
the task of finding the $\Gamma$-periodic eigenmodes of $\ETR$.  In
this section we investigate how an isometry $\gamma \in \Gamma$ acts
on the space of eigenmodes.  The two lemmas we obtain will make it
easy to determine the eigenmodes of specific multi-connected spaces in
subsequent sections.

Every isometry $\gamma$ of Euclidean space $\ETR$ factors as a
rotation/reflection followed by a translation.  If we write the
rotation/reflection as a $3 \times 3$ matrix $M$ in the orthogonal
group ${\rm O} (3)$ and write the translation as a vector $\bT$, then
$\gamma$ acts on $\ETR$ as
\begin{equation}
\VECTROISD{x}{y}{z}
\quad \mapsto \quad
   \MATTROISD{M_{x x}}{M_{x y}}{M_{x z}}
             {M_{y x}}{M_{y y}}{M_{y z}}
             {M_{z x}}{M_{z y}}{M_{z z}}
   \VECTROISD{x}{y}{z}
 + \VECTROISD{T_x}{T_y}{T_z} .
\end{equation}
This isometry of $\ETR$ induces a natural action $\Upsilon_\bk (\bx)
\mapsto \Upsilon_\bk (M \bx + \bT)$ on the space of eigenmodes. \\

\noindent {\bf Lemma 1 (Action Lemma).}  {\it The natural action of
an isometry $\gamma$ of $\ETR$ takes a planar eigenmode $\Upsilon_\bk$
to another planar eigenmode $\ee{i \bk \cdot \bT} \Upsilon_{\bk
M}$.} \\

\noindent {\it Proof.}  Keeping in mind that $\bk$ is a row vector
while $\bx$ is a column vector, the proof is an easy computation:
\begin{eqnarray}
\label{Lemma1Proof}
\Upsilon_\bk(\bx)
 & = &       \ee{i \bk \cdot \bx} \nonumber \\
 & \mapsto & \ee{i \bk \cdot (M\bx + \bT)} \nonumber \\
 & = &       \ee{i \bk \cdot \bT} \; \ee{i \bk M \bx} \nonumber \\
 & = &       \ee{i \bk \cdot \bT} \; \Upsilon_{\bk M}(\bx) .
\end{eqnarray}
\QED \\

\noindent {\bf Lemma 2 (Invariance Lemma).}  {\it If $\gamma$ is an
isometry of $\ETR$ with matrix part $M$ and translational part $T$,
the mode $\Upsilon_\bk$ is a planar wave, and $n$ is the smallest
positive integer such that $\bk = \bk M^n$ (typically $n$ is simply
the order of the matrix $M$), then the action of $\gamma$

\begin{enumerate}

\item preserves the $n$-dimensional space of eigenmodes spanned by
$\lbrace \Upsilon_\bk, \Upsilon_{\bk M}, \ldots, \Upsilon_{\bk
M^{n-1}} \rbrace$ as a set, and

\item fixes a specific element
\begin{equation}
   a_0 \Upsilon_\bk
 + a_1 \Upsilon_{\bk M}
 + \ldots
 + a_{n - 1} \Upsilon_{\bk M^{n - 1}} ,
\end{equation}
if and only if for each $j \MOD{n}$
\begin{equation}
a_{j + 1} = \ee{i \bk M^j \bT} a_j .
\end{equation}

\end{enumerate}
}

\noindent {\it Proof.} Both parts are immediate corollaries of Lemma
1. Specifically, the action of $\gamma$ takes a linear combination
\begin{equation}
\label{Lemma1ProofBefore}
   a_0 \Upsilon_\bk
 + a_1 \Upsilon_{\bk M}
 + \ldots
 + a_{n - 2} \Upsilon_{\bk M^{n - 2}}
 + a_{n - 1} \Upsilon_{\bk M^{n - 1}} ,
\end{equation}
to
\begin{equation}
\label{Lemma1ProofAfter}
   a_0 \ee{i \bk \bT} \Upsilon_{\bk M}
 + a_1 \ee{i \bk M \bT} \Upsilon_{\bk M^2}
 + \ldots
 + a_{n - 2} \ee{i \bk M^{n - 2} \bT} \Upsilon_{\bk M^{n - 1}}
 + a_{n - 1} \ee{i \bk M^{n - 1} \bT} \Upsilon_{\bk} ,
\end{equation}
so it's clear that the $n$-dimensional subspace spanned by $\lbrace
\Upsilon_\bk, \Upsilon_{\bk M}, \ldots, \Upsilon_{\bk M^{n-1}}
\rbrace$ is preserved as a set. Equating (\ref{Lemma1ProofBefore}) to
(\ref{Lemma1ProofAfter}) and comparing coefficients proves the second
part. \QED \\

\section{Compact Flat Three-Manifolds}
\label{SectionCompactSpaces}

We will first find the eigenmodes of the 3-torus, and then use them to
find the eigenmodes of the remaining compact flat three-manifolds.

\subsection{3-Torus}
\label{SubsectionThreeTorus}

\begin{figure}
\centerline{\psfig{file=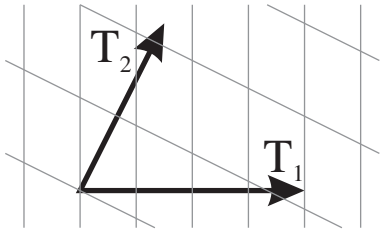,width=5cm}}
\caption{An eigenmode $\Upsilon_\bk$ of $\EDE$ is invariant under the
translation $\bT_i$ if and only if $\bk$ lies on a family of parallel
lines orthogonal to $\bT_i$, defined by $\bk \cdot \bT_i \in 2 \pi Z$.
The mode $\Upsilon_\bk$ is invariant under both translations $\bT_1$
and $\bT_2$ if and only if it lies in the lattice of intersection
points of the two families of parallel lines.  The construction in
three dimensions is similar, but with three families of planes instead
of two families of lines.  Strictly speaking we should not draw the
parallel lines in the same space as the translation vectors $\bT_i$
because the wave vectors $\bk$ live in the dual space (with units of
$\UUNIT{length}{-1}$) while the $\bT_i$ live in the primary space
(with units of $\UUNIT{length}{}$), but nevertheless it's visually
helpful to do so.}
\label{FigureTorusLattice}
\end{figure}

The {\it 3-torus} is the quotient of Euclidean space $\ETR$ under the
action of three linearly independent translations $\bT_1$, $\bT_2$ and
$\bT_3$. Its fundamental domain is a parallelepiped.  Its eigenmodes
are the eigenmodes of $\ETR$ invariant under the translations $\bT_1$,
$\bT_2$ and $\bT_3$ (recall from
Section~\ref{SectionEigenmodesOfMulticonnected} the convention that
eigenmodes of the quotient are represented as periodic eigenmodes of
$\ETR$). The Invariance Lemma (with $n = 1$) implies that an eigenmode
$\Upsilon_\bk$ of $\ETR$ is invariant under the translation $\bT_1$ if
and only if $\ee{i \bk \cdot \bT_1} = 1$ which occurs precisely when
$\bk \cdot \bT_1$ is an integer multiple of $2 \pi$. Thus
geometrically the allowed values of the wave vector $\bk$ form a
family of parallel planes orthogonal to $\bT_1$. Similarly, the
eigenmode $\Upsilon_\bk$ is invariant under the translation $\bT_2$
(resp.\ $\bT_3$) if and only if $\bk$ lies on a family of parallel
planes orthogonal to $\bT_2$ (resp.\ $\bT_3$), defined by $\bk \cdot
\bT_2 \in 2 \pi Z$ (resp.\ $\bk \cdot \bT_3 \in 2 \pi Z$).  The
eigenmode $\Upsilon_\bk$ is invariant under all three translations
$\bT_1$, $\bT_2$ and $\bT_3$ if and only if it lies on all three
families of parallel planes simultaneously. The intersection of the
three families forms a lattice of discrete points.
Fig.~\ref{FigureTorusLattice} illustrates the construction for the
2-torus; the construction for the 3-torus is analogous. This lattice
of points defines the standard basis for the eigenspace of a torus.\\

\noindent {\bf Definition.}  The {\it standard basis} for a 3-torus
$\TTR$ generated by three linearly independent translations $\bT_1$,
$\bT_2$ and $\bT_3$ is the set $B = \lbrace \Upsilon_\bk \; | \; \bk
\cdot \bT_i \in 2 \pi Z \mbox{ for } i = 1, 2, 3 \rbrace$. \\

\begin{figure}
\centerline{\psfig{file=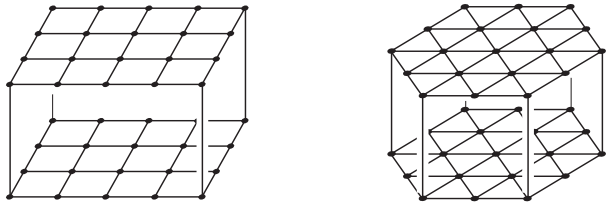,width=8cm}} 
\caption{In a rectangular 3-torus (left) the allowed wave vectors
$\bk$ form a rectangular lattice.  In a hexagonal 3-torus (right) the
lattice is hexagonal within each layer.}
\label{FigureHoneycomb}
\end{figure}

The most important special case of a 3-torus is the {\it rectangular
3-torus} generated by three mutually orthogonal translations
\begin{eqnarray}
\bT_1 & = & (L_x, 0, 0) \nonumber \\
\bT_2 & = & (0, L_y, 0) \nonumber \\
\bT_3 & = & (0, 0, L_z) ,
\end{eqnarray}
in which case the allowed wave vectors $\bk$ take the form
\begin{equation}
\label{RectangularTorusBasis}
\bk = 2 \pi \, \left(\frac{n_x}{L_x}, \, 
                     \frac{n_y}{L_y}, \, 
                     \frac{n_z}{L_z} \right) ,
\end{equation}
for integer values of $n_x$, $n_y$ and $n_z$, thus forming a
rectangular lattice (Fig.~\ref{FigureHoneycomb} left).

The second most important special case is the {\it hexagonal 3-torus}
generated by
\begin{eqnarray}
\label{HexagonalGenerators}
\bT_1 & = & \left(L, 0, 0 \right) \nonumber \\
\bT_2 & = & \left(- \frac{1}{2} L, 
                  + \frac{\sqrt{3}}{2} L, 
                  0 \right) \nonumber \\
\bT_3 & = & \left(- \frac{1}{2} L,
                  - \frac{\sqrt{3}}{2} L, 
                  0 \right) \nonumber \\
\bT_4 & = & \left(0, 0, L_z\right) .
\end{eqnarray}
These four generators and their inverses define a fundamental domain
as a hexagonal prism.  The first three generators are linearly
dependent ($\bT_1 + \bT_2 + \bT_3 = 0$); eliminating any one of them
suggests an alternative fundamental domain as a prism with a rhombic
base.  Even though the hexagonal and rhombic prisms look different,
they define the same manifold.  Either way, the allowed wave vectors
$\bk$ form a hexagonal lattice (Fig.~\ref{FigureHoneycomb} right) and
may be parameterized as
\begin{equation}
\label{HexagonalTorusBasis}
\bk \quad = \quad 2 \pi \, \left(- \frac{n_1}{L}, 
                                   \frac{2 n_1 - n_2}{\sqrt{3} L},
                                   \frac{n_3}{L_z} \right) ,
\end{equation}
for integer values of $n_1$, $n_2$ and $n_3$.

In the case of a general 3-torus, one writes the translations $\bT_1$,
$\bT_2$ and $\bT_3$ as the columns of a $3 \times 3$ matrix $T$ and
solves the equation $\bk T = 2 \pi (n_1, n_2, n_3)$ to find the
allowable wave vectors $\bk = 2 \pi (n_1, n_2, n_3) T^{-1}$ for
integers $n_1$, $n_2$ and $n_3$. \\

When one wants to simulate CMB maps, one needs to know not only the
modes themselves but also how the modes are paired under complex
conjugation. The reason is that the cosmological fields are in fact
real-valued stochastic variables. Any such field can be decomposed
into Fourier modes as
\begin{equation}
\phi(\bx, t) = \int \frac{\dd^3 \bk}{(2 \pi)^{3 / 2}} \phi_\bk (t)
\ee{i \bk . \bx} \hat e_\bk ,
\end{equation}
where $\hat e_\bk$ is a complex, usually Gaussian, random variable
satisfying
\begin{equation}
\left<\hat e_\bk \hat e_{\bk'}^* \right> = \DIR{\bk - \bk'} .
\end{equation}
The evolution equations of the cosmological perturbations involve time
derivatives and a Laplacian so that the coefficient $\phi_\bk (t)$ can be
decomposed as
\begin{equation}
\phi_\bk (t) = \phi_k (t) \ee{i \theta_\bk} ,
\end{equation}
where $\theta_\bk$ is a phase that is constant throughout the
evolution. By absorbing the phase into the random variable, we can
always choose $\phi_\bk$ to be a real function of $k$ only, i.e.,
$\phi_k(t)$.  Since $\ee{i \bk \cdot \bx}$ and $\ee{- i \bk \cdot
\bx}$ are conjugate in the preceding decomposition, the fact that
$\phi (\bx, t)$ is real implies that
\begin{equation}
\label{random_relation}
\hat e_\bk^* = \hat e_{- \bk} .
\end{equation}
This latter constraint may not hold for all the other spaces studied
in this article and we will need to give its analog for each case.

\subsection{Quotients of the 3-Torus}

\begin{figure}
\centerline{\psfig{file=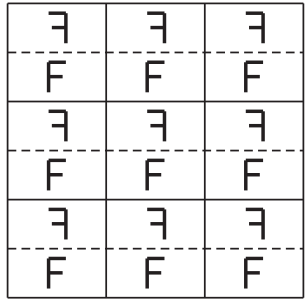,width=4cm}}
\caption{The Klein bottle's holonomy group contains glide reflections
as well as translations.  The translations alone form an index 2
subgroup defining a torus.}
\label{FigureKleinBottle}
\end{figure}

For ease of illustration we first explain our general method for the
two-dimensional Klein bottle.  Figure~\ref{FigureKleinBottle} shows a
portion of the Klein bottle's universal cover, in which alternate
images of the fundamental domain appear mirror reversed.  Half the
holonomies are pure translations while the other half are glide
reflections.  In other words, the holonomy group $\Gamma$ contains an
index 2 subgroup $\Gamma' \subset \Gamma$ comprising the pure
translations.  While $\EDE / \Gamma$ gives the original Klein bottle,
$\EDE / \Gamma'$ gives a torus whose fundamental domain is the square
formed by the solid lines in Figure~\ref{FigureKleinBottle} (ignoring
the dotted lines).  According to the convention introduced in
Section~\ref{SectionEigenmodesOfMulticonnected}, the Klein bottle's
eigenmodes are represented as $\Gamma$-periodic functions on $\EDE$.
But every $\Gamma$-periodic function is automatically a
$\Gamma'$-periodic function as well, because $\Gamma'$ is a subgroup
of $\Gamma$.  Thus every eigenmode of the Klein bottle is {\it a
priori} an eigenmode of the 2-torus.  The task in finding the
eigenspace of the Klein bottle is to start with the eigenspace of the
torus and find the subspace that is invariant under the glide
reflection (the one taking the lower half of a square to the upper
half).  In practice this is quite simple.  A rectangular torus has
holonomy group $\Gamma'$ generated by the two translations
\begin{equation}
\VECDEUXD{x}{y} \; \mapsto \; \VECDEUXD{x}{y} + \VECDEUXD{L_x}{0}
\qquad \mbox{and} \qquad
\VECDEUXD{x}{y} \; \mapsto \; \VECDEUXD{x}{y} + \VECDEUXD{0}{L_y} .
\end{equation}
The standard eigenbasis for this torus takes the
form~(\ref{RectangularTorusBasis}), namely $B = \lbrace \Upsilon_\bk
\; | \; \bk = 2 \pi (\frac{n_x}{L_x}, \frac{n_y}{L_y}) \mbox{ for }
n_x, n_y \in Z \rbrace$.  To extend this $\Gamma'$ to the full
holonomy group $\Gamma$ of the Klein bottle, we add the glide
reflection
\begin{equation}
\VECDEUXD{x}{y} 
\; \mapsto \; 
\MATDEUXD{- 1}{0}{0}{1} \VECDEUXD{x}{y} + \VECDEUXD{0}{L_y / 2} , 
\end{equation}
and ask which elements of the basis $B$ it preserves.  The Invariance
Lemma provides a ready answer: when $k_x \neq 0$ the two-dimensional
subspace $\lbrace \Upsilon_{k_x, k_y}, \Upsilon_{- k_x, k_y} \rbrace$
is preserved as a set (Part~1 of the Invariance Lemma) while the mode
$\Upsilon_{k_x, k_y} + (-1)^{n_y} \Upsilon_{- k_x, k_y}$ is fixed
exactly (Part~2 of the Invariance Lemma). In the exceptional case that
$k_x = 0$, the one-dimensional subspace $\lbrace \Upsilon_{0, k_y}
\rbrace$ is preserved as a set, while $\Upsilon_{0, k_y}$ is fixed if
and only if $n_y$ is even (because when $n = 1$, Part~2 of the
Invariance Lemma requires $a_0 = \ee{i \bk \cdot \bT} a_0$ which
implies $\bk \cdot \bT / (2 \pi) = n_z / 2 \in Z$). In summary, an
orthonormal basis for the space of eigenmodes of the Klein bottle is
the union of
\begin{equation}
\begin{array}{lcl}
\left[  \Upsilon_{2 \pi (\frac{n_x}{L_x}, \frac{n_y}{L_y})}
      + (- 1)^{n_y} \Upsilon_{2 \pi (- \frac{n_x}{L_x}, \frac{n_y}{L_y})}
\right] / \sqrt{2}
 & \quad \mbox{for} \quad
 & n_x \in Z^+, n_y \in Z, \\
\Upsilon_{2 \pi (0, \frac{n_y}{L_y})}
 & \quad \mbox{for} \quad
 & n_y \in 2 Z . \\
\end{array}
\end{equation} \\

Let us now apply this same method to each of the nine quotients of the
3-torus.

\subsubsection{Half turn space}

The analysis of the half turn space closely follows that of the Klein
bottle given immediately above.  The only difference is that the Klein
bottle's holonomy group contained translations and glide reflections,
while the half turn space's holonomy group contains translations and
corkscrew motions.  Specifically, we begin with the generators for the
holonomy group $\Gamma'$ of a rectangular 3-torus
\begin{equation}
\VECTROISD{x}{y}{z} \; \mapsto \; \VECTROISD{x}{y}{z} + \VECTROISD{L_x}{0}{0}
, \qquad
\VECTROISD{x}{y}{z} \; \mapsto \; \VECTROISD{x}{y}{z} + \VECTROISD{0}{L_y}{0}
, \qquad
\VECTROISD{x}{y}{z} \; \mapsto \; \VECTROISD{x}{y}{z} + \VECTROISD{0}{0}{L_z} .
\end{equation}
and add a generator for the half-turn corkscrew motion
\begin{equation}
\VECTROISD{x}{y}{z}
\; \mapsto \;
\MATTROISD{- 1}{0}{0}
          {0}{- 1}{0}
          {0}{0}{\;\;\; 1} \VECTROISD{x}{y}{z} + \VECTROISD{0}{0}{L_z / 2} , 
\end{equation}
to get the full holonomy group $\Gamma$ of the half turn space.  The
Invariance Lemma shows that when $(k_x, k_y) \neq (0, 0)$ the
two-dimensional subspace $\lbrace \Upsilon_{k_x, k_y, k_z},
\Upsilon_{- k_x, - k_y, k_z} \rbrace$ is preserved as a set (Part~1 of
the Invariance Lemma) while the mode $\Upsilon_{k_x, k_y, k_z} +
(-1)^{n_z} \Upsilon_{- k_x, - k_y, k_z}$ is fixed exactly (Part~2 of
the Invariance Lemma). In the exceptional case that $(k_x, k_y) = (0,
0)$, the one-dimensional subspace $\lbrace \Upsilon_{0, 0, k_z}
\rbrace$ is preserved as a set, while $\Upsilon_{0, 0, k_z}$ is fixed
if and only if $n_z$ is even. In summary, an orthonormal basis for the
space of eigenmodes of the half turn space, $\Upsilon_{k_x, k_y,
k_z}^{[E_2]}$, is the union of
\begin{equation}
   \begin{array}{lcl}
     \frac{1}{\sqrt{2}}
       \left[\Upsilon_{2\pi(\frac{n_x}{L_x},\frac{n_y}{L_y},\frac{n_z}{L_z})}
        + (-1)^{n_z} \Upsilon_{2\pi(-\frac{n_x}{L_x},-\frac{n_y}{L_y},\frac{n_z}{L_z})}
       \right]
       & \quad \mbox{for} \quad
       & (n_x \in Z^+, n_y, n_z \in Z) \mbox{ or } (n_x = 0, n_y \in Z^+, n_z \in Z), \\
     \Upsilon_{2\pi(0,0,\frac{n_z}{L_z})}
       & \quad \mbox{for} \quad
       & n_z \in 2Z . \\
   \end{array}
\end{equation}
In terms of the notations used in Ref.~\cite{rulw02}, it leads to the
coefficients
\begin{equation}
   \xi_{k\ell m}^{\hat\bk}=\left\lbrace
   \begin{array}{lcl}
     \frac{i^\ell}{\sqrt{2}}
       \left[Y_\ell^{m*}(\hat\bk)+ (-1)^{n_z}Y_\ell^{m*}(\hat\bk M)
       \right]
       & \quad \mbox{for} \quad
       & (n_x \in Z^+, n_y, n_z \in Z) \mbox{ or } (n_x = 0, n_y \in Z^+, n_z \in Z), \\
       i^\ell Y_\ell^{m*}(\hat\bk)
       & \quad \mbox{for} \quad
       & (n_x,n_y)=(0,0),\quad n_z \in 2Z . \\
   \end{array}\right.,
\end{equation}
$\bk$ being given by Eq.~(\ref{RectangularTorusBasis}). Passing from
the expression of the modes to the coefficients $\XIKSLM{k}{\hat
\bk}{\ell}{m}$ is straightforward and in the following we will give
only the expressions of the modes.

If desired, one could construct a more general half turn space from a
right prism with a parallelogram base, instead of a rectangular box.

To find the analog of Eq.~(\ref{random_relation}), one simply
needs to check that
\begin{equation}
\Upsilon_{k_x, k_y, k_z}^{[E_2]*}
 = (- 1)^{n_z} \Upsilon_{k_x, k_y, - k_z}^{[E_2]} ,
\end{equation}
so that it follows that
\begin{enumerate}

\item when $k_z \not = 0$, $\hat e_\bk$ is a complex random
variable satisfying
\begin{equation}
\hat e_{k_x, k_y, k_z}^* = (- 1)^{n_z} \hat e_{k_x, k_y, - k_z} ,
\end{equation}

\item when $k_z = 0$, $\hat e_\bk$ is a real random variable.

\end{enumerate}
\begin{figure}
\centerline{\psfig{file=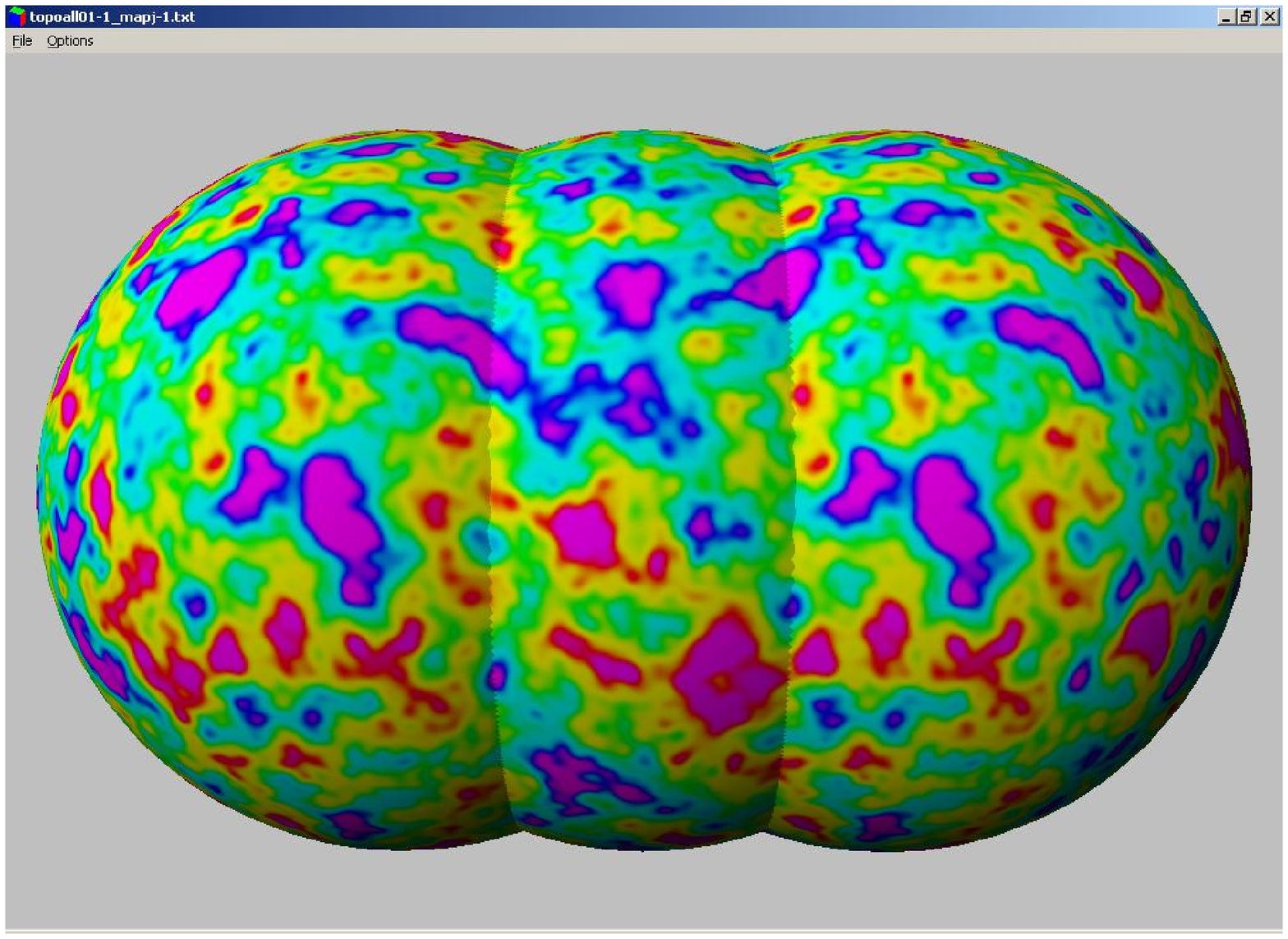,width=3.5in}
            \psfig{file=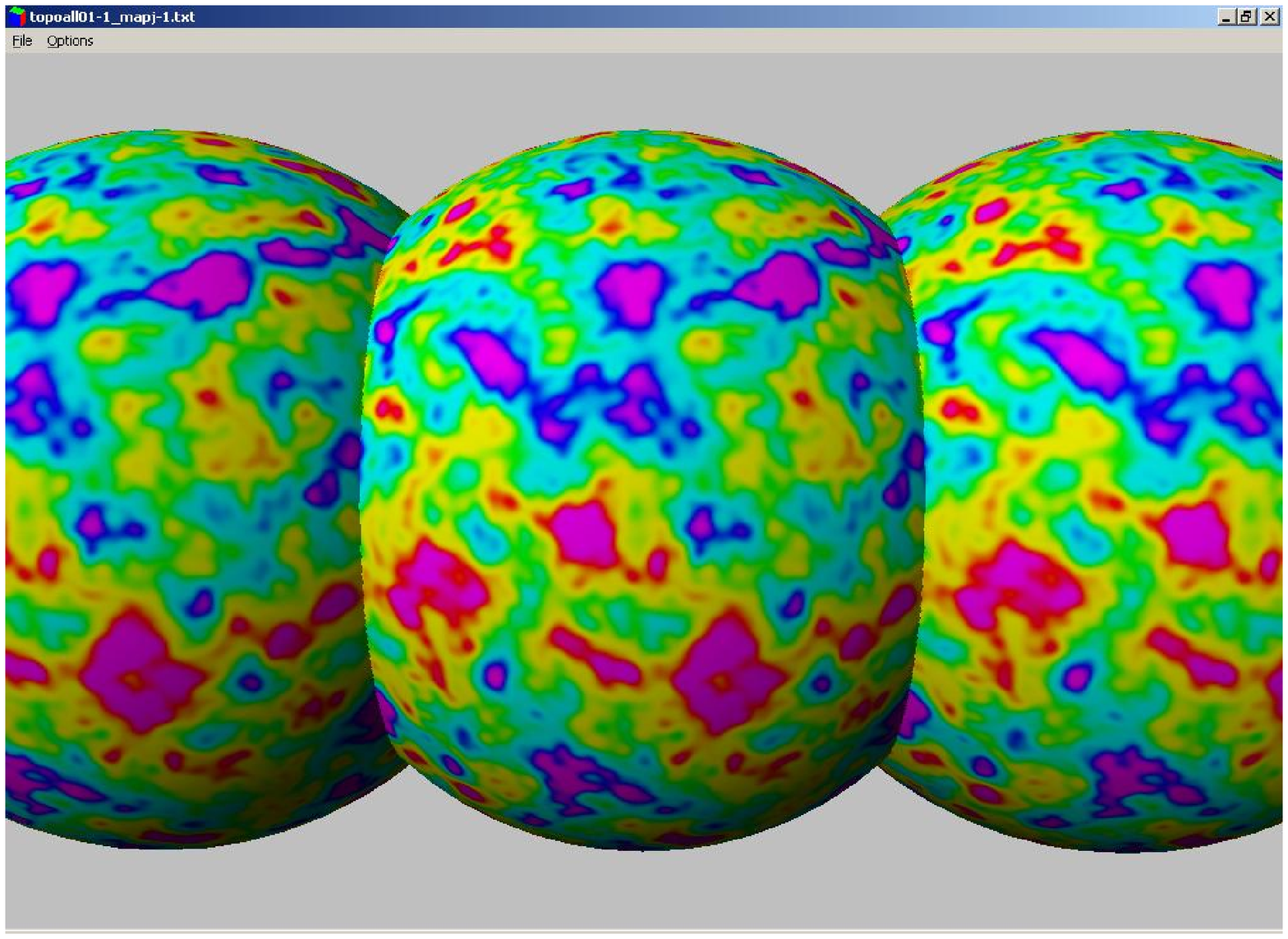,width=3.5in}}
\centerline{\psfig{file=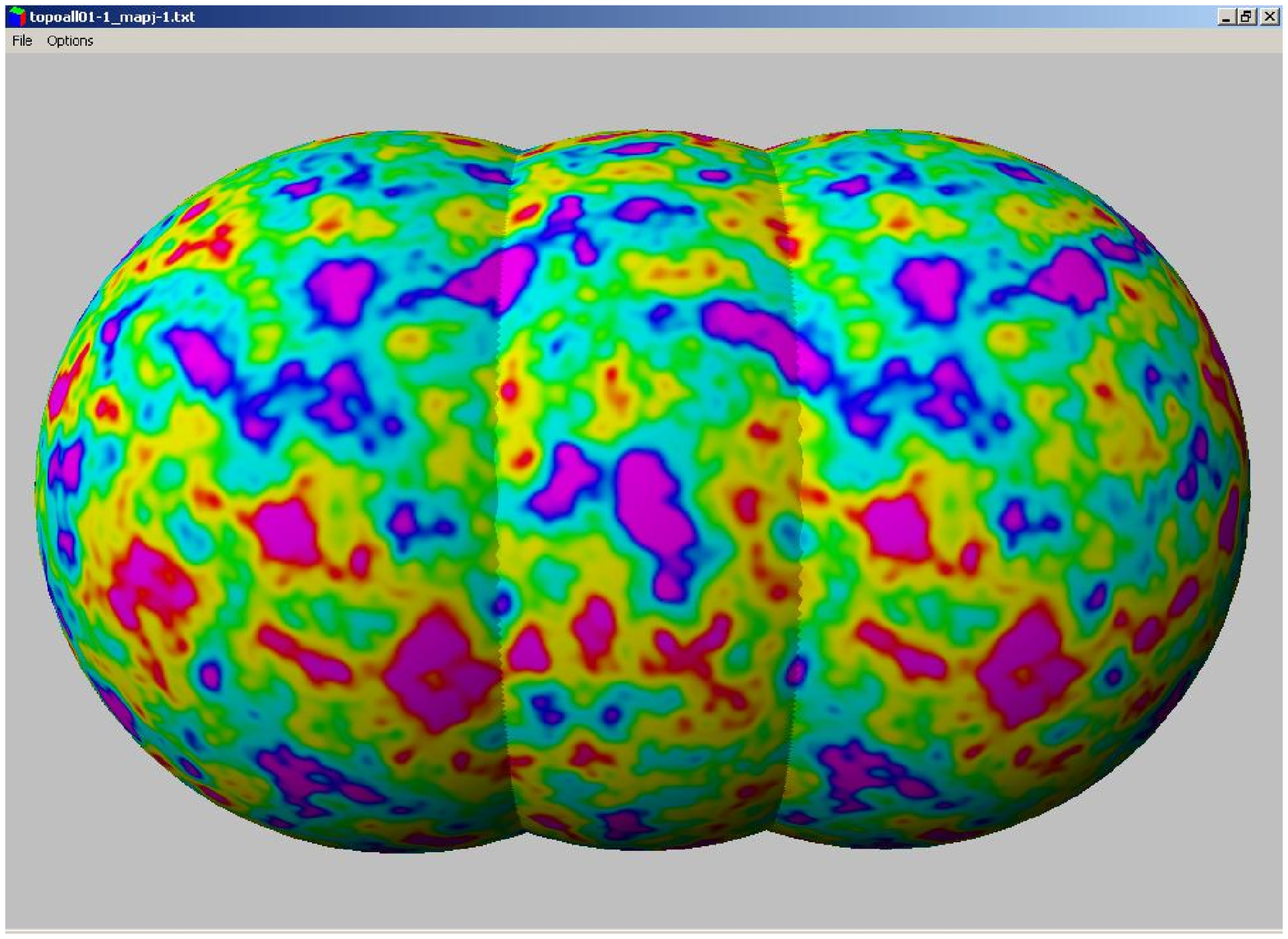,width=3.5in}
            \psfig{file=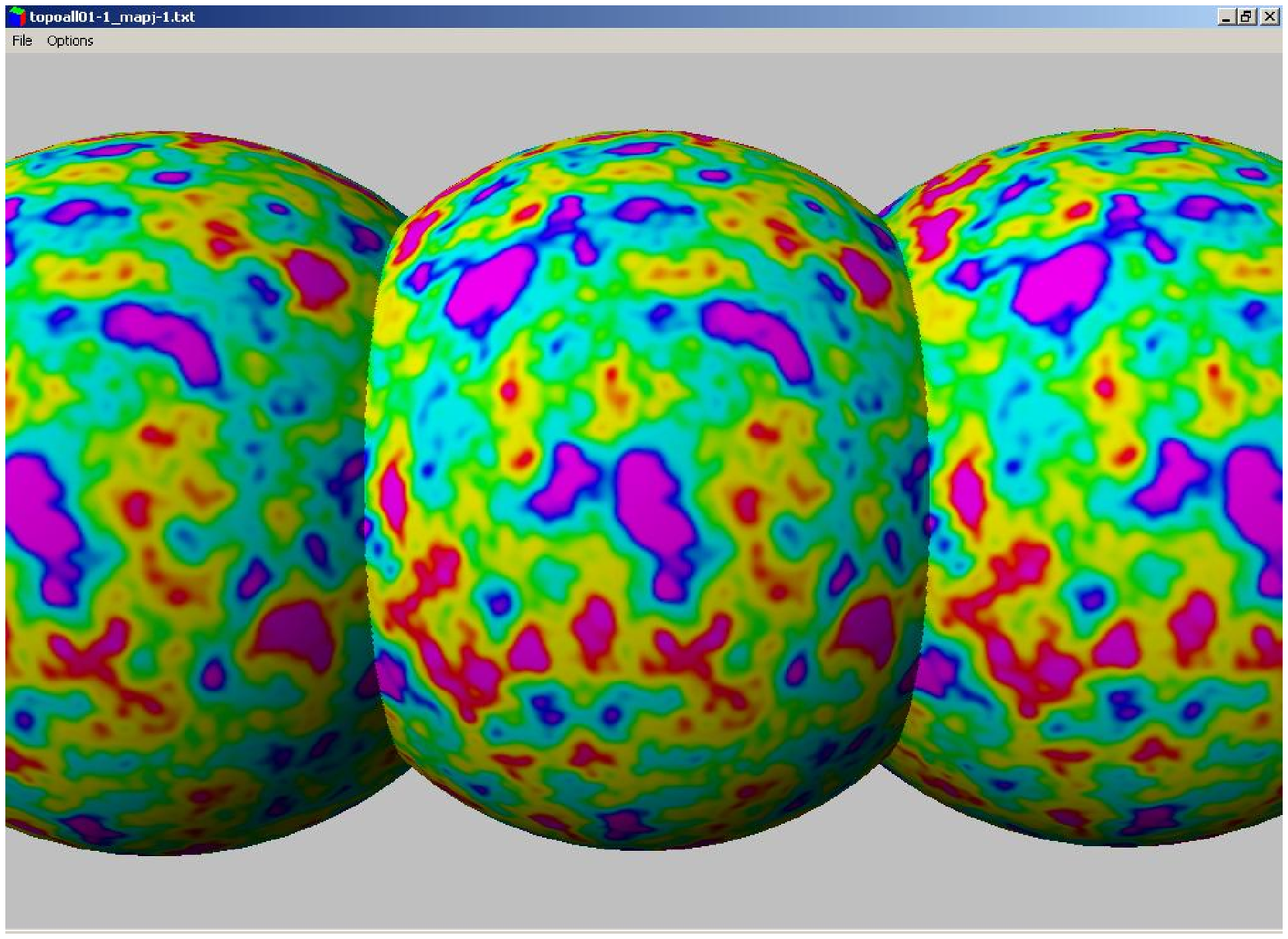,width=3.5in}}
\caption{The last scattering surface seen from outside for a
half-turn space $E_2$ with $L_x = L_y = 0.64$, $L_z = 1.28$ in units
of the radius of the last scattering surface. Each row presents the
two pairs of matching circles along the $z$ direction, but seen from
opposite directions. The left panels show circles which match with a
$\pi$ rotation while the right panel shows circles which match without
rotation. We recover the invariance by a translation of $L_z$ due to
the fact that the cubic torus of size $L_z = 1.28$ is the double cover
of the half-turn space considered here. Only the Sachs-Wolfe
contribution has been depicted here.}
\label{plot1}
\end{figure}

\subsubsection{Quarter turn space}

The quarter turn space is similar to the half turn space, but with
with a quarter turn corkscrew motion
\begin{equation}
\VECTROISD{x}{y}{z}
\; \mapsto \;
\MATTROISD{0}{- 1}{0}
          {1}{0}{0}
          {0}{0}{1} \VECTROISD{x}{y}{z} + \VECTROISD{0}{0}{L_x / 4} .
\end{equation}
In particular this implies that $L_x = L_y$. The Invariance Lemma
shows that when $(k_x, k_y) \neq (0, 0)$ the four-dimensional subspace
$\lbrace \Upsilon_{k_x, k_y, k_z}$, $\Upsilon_{k_y, - k_x, k_z}$,
$\Upsilon_{- k_x, - k_y, k_z}$, $\Upsilon_{- k_y, k_x, k_z} \rbrace$
is preserved as a set, while the mode $\Upsilon_{k_x, k_y, k_z} +
i^{n_z} \Upsilon_{k_y, - k_x, k_z} + (-1)^{n_z} \Upsilon_{- k_x, -
k_y, k_z} + (-i)^{n_z} \Upsilon_{- k_y, k_x, k_z}$ is fixed exactly.
In the exceptional case that $(k_x, k_y) = (0, 0)$, the
one-dimensional subspace $\lbrace \Upsilon_{0, 0, k_z} \rbrace$ is
preserved as a set, while $\Upsilon_{0, 0, k_z}$ is fixed if and only
if $n_z$ is a multiple of 4. In summary, an orthonormal basis for the
space of eigenmodes of the quarter turn space, $\Upsilon_{k_x, k_y, -
k_z}^{[E_3]}$, is the union of
\begin{eqnarray}
     & & \frac{1}{2}
       \left[        \Upsilon_{2\pi ( \frac{n_x}{L_x}, \frac{n_y}{L_y}, \frac{n_z}{L_z})}
          + i^{  n_z} \Upsilon_{2\pi ( \frac{n_y}{L_y},-\frac{n_x}{L_x}, \frac{n_z}{L_z})}
          + i^{2 n_z} \Upsilon_{2\pi (-\frac{n_x}{L_x},-\frac{n_y}{L_y}, \frac{n_z}{L_z})}
          + i^{3 n_z} \Upsilon_{2\pi (-\frac{n_y}{L_y}, \frac{n_x}{L_x}, \frac{n_z}{L_z})}
       \right]  \nonumber\\
     & & \qquad\qquad \mbox{for} \quad
         n_x \in Z^+, n_y \in Z^+ \cup \lbrace 0 \rbrace, n_z \in Z, \nonumber\\
     & & \Upsilon_{2\pi (0,0,\frac{n_z}{L_z})}
        \quad  \mbox{for}  \quad n_z \in 4Z.
\end{eqnarray}

As in the case of the half turn space, one can easily check that
\begin{equation}
\Upsilon_{k_x, k_y, k_z}^{[E_3]*}
 = (- 1)^{n_z} \Upsilon_{k_x, k_y, - k_z}^{[E_3]} ,
\end{equation}
so that the analog of Eq.~(\ref{random_relation}) is given by
\begin{enumerate}

\item when $k_z \not = 0$, $\hat e_\bk$ is a complex random variable
satisfying
\begin{equation}
\hat e_{k_x, k_y, k_z}^* = (- 1)^{n_z} \hat e_{k_x, k_y, - k_z} ,
\end{equation}

\item when $k_z = 0$, $\hat e_\bk$ is a real random variable.

\end{enumerate}
\begin{figure}
\centerline{\psfig{file=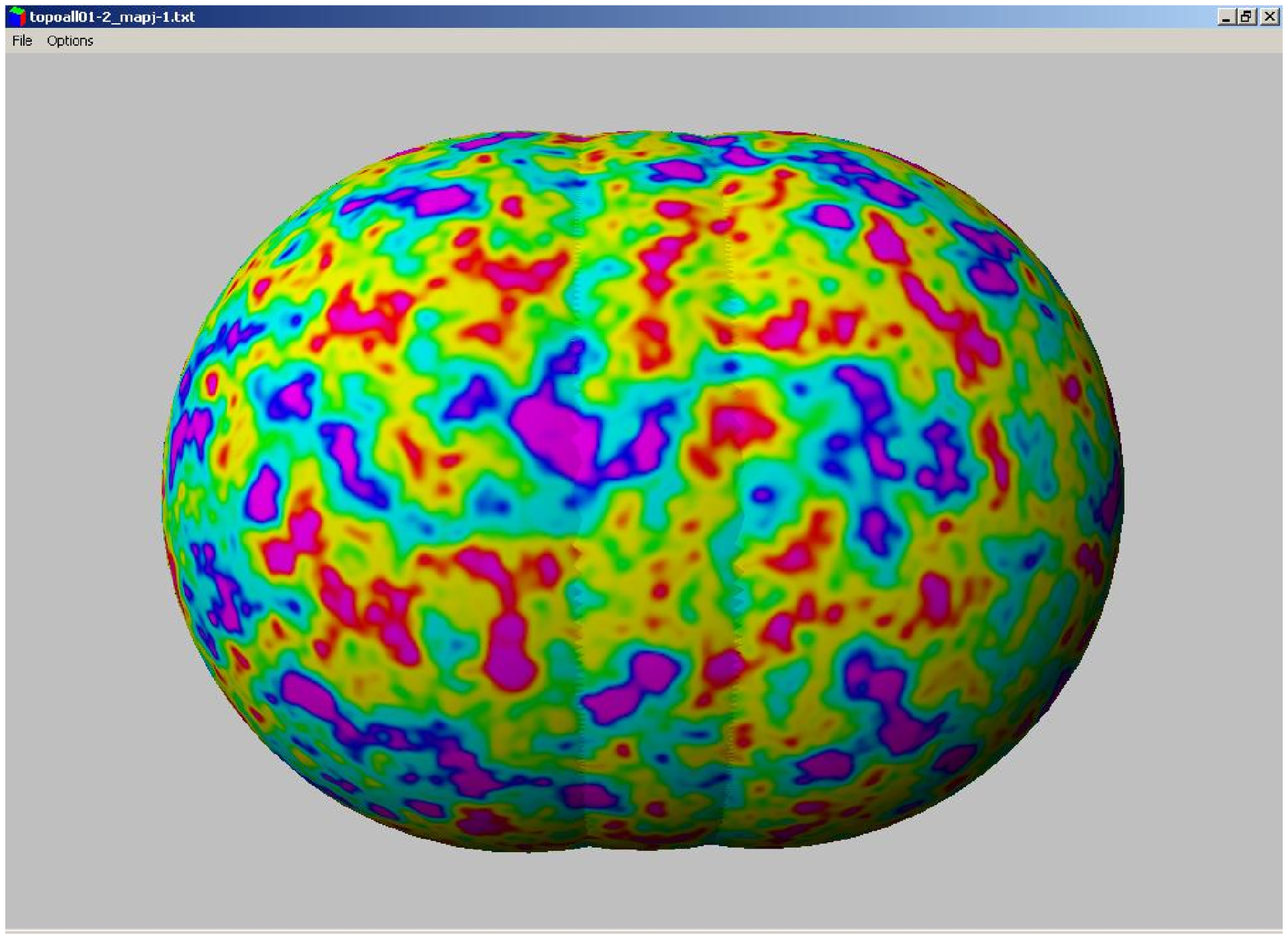,width=3.5in}
            \psfig{file=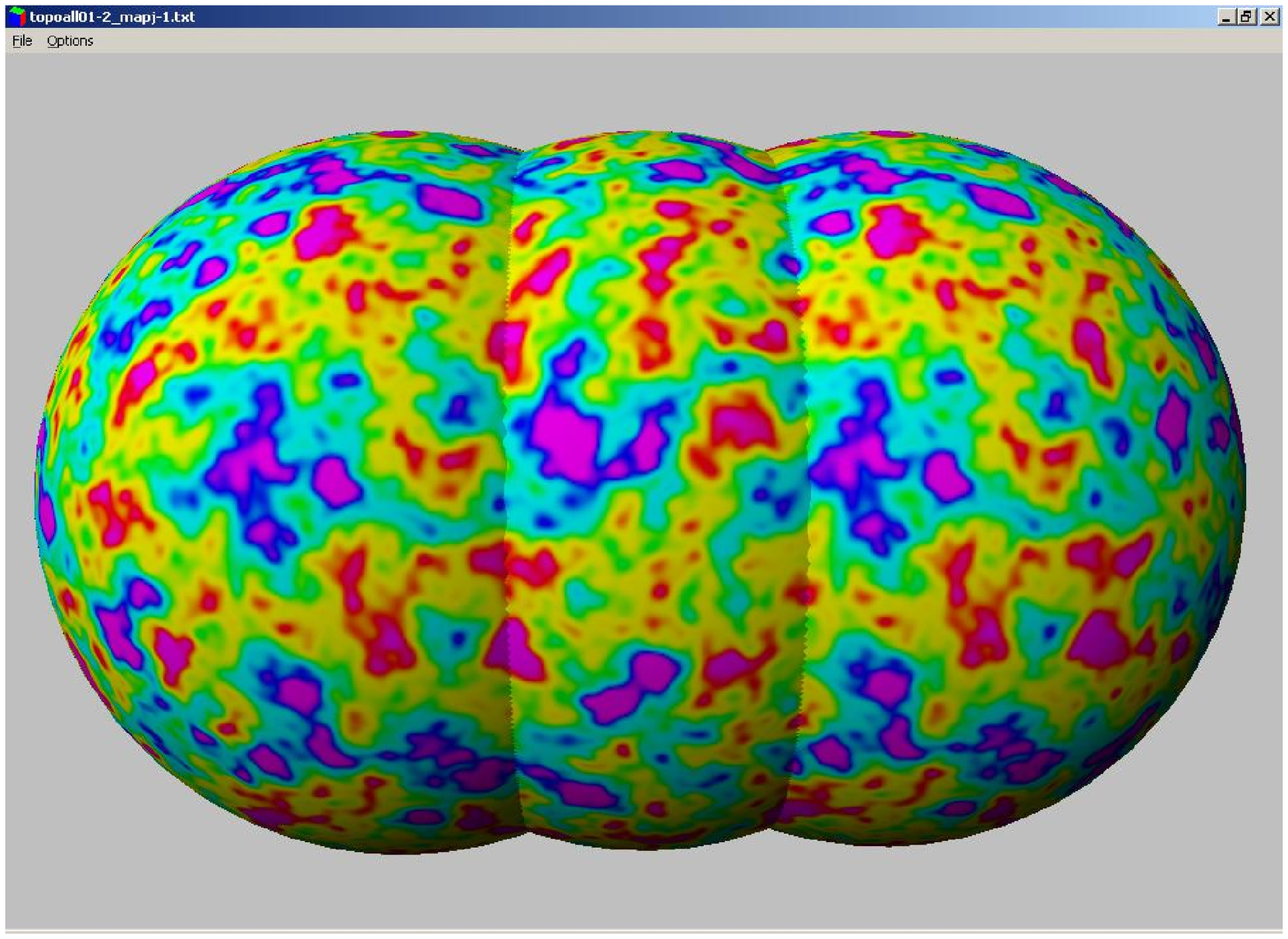,width=3.5in}}
\centerline{\psfig{file=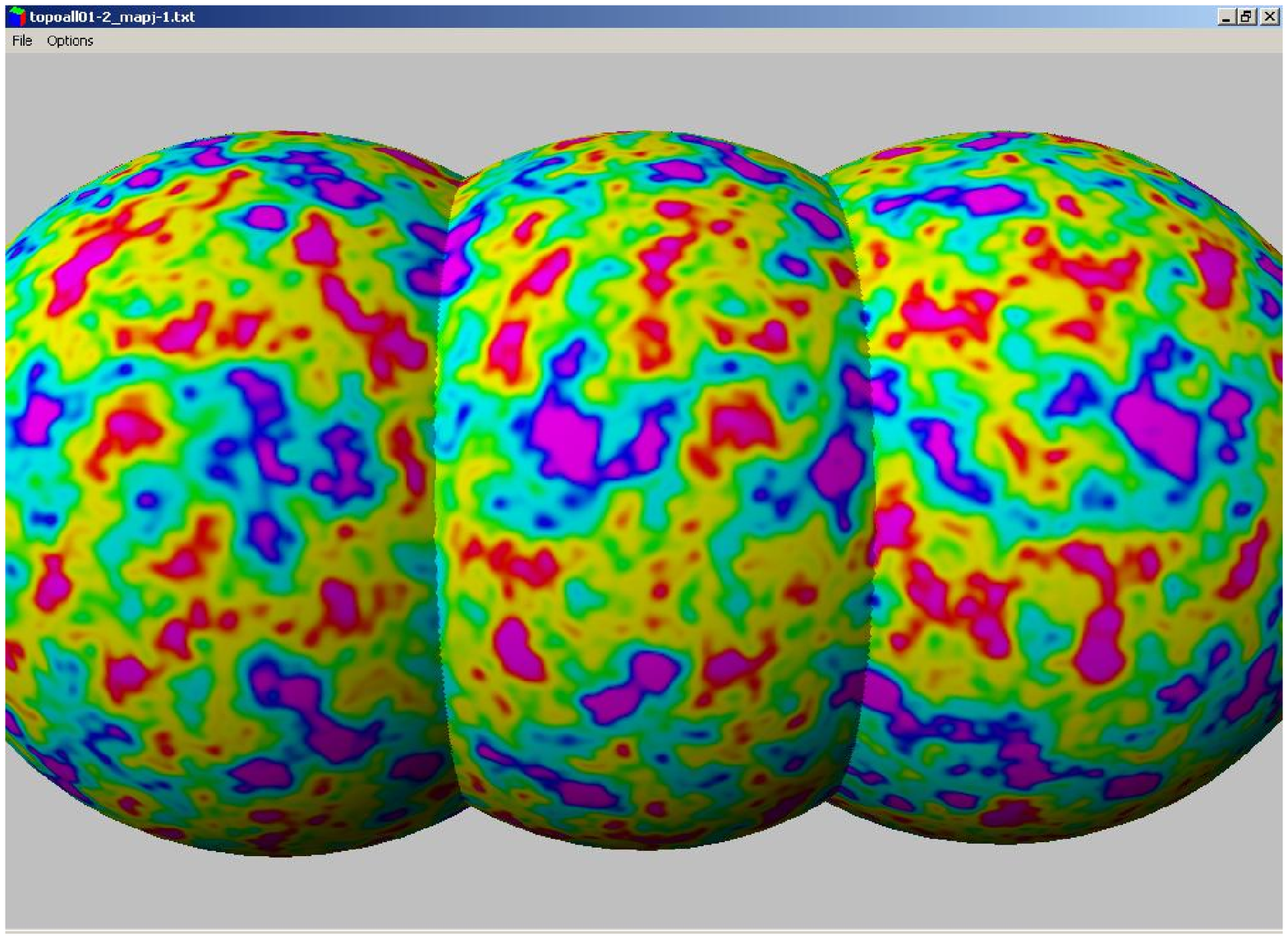,width=3.5in}
            \psfig{file=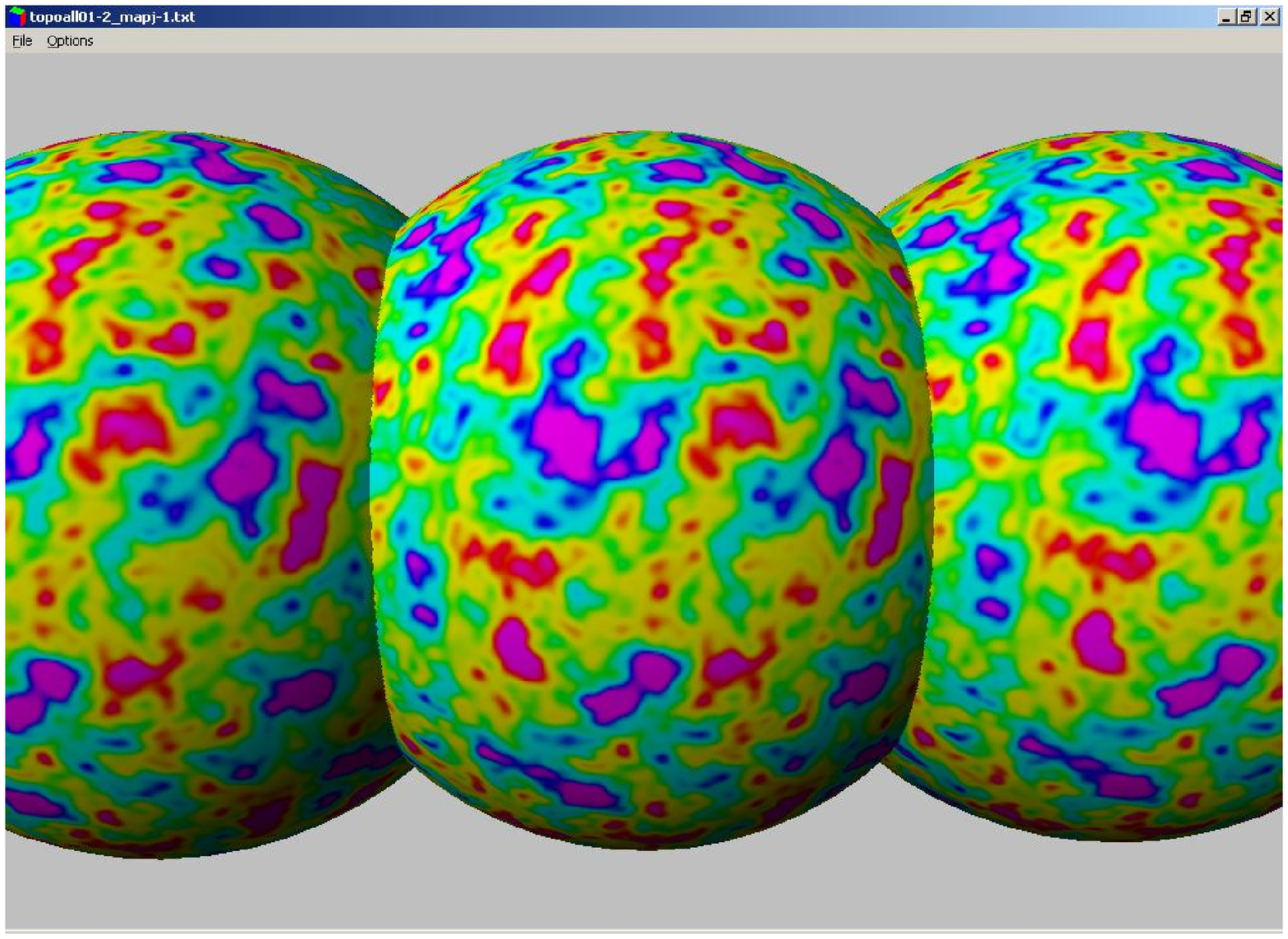,width=3.5in}}
\caption{The last scattering surface seen from outside for a
quarter-turn space $E_3$ with $L_x = L_y = 0.64$ and $L_z = 1.28$. We
present the four pairs of matched circles along the $z$-axis, which
match with a twist of $\pi / 2$, $\pi$, $3 \pi / 2$ and $2 \pi$,
respectively.  We recover the invariance by a translation of $L_z$ for
the last pair due to the fact that the cubic torus of size $L_z =
1.28$ is the four-fold cover of the quarter-turn space considered
here. Only the Sachs-Wolfe contribution has been depicted
here.}
\label{plot2}
\end{figure}

\subsubsection{Third turn space}

The third turn space is a three-fold quotient of a hexagonal 3-torus,
not a rectangular one.  To the generators (\ref{HexagonalGenerators})
of the hexagonal 3-torus we add a one-third turn corkscrew motion
\begin{equation}
\label{ThirdTurnGenerator}
\VECTROISD{x}{y}{z}  \; \mapsto \;
\MATTROISD{- \frac{1}{2}}{- \frac{\sqrt{3}}{2}}{0}
          {\frac{\sqrt{3}}{2}}{- \frac{1}{2}}{0}
          {0}{0}{1} \VECTROISD{x}{y}{z}  + \VECTROISD{0}{0}{L_z / 3} .
\end{equation}
The eigenmodes $\Upsilon_\bk$ of the hexagonal 3-torus are already
known from Eqn.~(\ref{HexagonalTorusBasis}) (and illustrated in
Figure~\ref{FigureHoneycomb}).  Applying the Invariance Lemma to them
with the additional generator~(\ref{ThirdTurnGenerator}) yields the
eigenbasis, $\Upsilon_{k_x, k_y, k_z}^{[E_4]}$,
\begin{equation}
   \begin{array}{lcl}
     \frac{1}{\sqrt{3}}
       \left[           \Upsilon_{\bk}
          + \zeta^{n_3}   \Upsilon_{\bk M}
          + \zeta^{2 n_3} \Upsilon_{\bk M^2}
       \right]
       & \quad \mbox{for} \quad
       & n_1 \in Z^+,\; n_2 \in Z^+ \cup \lbrace 0 \rbrace, n_3 \in Z, \\
     \Upsilon_{2\pi (0,0,\frac{n_3}{L_z})}
       & \quad \mbox{for} \quad
       & n_3 \in 3Z, \\
   \end{array}
\end{equation}
where $\zeta = \ee{2 i \pi / 3}$ is a cube root of unity and it is
easily checked that
\begin{eqnarray}
\bk     & = & 2 \pi \left(\frac{- n_2}{L}, \;
                          \frac{2 n_1 - n_2}{\sqrt{3} L}, \;
                          \frac{n_3}{L_z} \right) \nonumber\\
\bk M   & = & 2 \pi \left(\frac{n_1}{L}, \;
                          \frac{2 n_2 - n_1}{\sqrt{3} L}, \;
                          \frac{n_3}{L_z} \right) \nonumber\\
\bk M^2 & = & 2 \pi \left(\frac{n_2 - n_1}{L}, \;
                          \frac{- n_1 - n_2}{\sqrt{3} L}, \;
                          \frac{n_3}{L_z} \right) .
\end{eqnarray}

One can check that
\begin{equation}
\Upsilon_{n_1, n_2, n_3}^*
 = \left\lbrace \begin{array}{ll}
                \zeta^{2 n_3} \Upsilon_{n_2, n_2 - n_1, - n_3} & 
                \hbox{when} \, n_2 > n_1 \\
                \zeta^{n_3} \Upsilon_{n_1 - n_2, n_1, - n_3} &
                \hbox{when}\, n_1 \geq n_2
                \end{array}\right. .
\end{equation}
It follows that the analog of Eq.~(\ref{random_relation}) is given by
\begin{enumerate}
\item \begin{equation}
\hat{e}_{n_1, n_2, n_3}^*
 = \left\lbrace \begin{array}{ll}
                \zeta^{2 n_3} \hat{e}_{n_2, n_2 - n_1, - n_3} &
                \hbox{when} \, n_2 > n_1 \\
                \zeta^{n_3} \hat{e}_{n_1 - n_2, n_1, - n_3} &
                \hbox{when}\, n_1 \geq n_2
                \end{array}\right. .
\end{equation}
\end{enumerate}
\begin{figure}
\centerline{\psfig{file=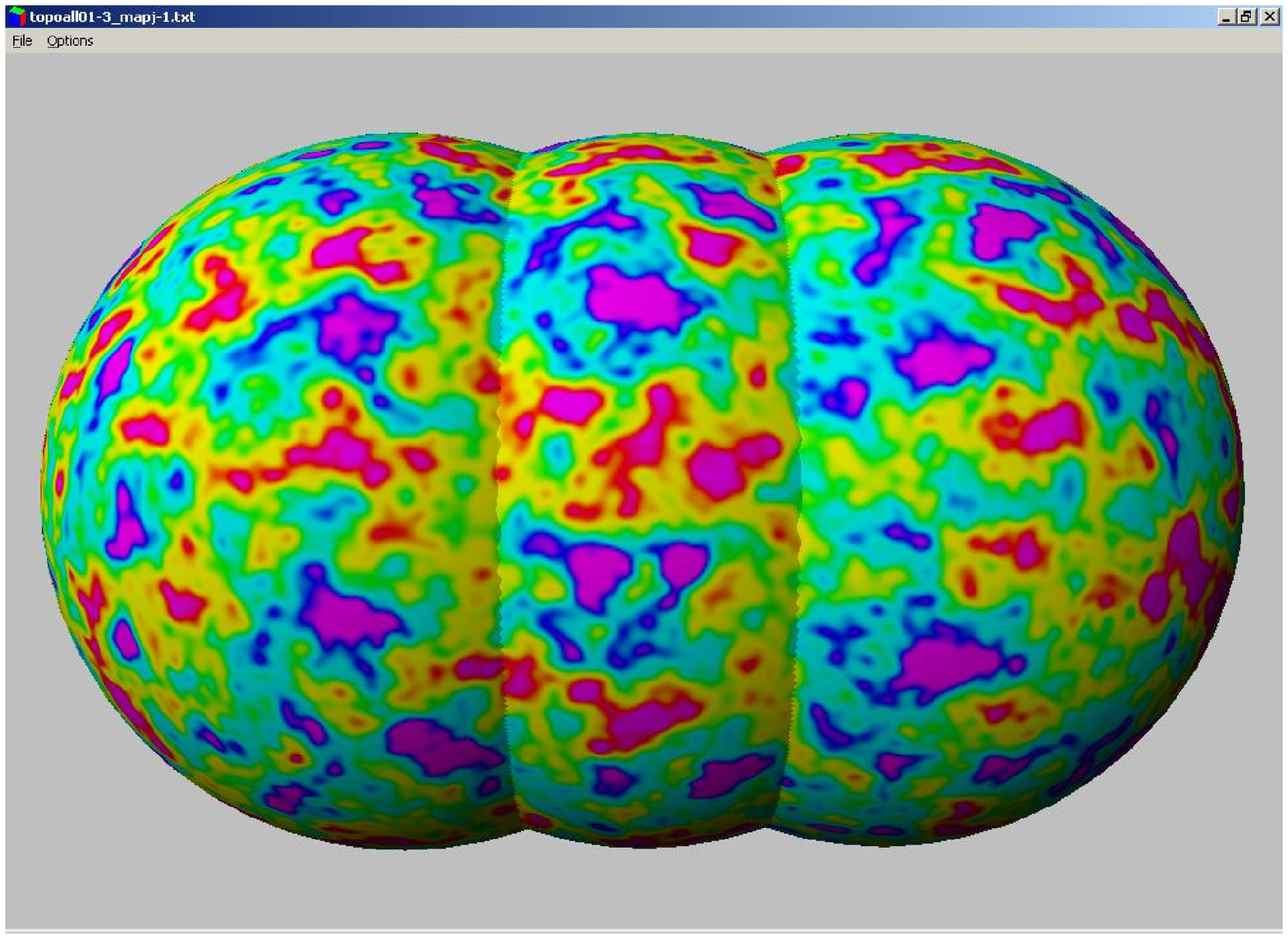,width=3.5in}
            \psfig{file=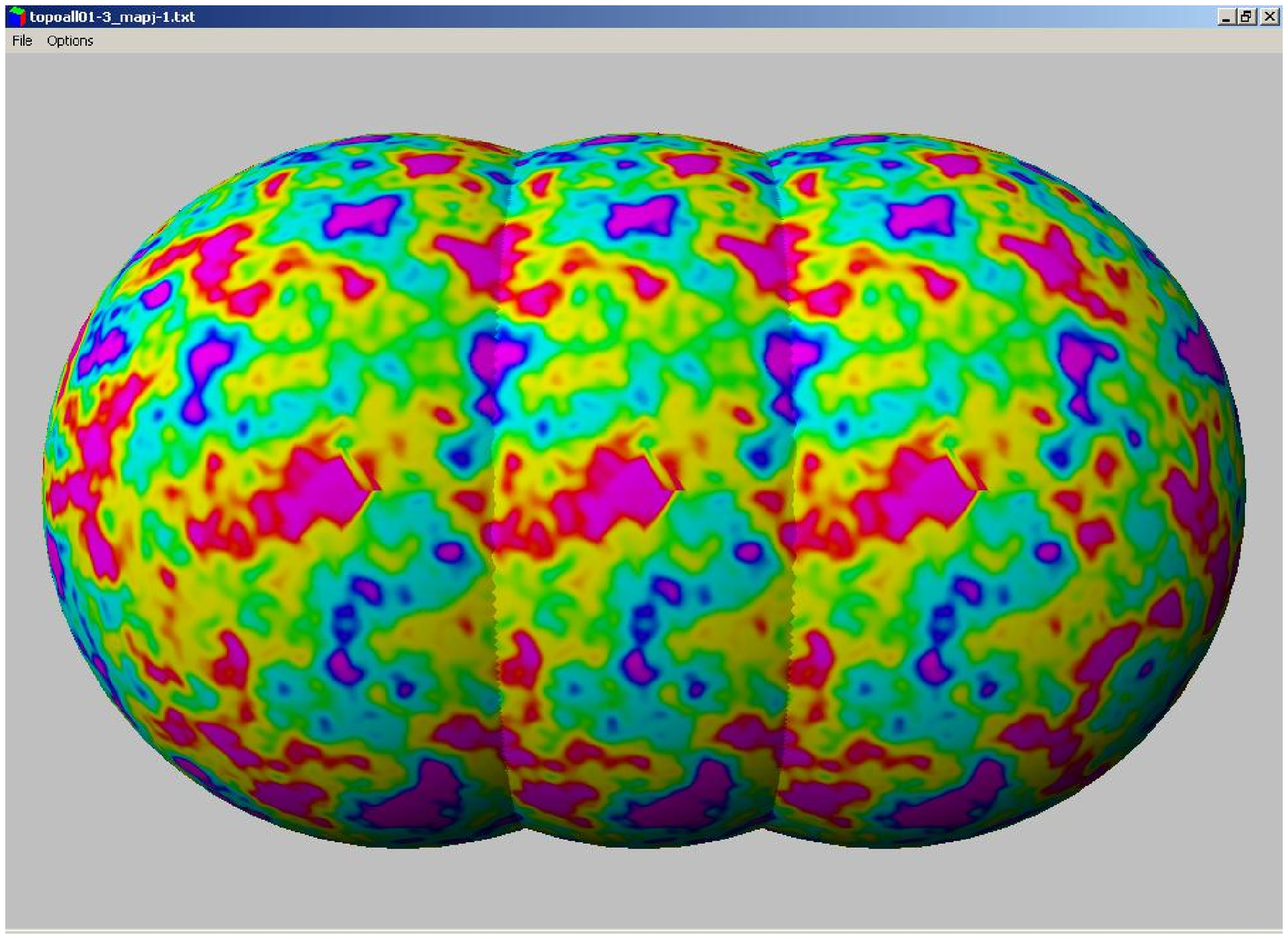,width=3.5in}}
\centerline{\psfig{file=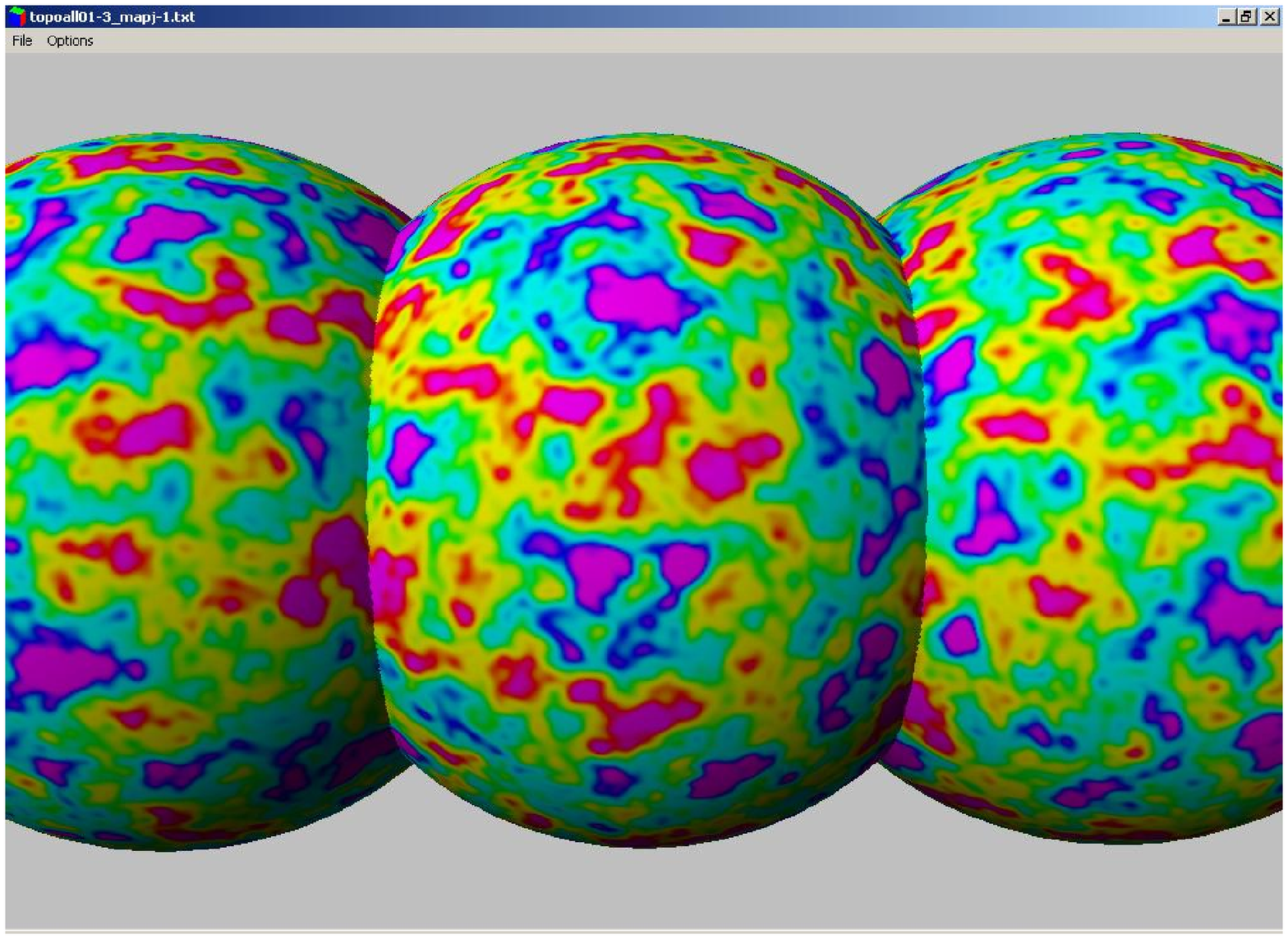,width=3.5in}
            \psfig{file=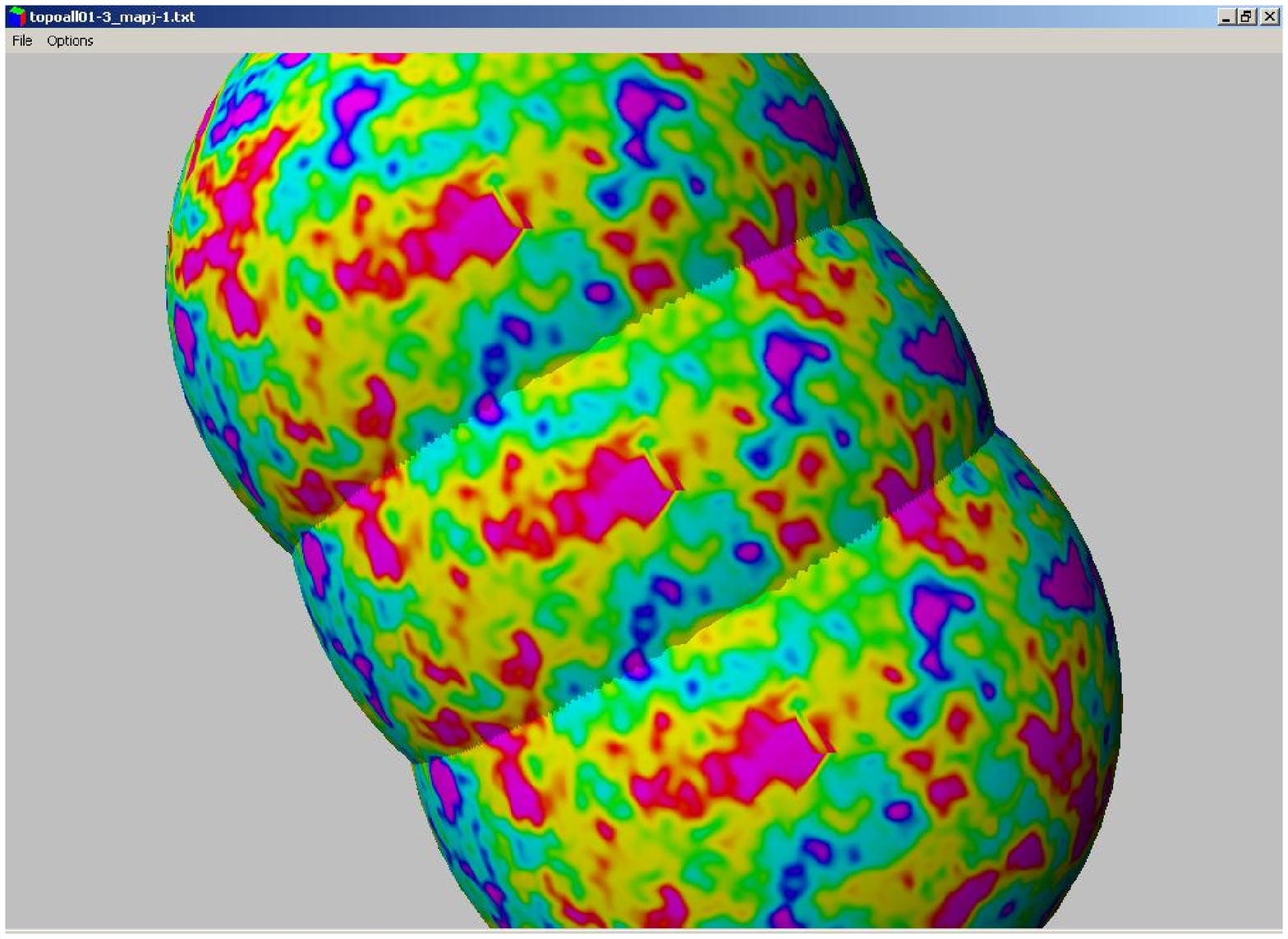,width=3.5in}}
\centerline{\psfig{file=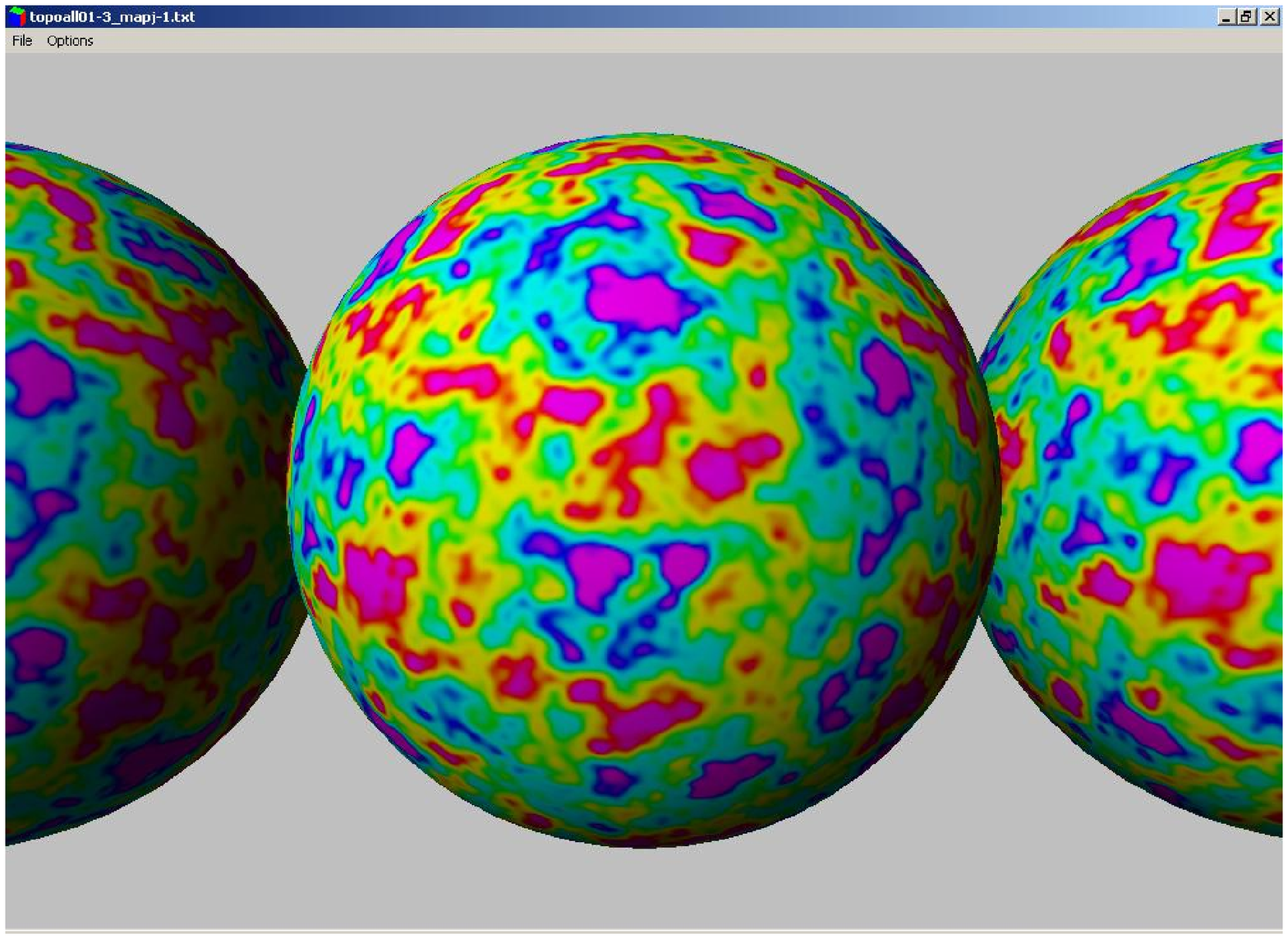,width=3.5in}
            \psfig{file=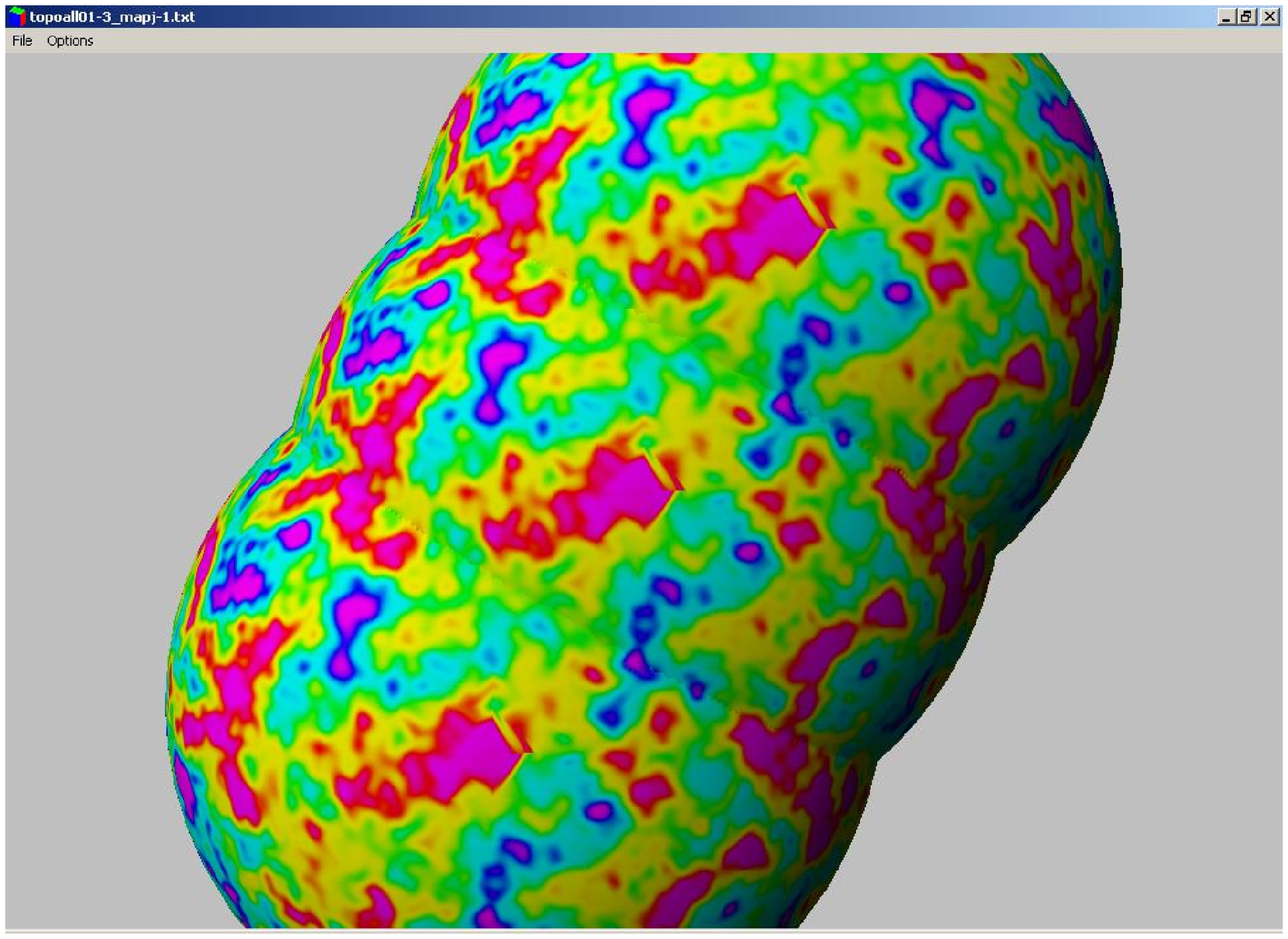,width=3.5in}}
\caption{The last scattering surface seen from outside for a
third-turn space $E_4$ with $L_x = L_y = 0.64$ and $L_z = 1.92$ in
units of the last scattering surface. The first column presents the
three pairs of circles alonq the $z$ axis which match after a rotation
of $2 \pi / 3$, $4 \pi / 3$ and $2 \pi$ respectively. We recover the
invariance by a translation of $L_z$ due to the fact that the cubic
torus of size $L_z = 1.92$ is the three-fold cover of the third-turn
space considered here.  The second column present the three pairs of
circles related by translations in the $xy$-plane. Only the
Sachs-Wolfe contribution has been depicted here.}
\label{plot3}
\end{figure}

\subsubsection{Sixth turn space}

The sixth turn space is like the third turn space, but with a
one-sixth turn corkscrew motion
\begin{equation}
\label{SixthTurnGenerator}
\VECTROISD{x}{y}{z}  
\; \mapsto \;
\MATTROISD{\frac{1}{2}}{- \frac{\sqrt{3}}{2}}{0}
          {\frac{\sqrt{3}}{2}}{\frac{1}{2}}{0}
          {0}{0}{1} \VECTROISD{x}{y}{z}  + \VECTROISD{0}{0}{L_z / 6} .
\end{equation}
The same reasoning as before shows the eigenbasis, $\Upsilon_{k_x,
k_y, k_z}^{[E_5]}$, to be
\begin{equation}
   \begin{array}{lcl}
     \frac{1}{\sqrt{6}}
       \left[           \Upsilon_{\bk}
          + \zeta^{n_3}   \Upsilon_{\bk M}
          + \zeta^{2 n_3} \Upsilon_{\bk M^2}
          + \zeta^{3 n_3} \Upsilon_{\bk M^3}
          + \zeta^{4 n_3} \Upsilon_{\bk M^4}
          + \zeta^{5 n_3} \Upsilon_{\bk M^5}
       \right]
       & \mbox{ for } \;
       & n_1 \in Z^+,\; n_2 \in Z^+ \cup \lbrace 0 \rbrace,\;n_2 < n_1, n_3 \in Z, \\
     \Upsilon_{2\pi (0,0,\frac{n_3}{L_z})}
       & \mbox{ for }
       & n_3 \in 6Z, \\
   \end{array}
\end{equation}
where $\zeta = \ee{2i \pi / 6}$ is a sixth root of unity and it is
easily checked that
\begin{eqnarray}
\bk     & = & 2 \pi \left(\frac{- n_2}{L}, \;
                          \frac{2 n_1 - n_2}{\sqrt{3} L}, \;
                          \frac{n_3}{L_z} \right) \nonumber\\
\bk M   & = & 2 \pi \left(\frac{n_1 - n_2}{L}, \;
                          \frac{n_1 + n_2}{\sqrt{3} L}, \;
                          \frac{n_3}{L_z} \right) \nonumber\\
\bk M^2 & = & 2 \pi \left(\frac{n_1}{L}, \;
                          \frac{2 n_2 - n_1}{\sqrt{3} L}, \;
                          \frac{n_3}{L_z} \right) \nonumber\\
\bk M^3 & = & 2 \pi \left(\frac{n_2}{L}, \;
                          \frac{n_2 - 2 n_1}{\sqrt{3} L}, \;
                          \frac{n_3}{L_z} \right) \nonumber\\
\bk M^4 & = & 2 \pi \left(\frac{n_2 - n_1}{L}, \;
                          \frac{- n_1 - n_2}{\sqrt{3} L}, \;
                          \frac{n_3}{L_z} \right) \nonumber\\
\bk M^5 & = & 2 \pi \left(\frac{- n_1}{L}, \;
                          \frac{n_1 - 2 n_2}{\sqrt{3} L}, \;
                          \frac{n_3}{L_z} \right).
\end{eqnarray}
One can check that
\begin{equation}
\Upsilon_{n_1, n_2, n_3}^* = (- 1)^{n_3} \Upsilon_{n_1, n_2, - n_3} ,
\end{equation}
so that when $n_3 \neq 0$ an analog relation exists with the
corresponding complex random variables, and when $n_3 = 0$, the random
variable $\hat{e}_{n_1, n_2, 0}$ is real.
\begin{figure}
\centerline{\psfig{file=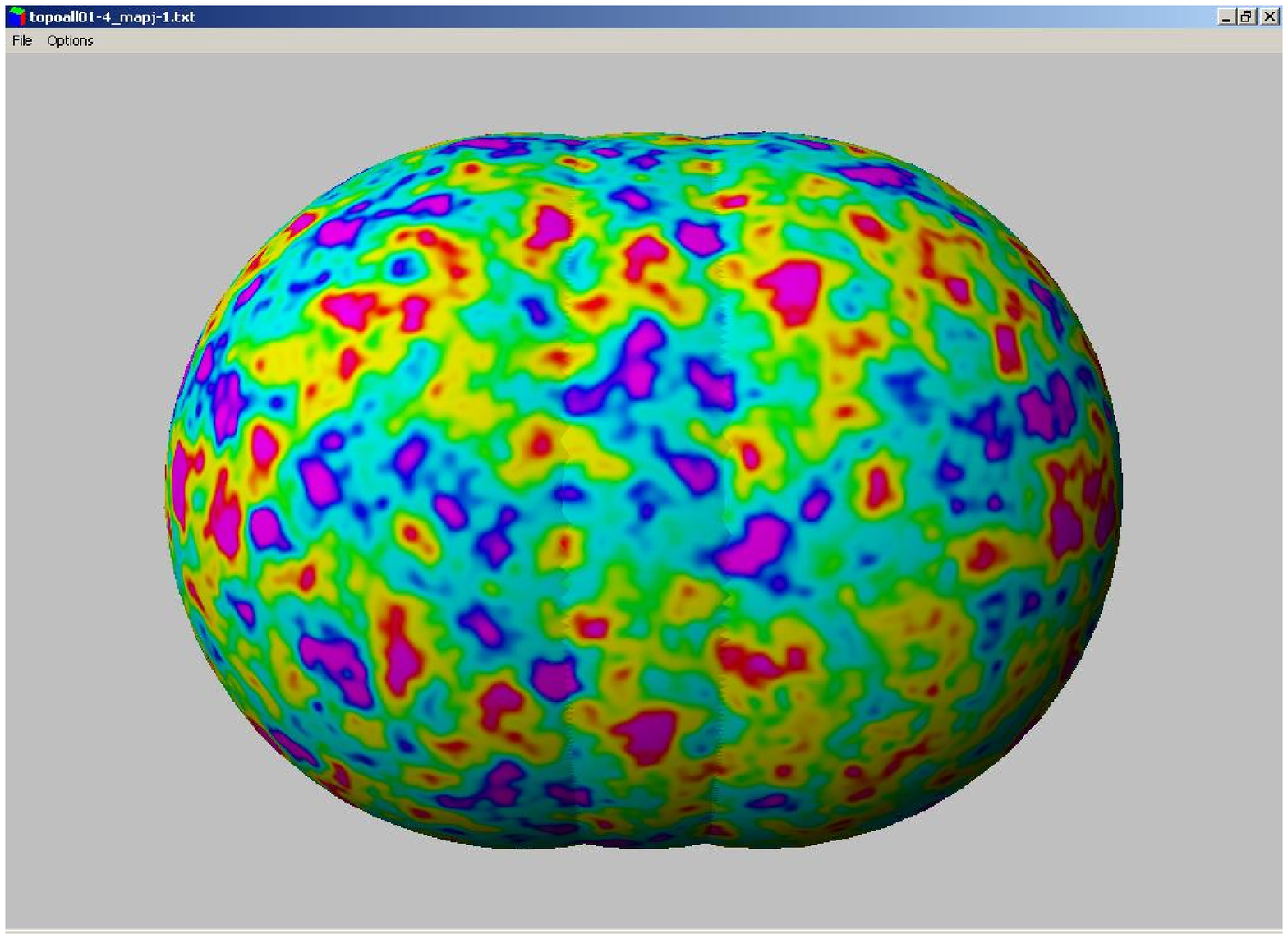,width=3.5in}
            \psfig{file=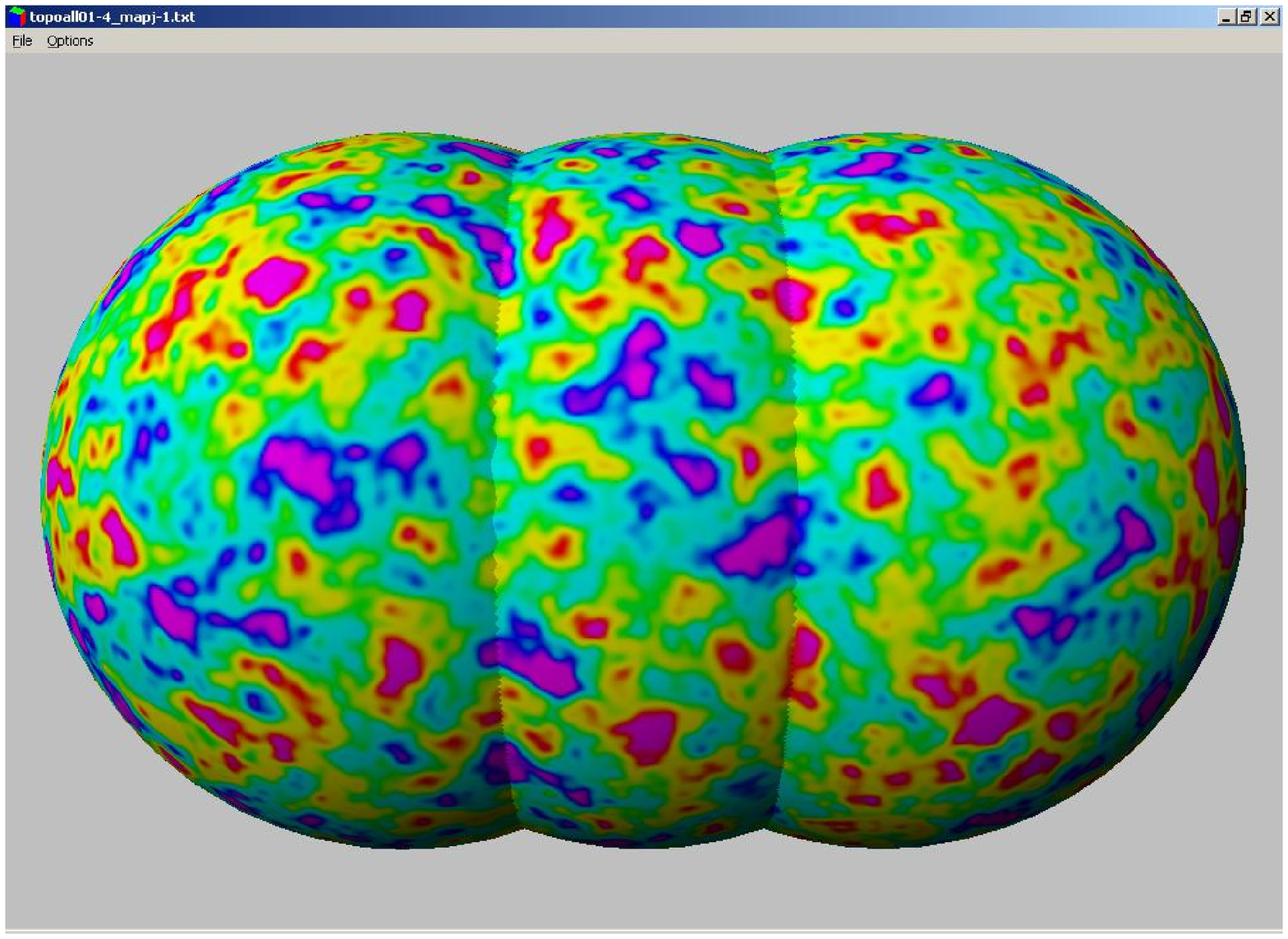,width=3.5in}}
\centerline{\psfig{file=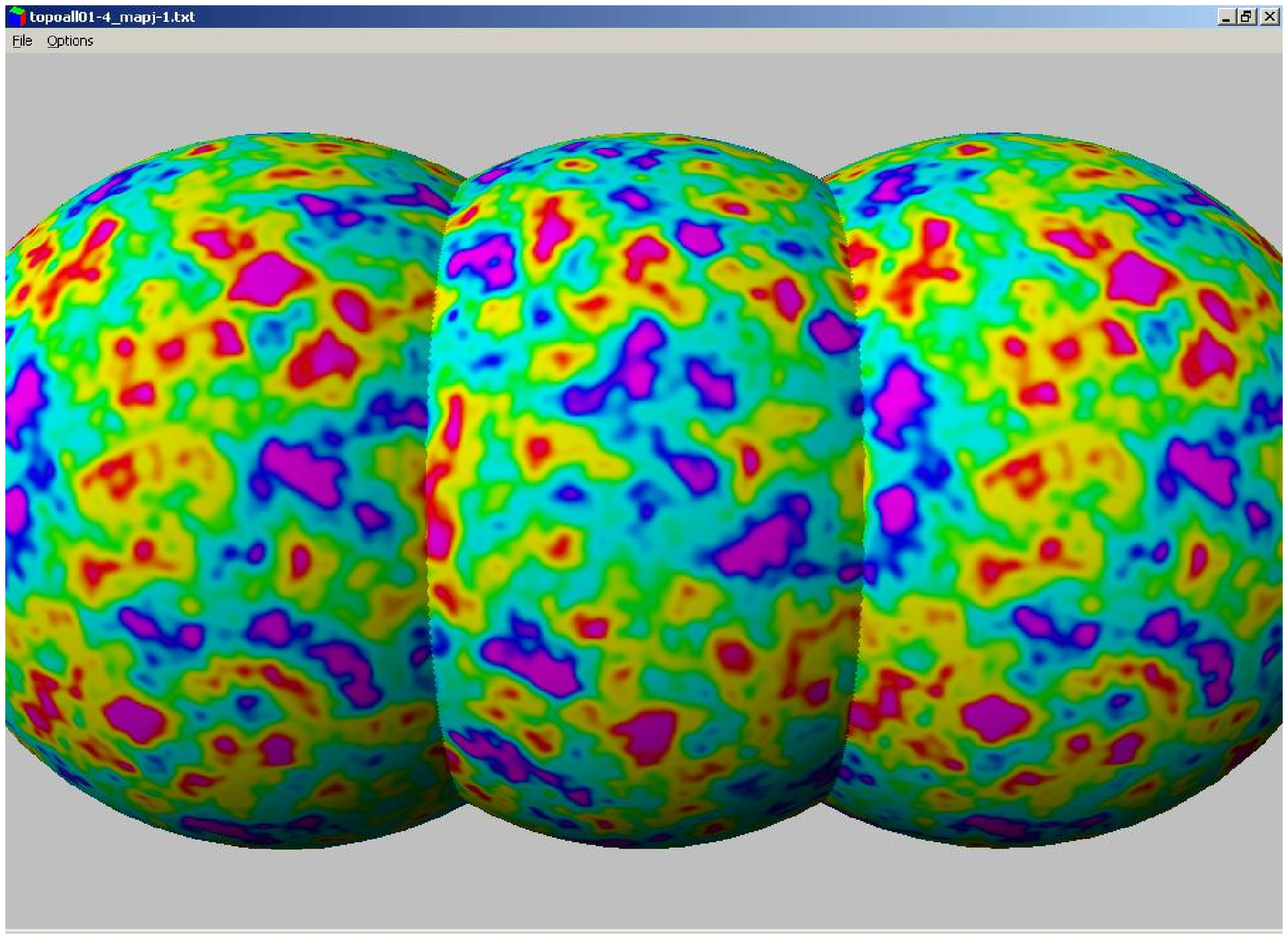,width=3.5in}
            \psfig{file=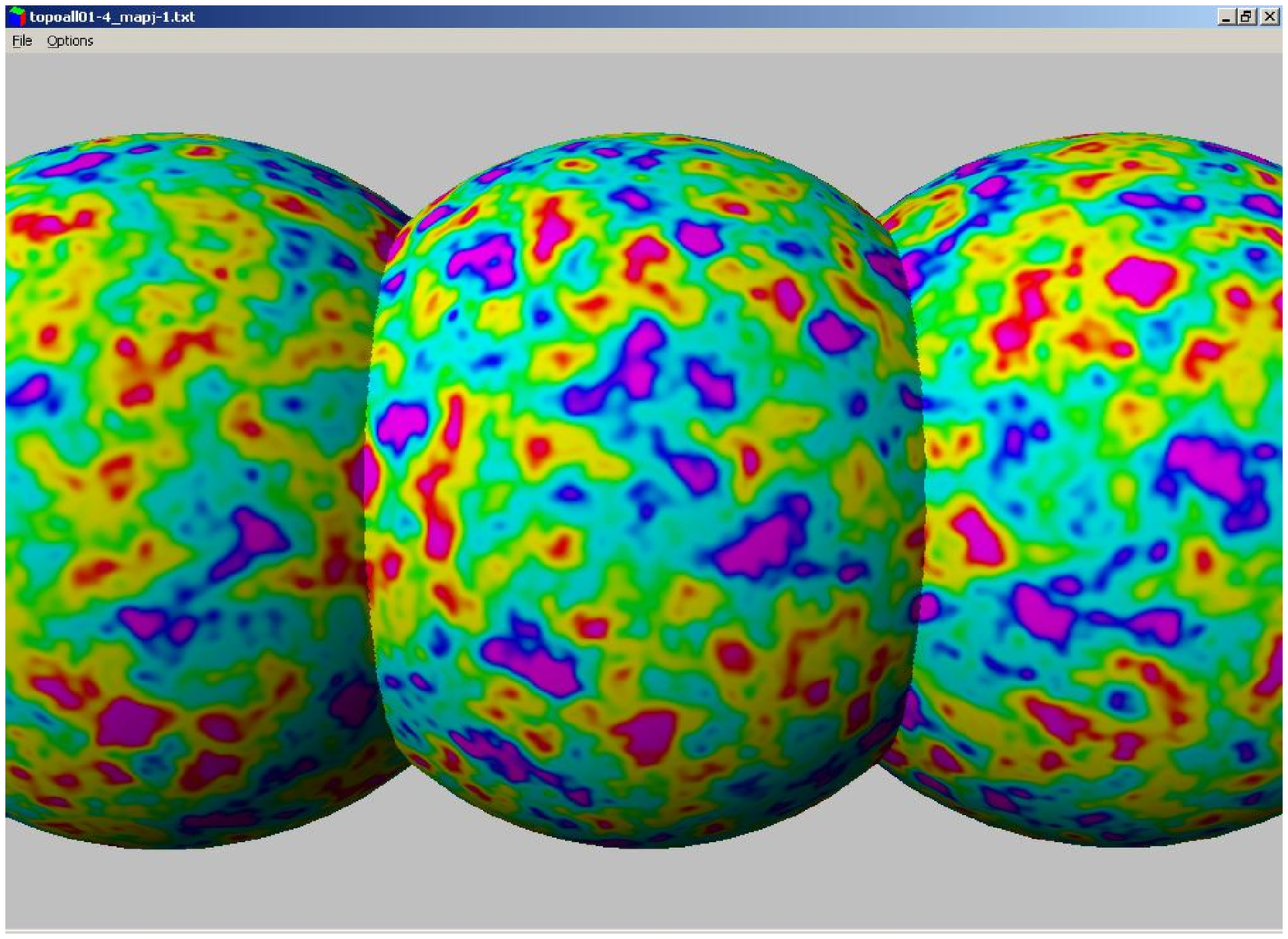,width=3.5in}}
\centerline{\psfig{file=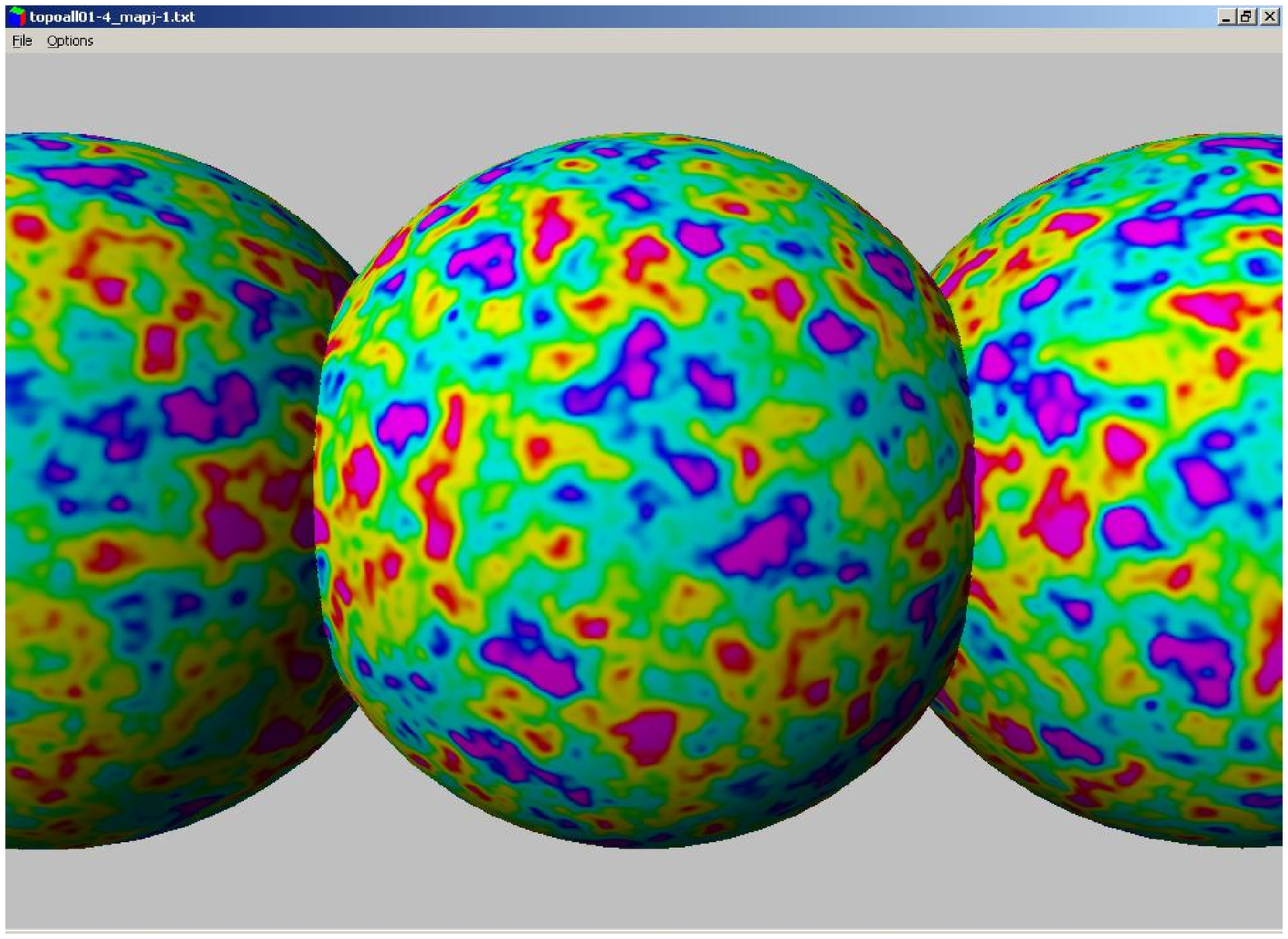,width=3.5in}
            \psfig{file=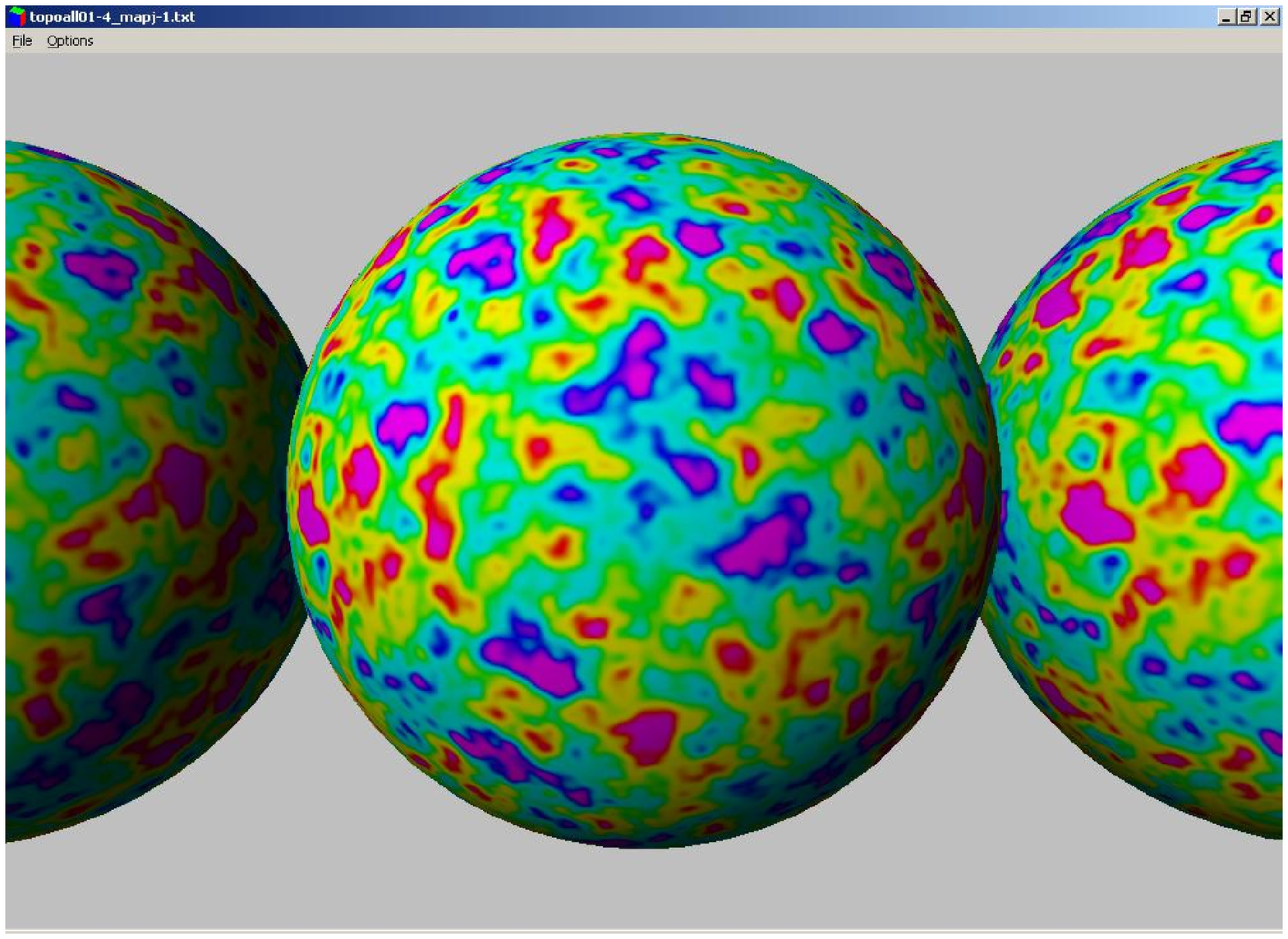,width=3.5in}} 
\caption{The last scattering surface seen from outside for a
sixth-turn space $E_5$ with $L_x = L_y = 0.64$ and $L_z = 1.92$ in
units of the last scattering surface. We presents the six pairs of
circles along the $z$ axis which match after a rotation of $\pi / 3$,
$2 \pi/ 3 \ldots 2 \pi$ respectively. We recover the invariance by a
translation of $L_z$ due to the fact that the cubic torus of size $L_z
= 1.92$ is the six-fold cover of the sixth-turn space considered
here. Only the Sachs-Wolfe contribution has been depicted here.}
\label{plot4}
\end{figure}

\subsubsection{Hantzsche-Wendt space}

The fundamental polyhedron of the Hantzsche-Wendt space is a rhombic
dodecahedron circumscribed about a rectangular box of size $(L_x / 2,
L_y / 2, L_z / 2)$. The holonomy group is generated by the three
half-turn corkscrew motions
\begin{eqnarray}
\VECTROISD{x}{y}{z}  
\; & \mapsto & \;
   \MATTROISD{\;\;\; 1}{0}{0}
             {0}{- 1}{0}
             {0}{0}{- 1} \VECTROISD{x}{y}{z}
 + \VECTROISD{L_x / 2}{L_y / 2}{0} \nonumber \\
\VECTROISD{x}{y}{z}
\; & \mapsto & \;
   \MATTROISD{- 1}{0}{0}
             {0}{\;\;\; 1}{0}
             {0}{0}{- 1} \VECTROISD{x}{y}{z}
 + \VECTROISD{0}{L_y / 2}{L_z / 2} \nonumber \\
\VECTROISD{x}{y}{z}
\; & \mapsto & \;
   \MATTROISD{- 1}{0}{0}
             {0}{- 1}{0}
             {0}{0}{\;\;\; 1} \VECTROISD{x}{y}{z}
 + \VECTROISD{L_x / 2}{0}{L_z / 2} .
\label{HantzscheWendtGenerators}
\end{eqnarray}
The composition of these three generators is the identity, so any two
suffice to generate the group.  Each element of the Hantzsche-Wendt
group has a rotational component $M \in \lbrace \diag (1, 1, 1)$,
$\diag (1, -1, -1)$, $\diag (-1, 1, -1)$, $\diag (-1, -1, 1)
\rbrace$. The pure translations (elements with $M = \diag (1, 1, 1)$)
form a subgroup of index 4; the corresponding four-fold cover of the
Hantzsche-Wendt space is a rectangular 3-torus of size $(L_x, L_y,
L_z)$, whose holonomy is generated by the squares of the above
corkscrew motions.  Thus we may begin with the eigenspace for a
rectangular 3-torus (\ref{RectangularTorusBasis}) and ask what
subspace remains fixed by the three Hantzsche-Wendt generators
(\ref{HantzscheWendtGenerators}).  The Invariance Lemma shows that all
three generators preserve the (typically four-dimensional) subspace
$\lbrace \Upsilon_{k_x, k_y, k_z}, \Upsilon_{k_x, - k_y, - k_z},
\Upsilon_{- k_x, k_y, - k_z}, \Upsilon_{- k_x, - k_y, k_z} \rbrace$ as
a set, and fix the linear combination
\begin{equation}
\Upsilon_{k_x,k_y,k_z}
 \; + \; (- 1)^{n_x - n_y} \Upsilon_{k_x, - k_y, - k_z}
 \; + \; (- 1)^{n_y - n_z} \Upsilon_{- k_x, k_y, - k_z}
 \; + \; (- 1)^{n_z - n_x} \Upsilon_{- k_x, - k_y, k_z} .
\label{HantzscheWendtFixedPoint}
\end{equation}
Visualizing the four wave vectors in the subscripts of
(\ref{HantzscheWendtFixedPoint}) as alternate corners of the cube
$(\pm k_x, \, \pm k_y, \, \pm k_z) = 2 \pi (\frac{\pm n_x}{L_x}, \,
\frac{\pm n_y}{L_y}, \, \frac{\pm n_z}{L_z})$, one sees that subspace
will be degenerate if and only if at least two of the indices $\lbrace
n_x, n_y, n_z \rbrace$ are zero.  In the degenerate case a
two-dimensional subspace like $\lbrace \Upsilon_{k_x, 0, 0},
\Upsilon_{- k_x, 0, 0} \rbrace$ is preserved as a set, while the mode
$\Upsilon_{k_x, 0, 0} + \Upsilon_{- k_x, 0, 0}$ is preserved by all
three generators if and only if $n_x$ is even.  Thus the
Hantzsche-Wendt's eigenspace has basis, $\Upsilon_{k_x, k_y,
k_z}^{[E_6]}$,
\begin{eqnarray}
     & & \frac{1}{2}
       \left[               \Upsilon_{2\pi ( \frac{n_x}{L_x}, \frac{n_y}{L_y}, \frac{n_z}{L_z})}
          + (-1)^{n_x - n_y} \Upsilon_{2\pi ( \frac{n_x}{L_x},-\frac{n_y}{L_y},-\frac{n_z}{L_z})}
          + (-1)^{n_y - n_z} \Upsilon_{2\pi (-\frac{n_x}{L_x}, \frac{n_y}{L_y},-\frac{n_z}{L_z})}
          + (-1)^{n_z - n_x} \Upsilon_{2\pi (-\frac{n_x}{L_x},-\frac{n_y}{L_y}, \frac{n_z}{L_z})}
       \right]  \nonumber\\
     & & \qquad\qquad \mbox{for} \quad
         (n_x, n_y \in Z^+, n_z \in Z) \mbox{ or }
         (n_x = 0, n_y, n_z \in Z^+) \mbox{ or }
         (n_y = 0, n_x, n_z \in Z^+),\nonumber\\
     & & \frac{1}{\sqrt{2}} \left[
             \Upsilon_{2\pi (\frac{n_x}{L_x},0,0)}
           + \Upsilon_{2\pi (-\frac{n_x}{L_x},0,0)}
         \right]
        \quad  \mbox{for}  \quad n_x \in 2Z^+,\nonumber\\
     & & \frac{1}{\sqrt{2}} \left[
             \Upsilon_{2\pi (0,\frac{n_y}{L_y},0)}
           + \Upsilon_{2\pi (0,-\frac{n_y}{L_y},0)}
         \right]
        \quad  \mbox{for}  \quad n_y \in 2Z^+,\nonumber\\
     & & \frac{1}{\sqrt{2}} \left[
             \Upsilon_{2\pi (0,0,\frac{n_z}{L_z})}
           + \Upsilon_{2\pi (0,0,-\frac{n_z}{L_z})}
         \right]
        \quad  \mbox{for}  \quad n_z \in 2Z^+.
\end{eqnarray}

One can easily check that
\begin{equation}
\Upsilon_{k_x, k_y, k_z}^{[E_6]*}
 = (- 1)^{n_x - n_z} \Upsilon_{k_x, k_y, - k_z}^{[E_6]} ,
\end{equation}
when $(n_x, n_y \in Z^+, n_z \in Z)$ or $(n_x = 0, n_y, n_z \in Z^+)$
or $(n_y = 0, n_x, n_z \in Z^+)$ and that otherwise $\Upsilon_{k_x,
k_y, k_z}^{[E_6]}$ is real. It follows that the analog of
Eq.~(\ref{random_relation}) is given by
\begin{enumerate}
\item when $n_x, n_y \in Z^+$ and $n_z \in Z$, $\hat e_\bk$ is a
complex random variable satisfying
\begin{equation}
\hat e_{k_x, k_y, k_z}^* = (- 1)^{n_x - n_z} \hat e_{k_x, k_y, - k_z} .
\end{equation}
It is thus a real random variable if $k_z = 0$ and $n_x \in 2 Z$ and a
purely imaginary random variable if $k_z = 0$ and $n_x \notin 2 Z$
\item when $n_x = 0$ and $n_y, n_z \in Z^+$, $\hat e_\bk$ is a random
variable satisfying
\begin{equation}
\hat e_{0, k_y, k_z}^* = (- 1)^{n_y} \hat e_{0, k_y, k_z} ,
\end{equation}
so that it is a real random variable when $n_y \in 2 Z^+$ and purely
imaginary otherwise,
\item when $n_y = 0$ and $n_x, n_z \in Z^+$, $\hat e_\bk$ is a random
variable satisfying
\begin{equation}
\hat e_{k_x, 0, k_z}^* = (- 1)^{n_z} \hat e_{k_x, 0, k_z} ,
\end{equation}
so that it is a real random variable when $n_z \in 2 Z^+$ and purely
imaginary otherwise,
\item when $n_x \in Z^+$, $n_y \in Z^+$ or $n_z \in Z^+$, $\hat
e_{k_x, 0, 0}$, $\hat e_{0, k_y, 0}$ and $\hat e_{0, 0, k_z}$ are real
random variables.
\end{enumerate}
\begin{figure}
\centerline{\psfig{file=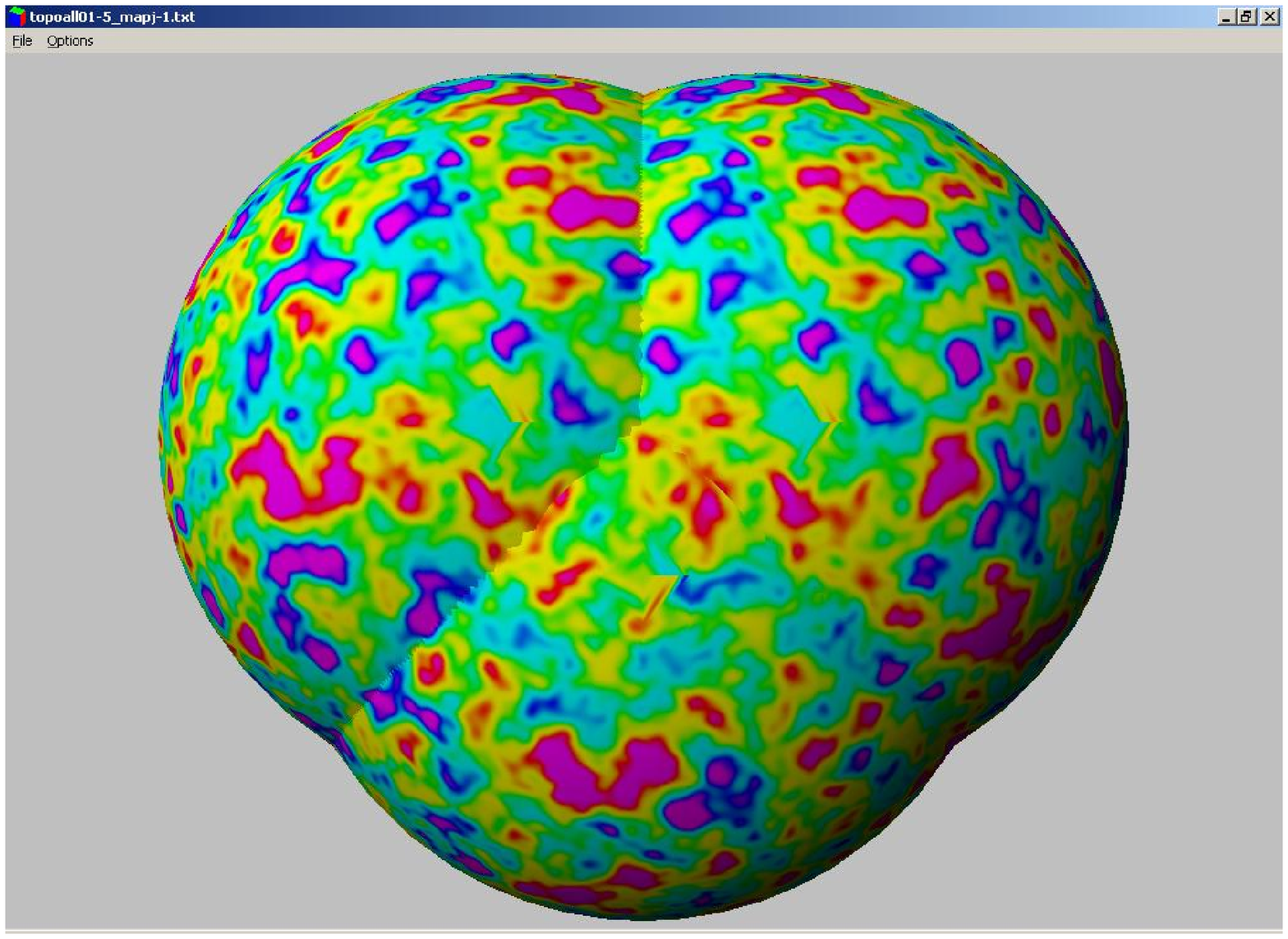,width=3.5in}
            \psfig{file=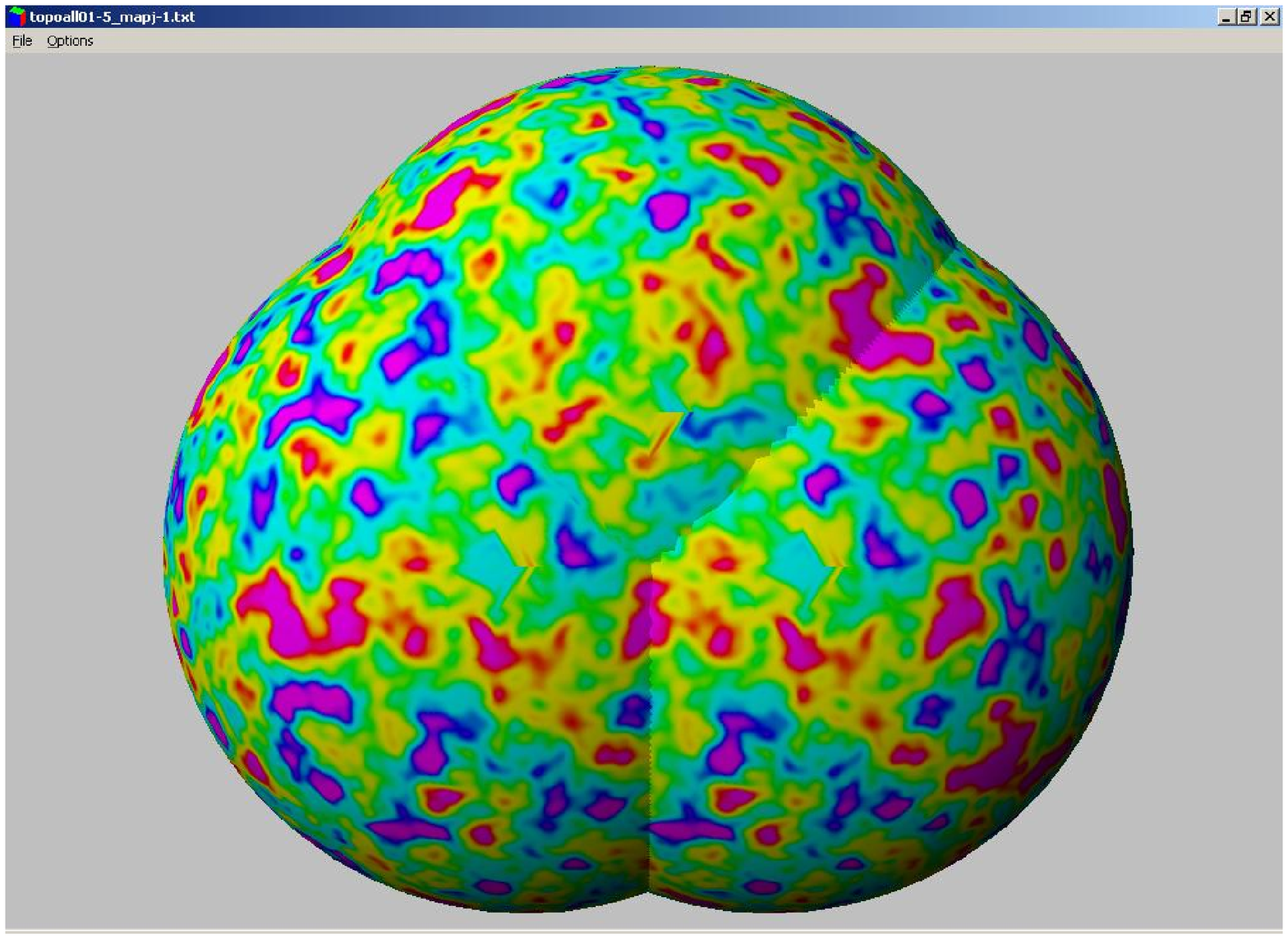,width=3.5in}}
\centerline{\psfig{file=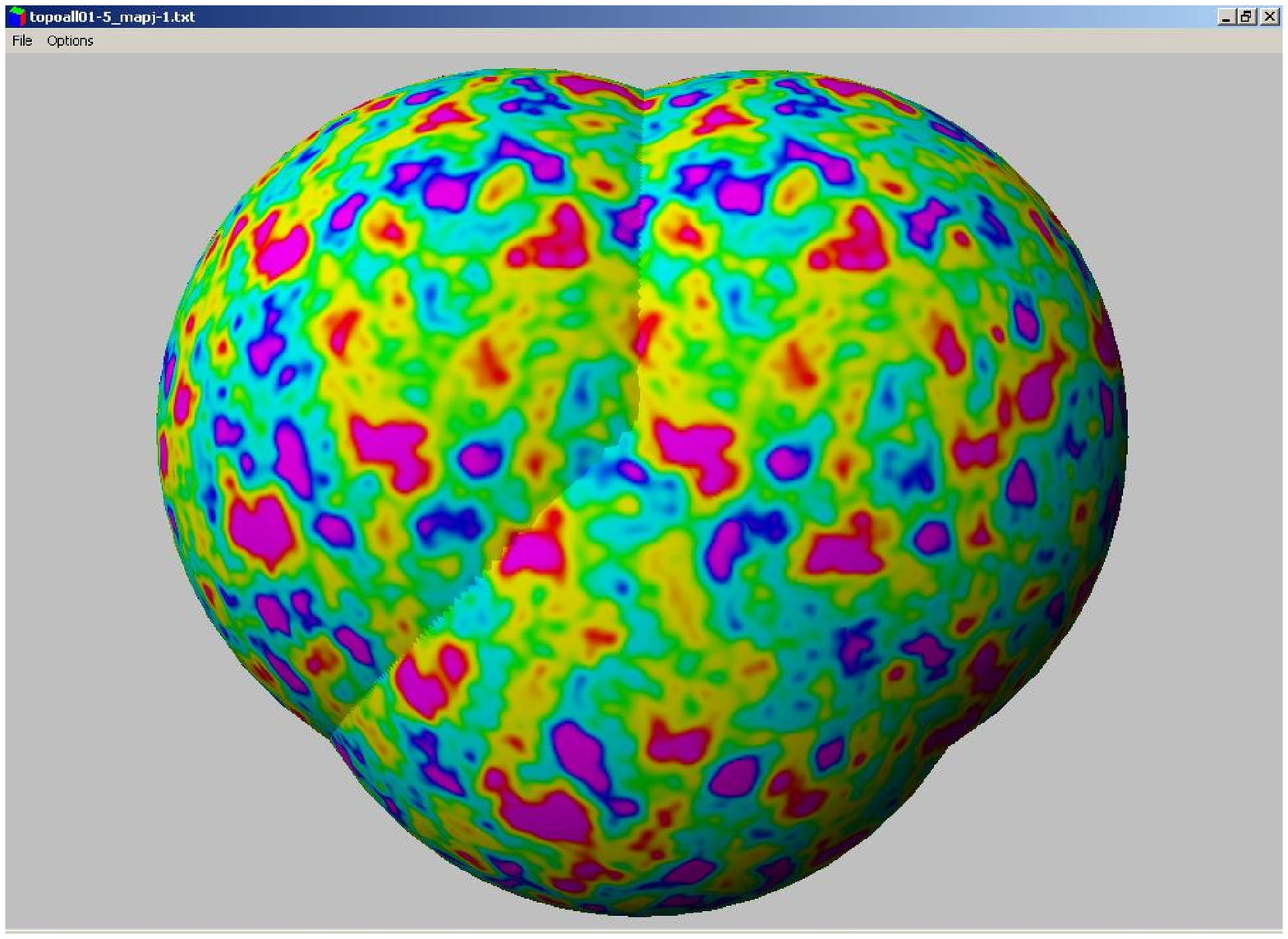,width=3.5in}
            \psfig{file=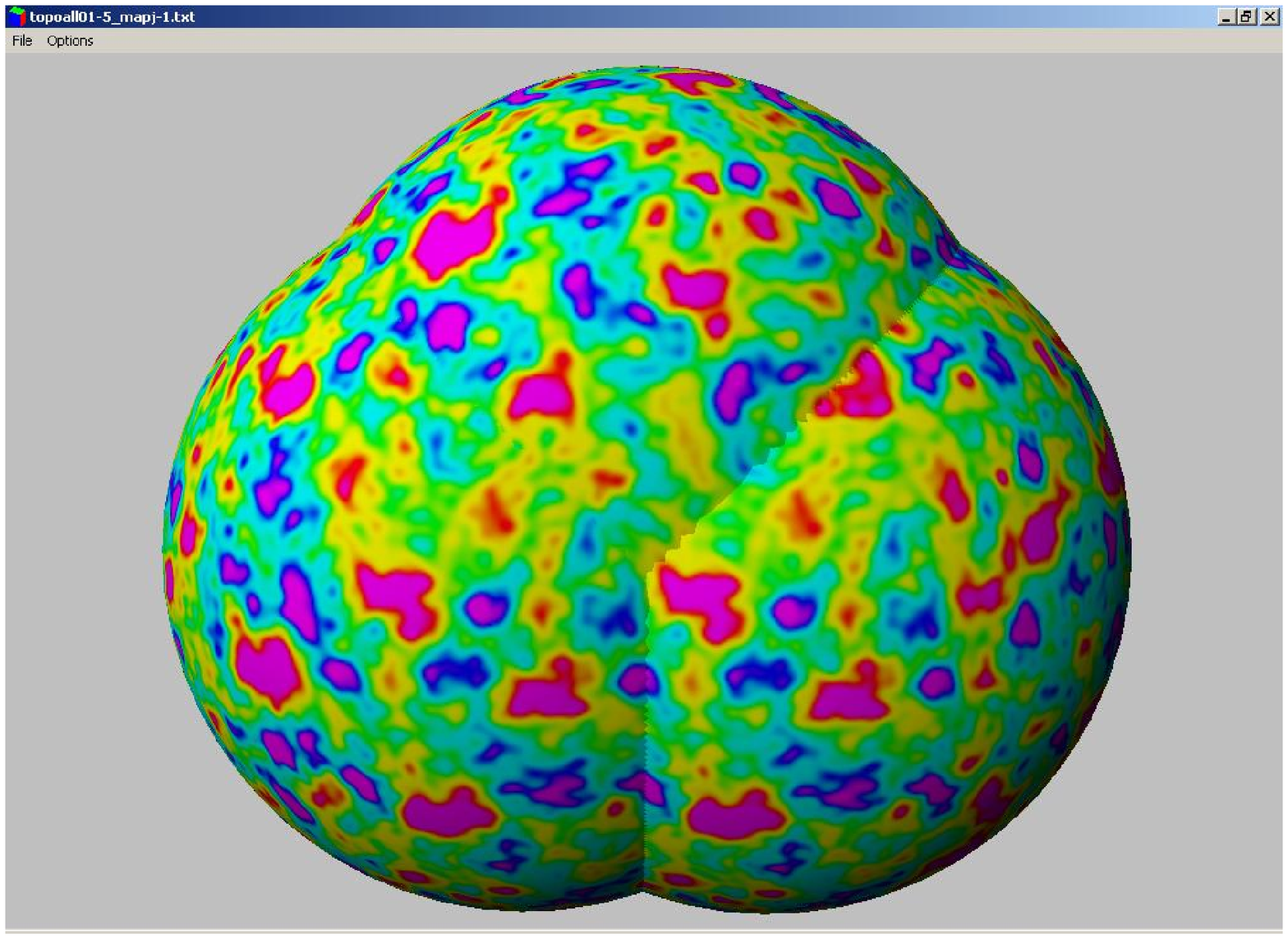,width=3.5in}}
\centerline{\psfig{file=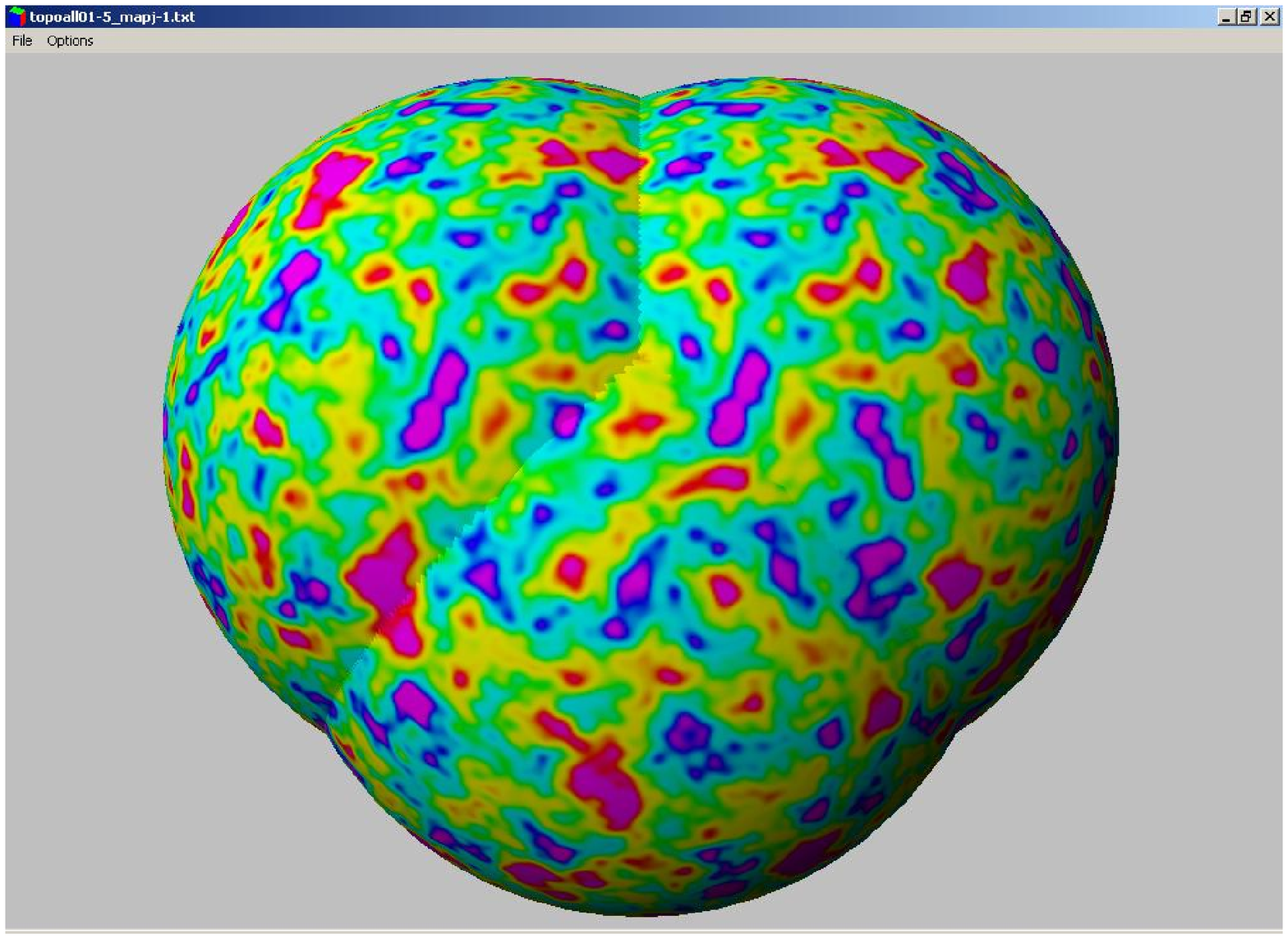,width=3.5in}
            \psfig{file=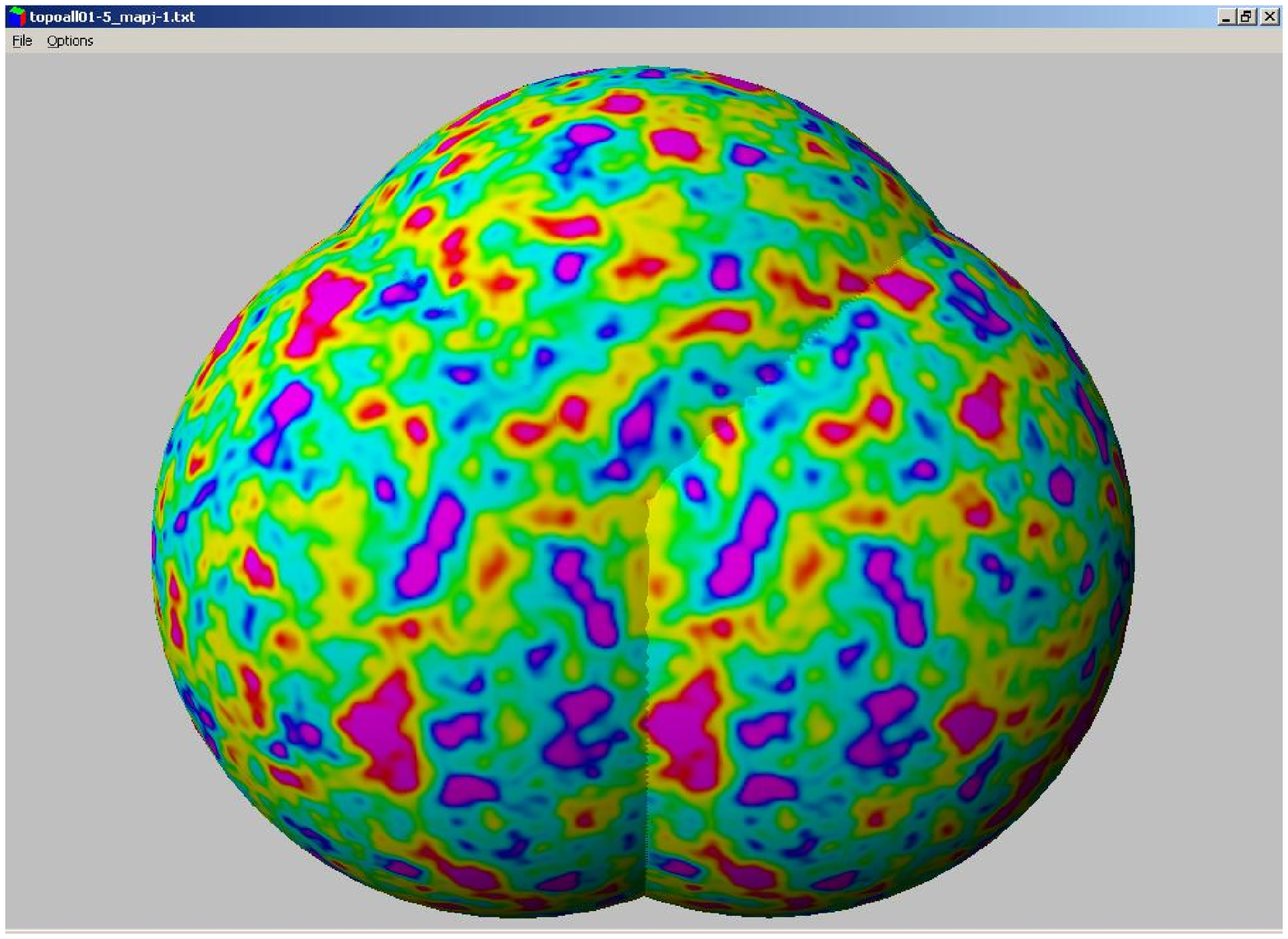,width=3.5in}}
\caption{The last scattering surface seen from outside for a
Hantzsche-Wendt space $E_6$ with $L_x = L_y = L_z = 0.64$ in units of
the last scattering surface. The geometry of the circle pairings
corresponds to the geometry of the fundamental polyhedron. The pairs
of matching circles are shown here looking from directions parallel to
the $x$-, $y$-, and $z$-axes, respectively.  Only the Sachs-Wolfe
contribution has been depicted here.}
\label{plot5}
\end{figure}

\subsubsection{Klein space}

Klein space is generated by two glide reflections
\begin{equation}
\VECTROISD{x}{y}{z}
\; \mapsto \;
   \MATTROISD{1}{0}{0}
             {0}{- 1}{0}
             {0}{0}{\;\;\; 1} \VECTROISD{x}{y}{z}
 + \VECTROISD{L_x / 2}{L_y / 2}{0} , \qquad
\VECTROISD{x}{y}{z}
\; \mapsto \;
   \MATTROISD{1}{0}{0}
             {0}{- 1}{0}
             {0}{0}{\;\;\; 1} \VECTROISD{x}{y}{z}
 + \VECTROISD{L_x / 2}{- L_y / 2}{0} , \qquad
\label{KleinGlideReflections}
\end{equation}
along with a simple translation
\begin{equation}
\VECTROISD{x}{y}{z} \; \mapsto \; \VECTROISD{x}{y}{z} + \VECTROISD{0}{0}{L_z} .
\label{KleinSimpleTranslation}
\end{equation}
The first (resp. second) glide reflection corresponds to the two
upper (resp. lower) faces of the hexagonal prism in
Figure~\ref{FigureCompactSpaces}, taking one to the other so that
the small dark-colored (resp. light-colored) windows match.  The
simple translation takes the front hexagonal face to the back
hexagonal face so that the doors match.

The square of either glide reflection is a horizontal translation $\bT
= (L_x, 0, 0)$ (taking the unmarked left wall to the unmarked right
wall in Figure~\ref{FigureCompactSpaces}), while the composition of
one glide reflection with the inverse of the other is a vertical
translation $\bT = (0, L_y, 0)$.  Thus the Klein space is the two-fold
quotient of a rectangular 3-torus of size $(L_x, L_y, L_z)$.

To find the Klein space's eigenmodes, we begin with the
modes~(\ref{RectangularTorusBasis}) of the rectangular 3-torus and
ask which remain invariant under the glide
reflections~(\ref{KleinGlideReflections}).  The Invariance Lemma
shows the Klein space's eigenmodes have the orthonormal basis
\begin{equation}
   \begin{array}{lcl}
     \frac{1}{\sqrt{2}}
       \left[\Upsilon_{2\pi(\frac{n_x}{L_x},\frac{n_y}{L_y},\frac{n_z}{L_z})}
        + (-1)^{n_x + n_y} \Upsilon_{2\pi(\frac{n_x}{L_x},-\frac{n_y}{L_y},\frac{n_z}{L_z})}
       \right]
       & \quad \mbox{for} \quad
       & n_y \in Z^+,\;n_x, n_z \in Z, \\
     \Upsilon_{2\pi(\frac{n_x}{L_x},0,\frac{n_z}{L_z})}
       & \quad \mbox{for} \quad
       & n_x \in 2Z,\;n_z \in Z. \\
   \end{array}
\label{PlainKleinBasis}
\end{equation}

One can easily check that
\begin{equation}
\Upsilon_{k_x, k_y, k_z}^{[E_7]*}
 = (- 1)^{n_x + n_y} \Upsilon_{- k_x, k_y, - k_z}^{[E_7]} ,
\end{equation}
when $n_y \in Z^+, \; n_x, n_z \in Z$ and that $\Upsilon_{k_x, k_y,
k_z}^{[E_7]}$ is real otherwise. It follows that the analog of
Eq.~(\ref{random_relation}) is given by
\begin{enumerate}
\item when $n_y \in Z^+, \; n_x, n_z \in Z$, $\hat e_\bk$ must satisfy
\begin{equation}
\hat e_{k_x, k_y, k_z}^* = (- 1)^{n_x + n_y} \hat e_{- k_x, k_y, - k_z} .
\end{equation}
It is thus a real random variable when $k_x = k_z = 0$ and $n_y \in 2
Z$ and a purely imaginary random variable when when $k_x = k_z = 0$
and $n_y \notin 2 Z$,
\item when $n_x \in 2 Z$, $n_y = 0$ and $n_z \in Z$, $\hat
e_{k_x, 0, k_z}$ is a real random variable.
\end{enumerate}
\begin{figure}
\centerline{\psfig{file=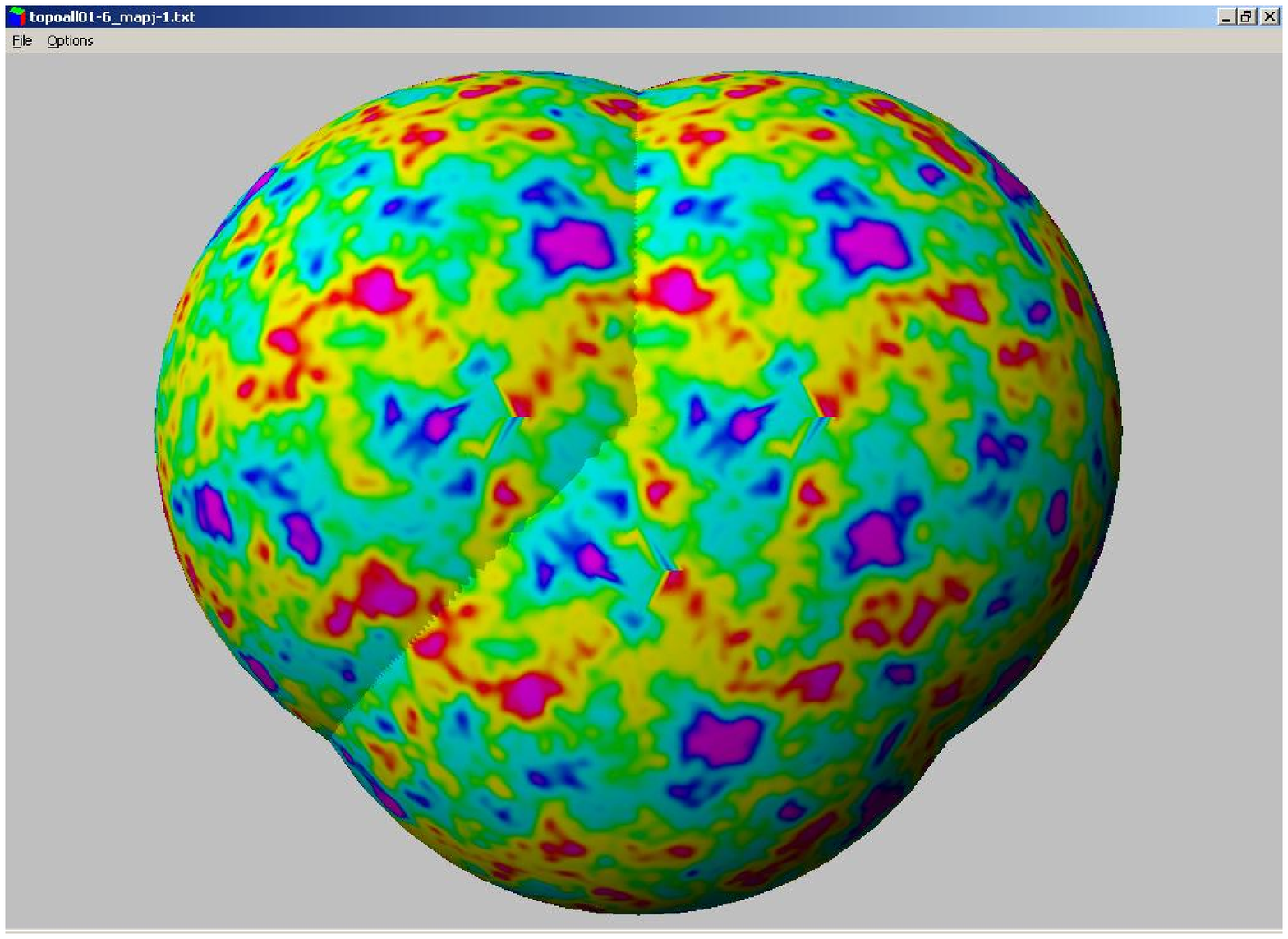,width=3.5in}
            \psfig{file=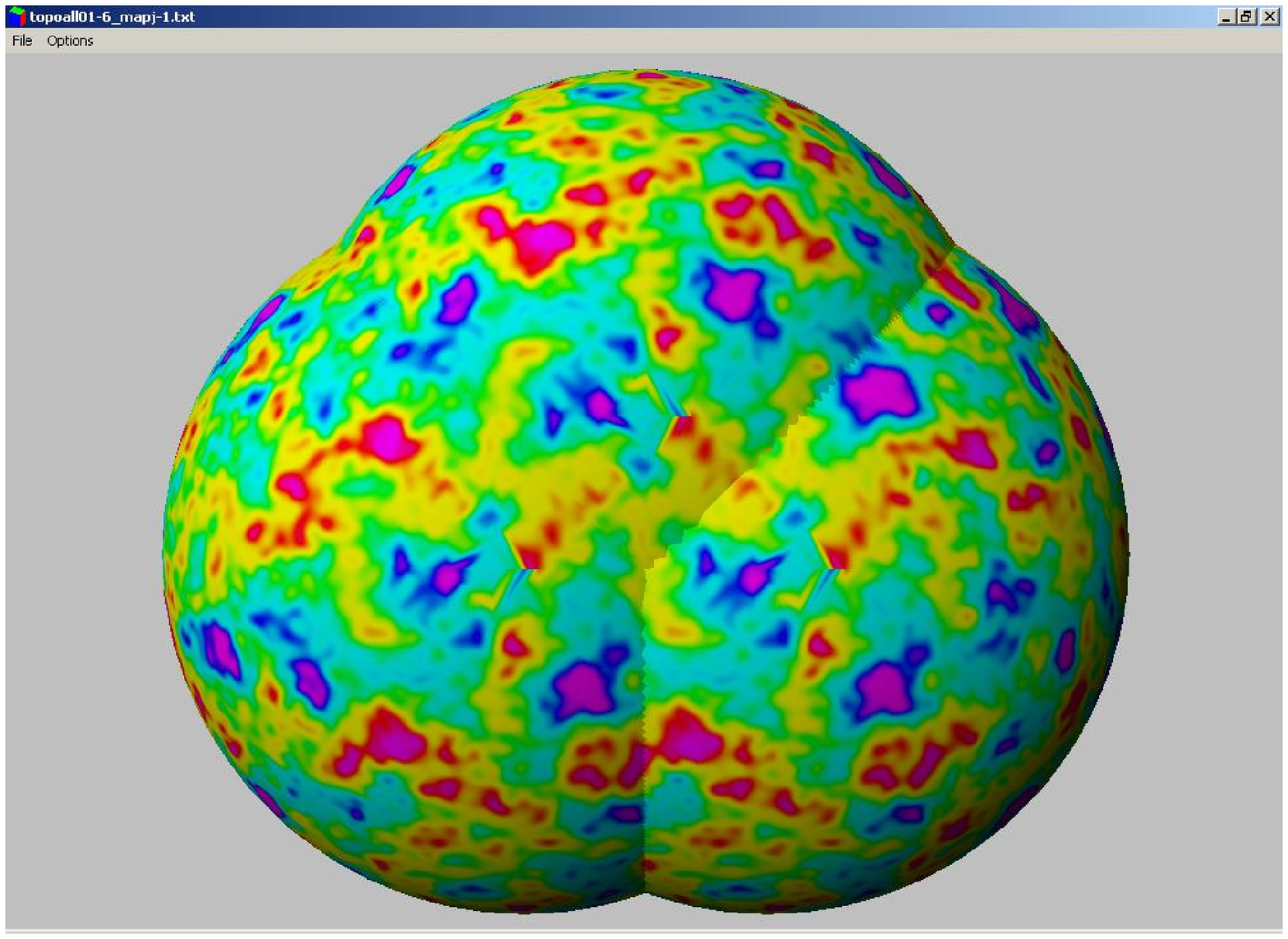,width=3.5in}}
\centerline{\psfig{file=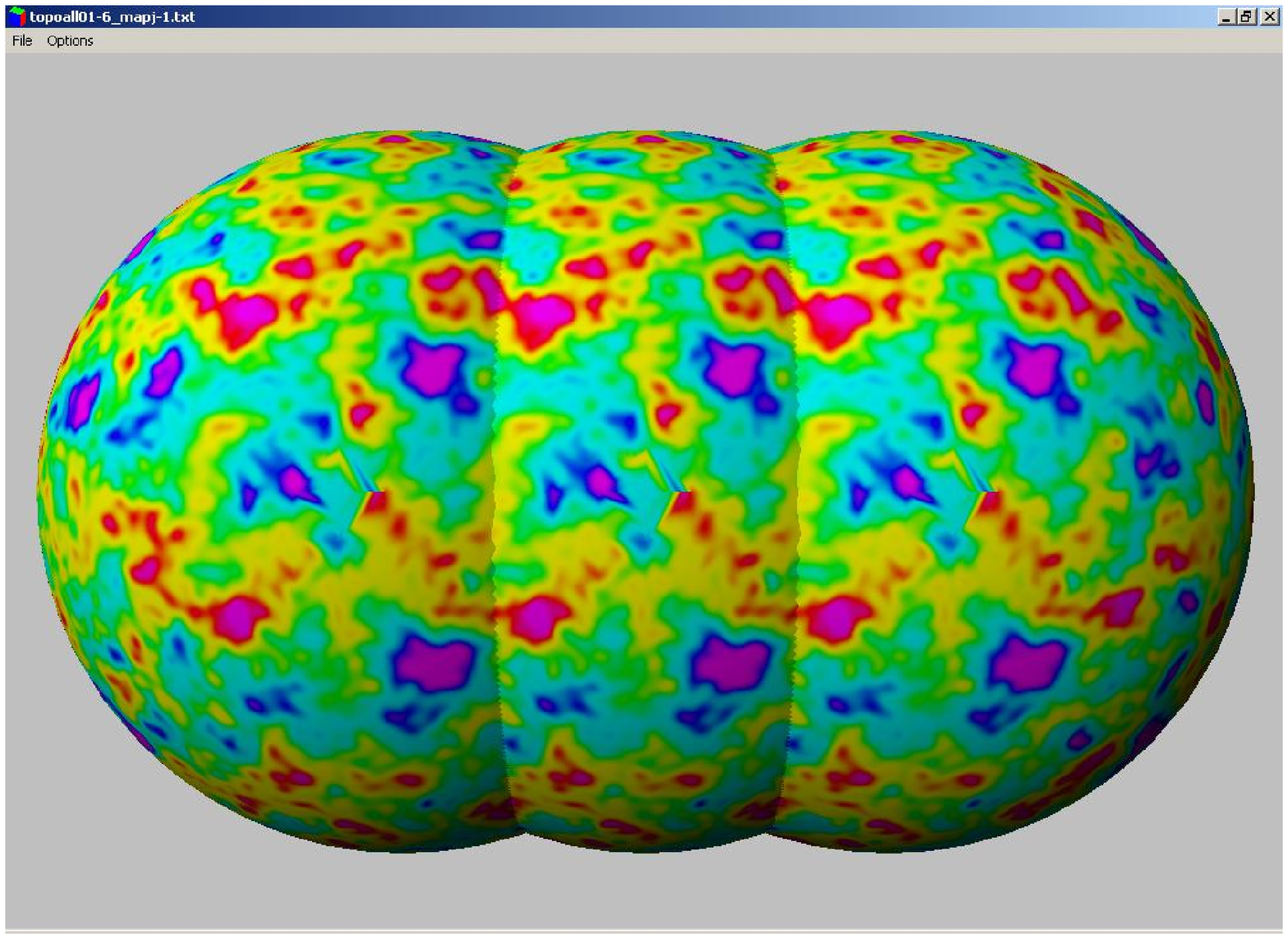,width=3.5in}
            \psfig{file=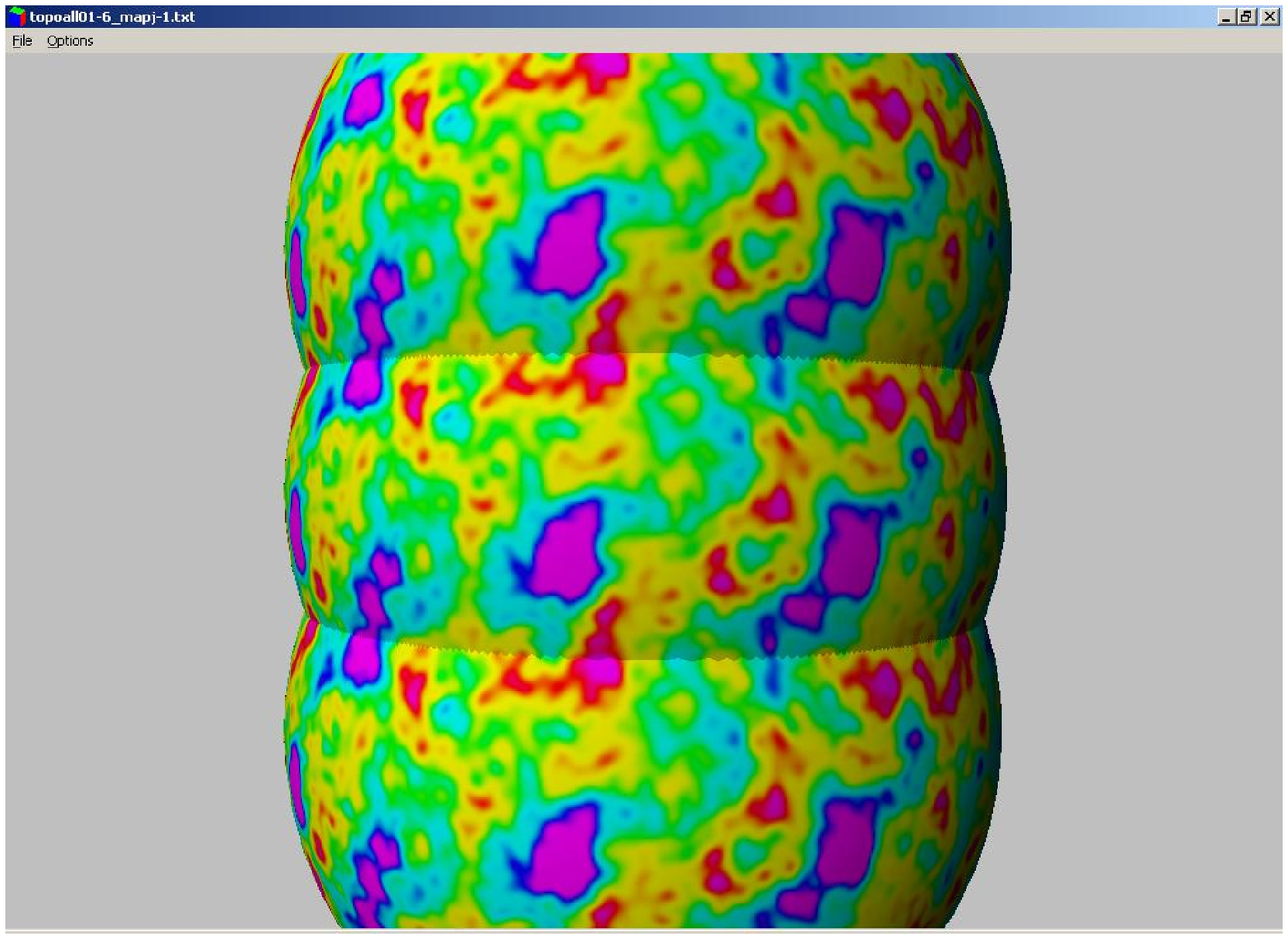,width=3.5in}}
\caption{The last scattering surface seen from outside for a Klein
space space $E_7$ with $L_x = L_y = L_z = 0.64$. We present four pairs
of matched circles.  The lower right figure corresponds to the
translation along the $z$ axis while the three other plots corresponds
to the holonomies acting in the $xy$-plane. Only the Sachs-Wolfe
contribution has been depicted here.}
\label{plot6}
\end{figure}

\subsubsection{Klein space with horizontal flip}

The Klein space with horizontal flip is a two-fold quotient of the
plain Klein space.  It includes the same two glide
reflections~(\ref{KleinGlideReflections}) as the Klein space, but
adds a square root
\begin{equation}
\VECTROISD{x}{y}{z}
\; \mapsto \;
\MATTROISD{- 1}{0}{0}
          {0}{\;\;\; 1}{0}
          {0}{0}{\;\;\; 1} \VECTROISD{x}{y}{z} + \VECTROISD{0}{0}{L_z / 2} , 
\label{KleinHorizontalFlip}
\end{equation}
of the Klein space's $L_z$-translation~(\ref{KleinSimpleTranslation}).
Because the Klein space with horizontal flip is a quotient of the
plain Klein space, every eigenmode of the former is automatically an
eigenmode of the latter (recall the reasoning of the first paragraph
of Section~\ref{SectionEigenmodesOfMulticonnected}).  Thus our task is
to decide which of the Klein space's
eigenmodes~(\ref{PlainKleinBasis}) are preserved by the new
generator~(\ref{KleinHorizontalFlip}).  The Invariance Lemma shows the
orthonormal basis to be
\begin{equation}
   \begin{array}{lcl}
     \frac{1}{2}
     \left[                       \Upsilon_{2\pi( \frac{n_x}{L_x}, \frac{n_y}{L_y}, \frac{n_z}{L_z})}
        + (-1)^{n_x + n_y}       \Upsilon_{2\pi( \frac{n_x}{L_x},-\frac{n_y}{L_y}, \frac{n_z}{L_z})}
        \right.& & \\
 \quad\left.  + (-1)^{n_z}             \Upsilon_{2\pi(-\frac{n_x}{L_x}, \frac{n_y}{L_y}, \frac{n_z}{L_z})}
        + (-1)^{n_x + n_y + n_z} \Upsilon_{2\pi(-\frac{n_x}{L_x},-\frac{n_y}{L_y}, \frac{n_z}{L_z})}
       \right]
       & \quad \mbox{for} \quad
       & n_x, n_y \in Z^+,\;n_z \in Z, \\
     \frac{1}{\sqrt{2}}
       \left[ \Upsilon_{2\pi(0,\frac{n_y}{L_y},\frac{n_z}{L_z})}
        + (-1)^{n_y} \Upsilon_{2\pi(0,-\frac{n_y}{L_y},\frac{n_z}{L_z})}
       \right]
       & \quad \mbox{for} \quad
       & n_y \in Z^+,\; n_z \in 2Z, \\
     \frac{1}{\sqrt{2}}
       \left[ \Upsilon_{2\pi(\frac{n_x}{L_x},0,\frac{n_z}{L_z})}
        + (-1)^{n_z} \Upsilon_{2\pi(-\frac{n_x}{L_x},0,\frac{n_z}{L_z})}
       \right]
       & \quad \mbox{for} \quad
       & n_x \in 2Z^+,\; n_z \in Z, \\
     \Upsilon_{2\pi(0,0,\frac{n_z}{L_z})}
       & \quad \mbox{for} \quad
       & n_z \in 2Z. \\
   \end{array}
\label{KleinHorizontalFlipModes}
\end{equation}

Following the same procedure as before, we obtain that the
analog of Eq.~(\ref{random_relation}) is given by
\begin{enumerate}
\item when $n_x, n_y \in Z^+, \; n_z \in Z$, $\hat e_\bk$ must
satisfy
\begin{equation}
\hat e_{k_x, k_y, k_z}^* = (- 1)^{n_x + n_y + n_z} \hat e_{k_x, k_y, - k_z} .
\end{equation}
It is thus a real random variable when $n_z = 0$ and $n_x + n_y \in 2
Z$ and a purely imaginary random variable when $n_z = 0$ and $n_x +
n_y \notin 2 Z$,
\item when $n_x = 0$, $n_y \in Z^+$ and $n_z \in Z$, $\hat e_{0, k_y,
k_z}$ must satisfy
\begin{equation}
\hat e_{0, k_y, k_z}^* = (- 1)^{n_y} \hat e_{0, k_y, - k_z} ,
\end{equation}
so that it is a real random variable when $k_z = 0$ and $n_y \in 2 Z$
and a purely imaginary random variable when $n_z = 0$ and $n_y \notin
2 Z$,
\item when $n_x \in Z^+$, $n_y = 0$ and $n_z \in Z$, $\hat e_{k_x, 0,
k_z}$ must satisfy
\begin{equation}
\hat e_{k_x, 0, k_z}^* = (- 1)^{n_z} \hat e_{k_x, 0, - k_z} ,
\end{equation}
so that it is a real random variable when $k_z = 0$,
\item when $n_z \in 2 Z$ it has to satisfy
\begin{equation}
\hat e_{0, 0, k_z}^* = \hat e_{0, 0, - k_z} .
\end{equation}
\end{enumerate}
\begin{figure}
\centerline{\psfig{file=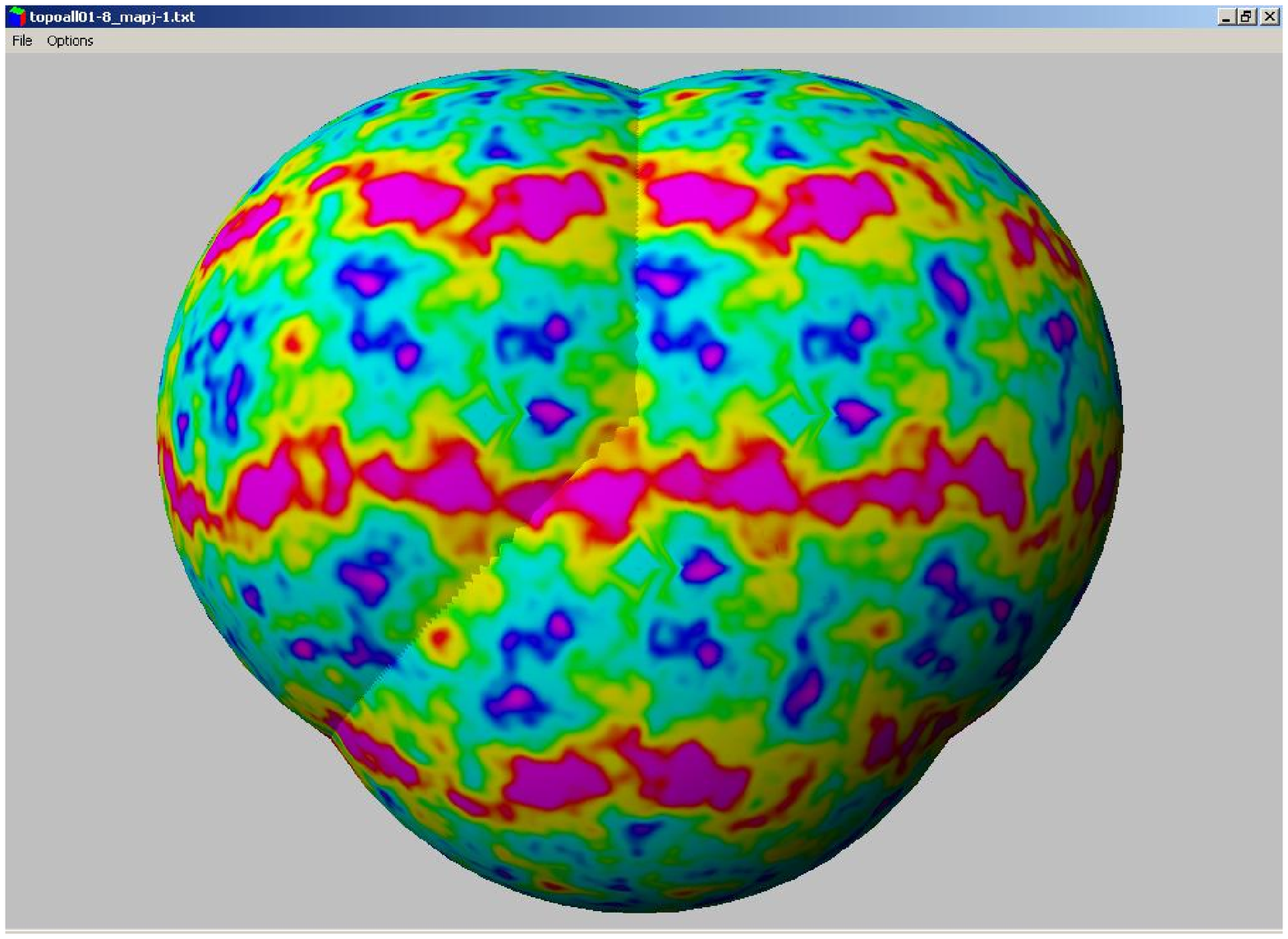,width=3.5in}
            \psfig{file=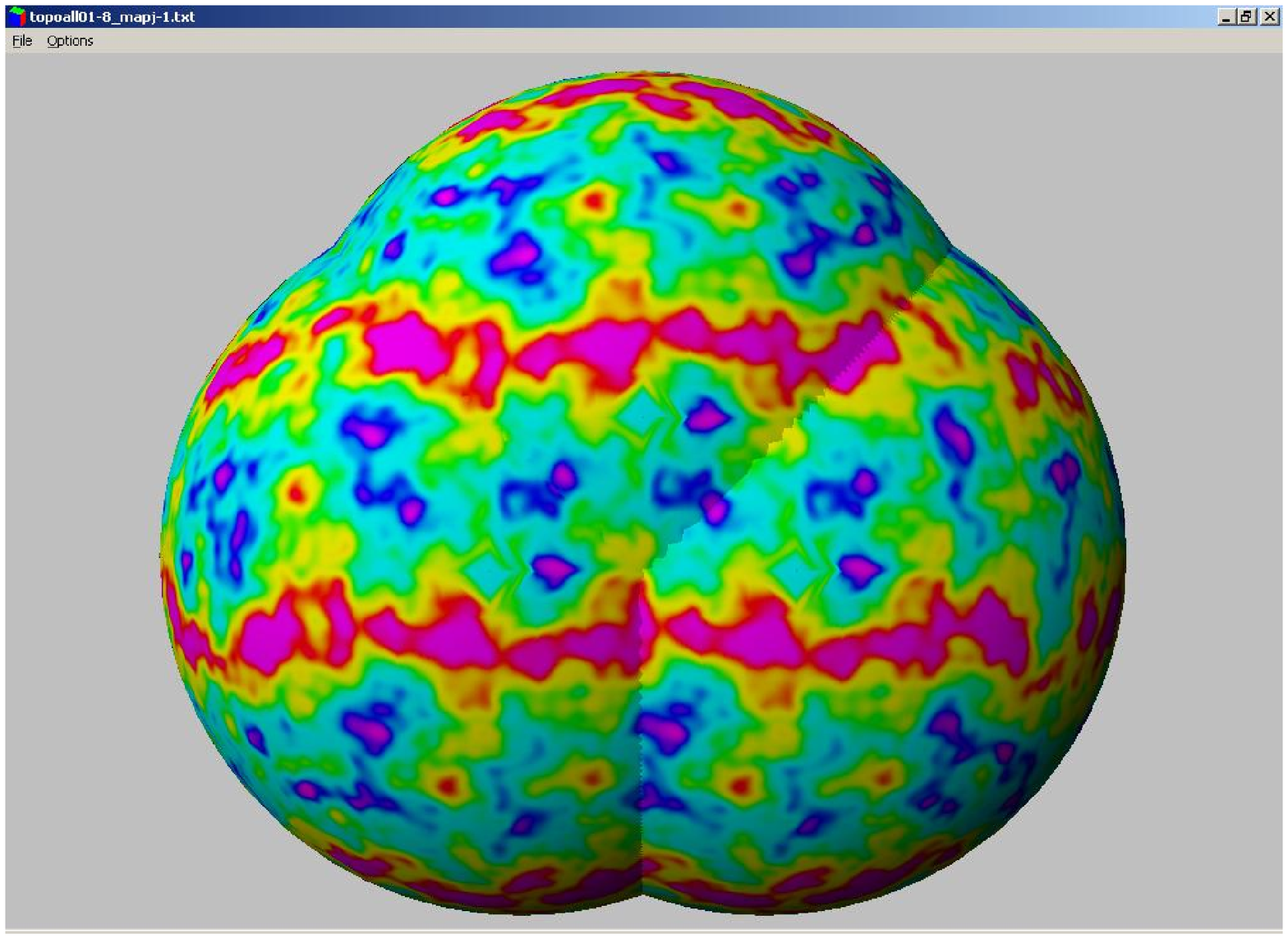,width=3.5in}}
\centerline{\psfig{file=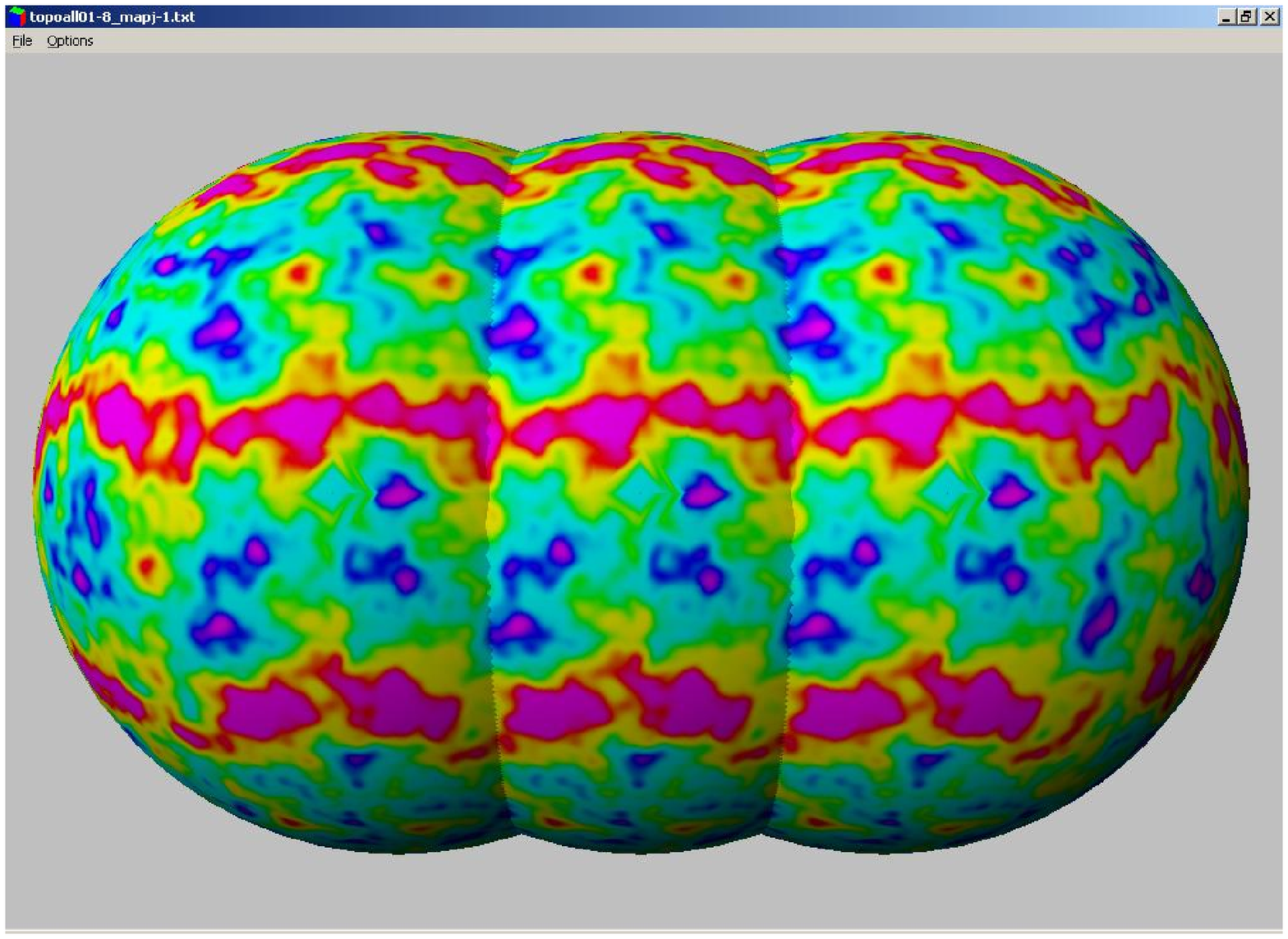,width=3.5in}
            \psfig{file=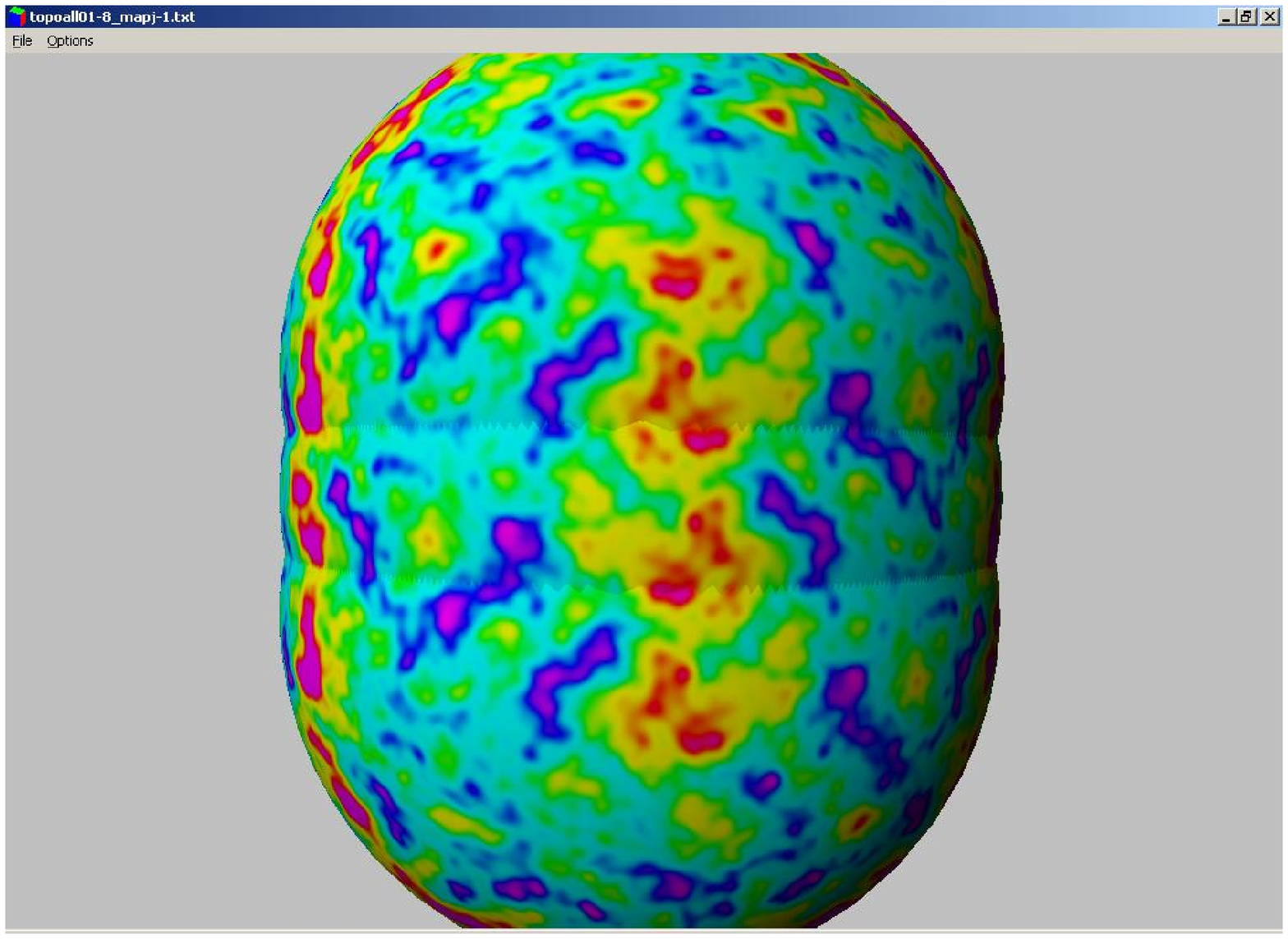,width=3.5in}}
\caption{The last scattering surface seen from outside for a Klein
space with horizontal flip $E_8$ with $L_x = L_y = L_z = 0.64$. We
present four pairs of matched circles.  The first three are identical
to those of Klein space, while the last one corresponds to the
transformation along the $z$-axis (the upper and lower circles are
mirror images of each other). Only the Sachs-Wolfe contribution has
been depicted here.}
\label{plot7}
\end{figure}

\subsubsection{Klein space with vertical flip}

The Klein space with vertical flip replaces
(\ref{KleinHorizontalFlip}) with
\begin{equation}
\VECTROISD{x}{y}{z}  
\; \mapsto \; 
\MATTROISD{1}{0}{0}
          {0}{- 1}{0}
          {0}{0}{\;\;\; 1} \VECTROISD{x}{y}{z} + \VECTROISD{0}{0}{L_z / 2} ,
\label{KleinVerticalFlip}
\end{equation}
whose action interchanges the modes $\Upsilon_{(k_x, k_y, k_z)}
\leftrightarrow (-1)^{n_z} \Upsilon_{(k_x, - k_y, k_z)}$. Consistency
with the glide reflections (\ref{KleinGlideReflections}), whose action
interchanges $\Upsilon_{(k_x, k_y, k_z)} \leftrightarrow (-1)^{n_x +
n_y} \Upsilon_{(k_x, - k_y, k_z)}$, requires $n_x + n_y \equiv n_z
\MOD{2}$.  Thus the orthonormal basis is
\begin{equation}
   \begin{array}{lcl}
     \frac{1}{\sqrt{2}}
       \left[\Upsilon_{2\pi(\frac{n_x}{L_x},\frac{n_y}{L_y},\frac{n_z}{L_z})}
        + (-1)^{n_x + n_y} \Upsilon_{2\pi(\frac{n_x}{L_x},-\frac{n_y}{L_y},\frac{n_z}{L_z})}
       \right]
       & \quad \mbox{for} \quad
       & n_y \in Z^+,\;n_x, n_z \in Z,\;n_x + n_y \equiv n_z \MOD{2}, \\
     \Upsilon_{2\pi(\frac{n_x}{L_x},0,\frac{n_z}{L_z})}
       & \quad \mbox{for} \quad
       & n_x,n_z \in 2Z. \\
   \end{array}
\end{equation}

Following the same procedure as before, we obtain that the
analog of Eq.~(\ref{random_relation}) is given by
\begin{enumerate}
\item when $n_y \in Z^+$ and $n_x, n_z \in Z$ with $n_x + n_y \equiv
n_z \MOD{2}$, $\hat e_\bk$ must satisfy
\begin{equation}
\hat e_{k_x, k_y, k_z}^* = (- 1)^{n_z} \hat e_{- k_x, k_y, - k_z} ,
\end{equation}
so that it is a real random variable when $k_x = k_z = 0$,
\item when $n_x, n_z \in 2 Z$, it has to be such that
\begin{equation}
\hat e_{k_x, 0, k_z}^* = \hat e_{- k_x, 0, - k_z} .
\end{equation}
\end{enumerate}
\begin{figure}
\centerline{\psfig{file=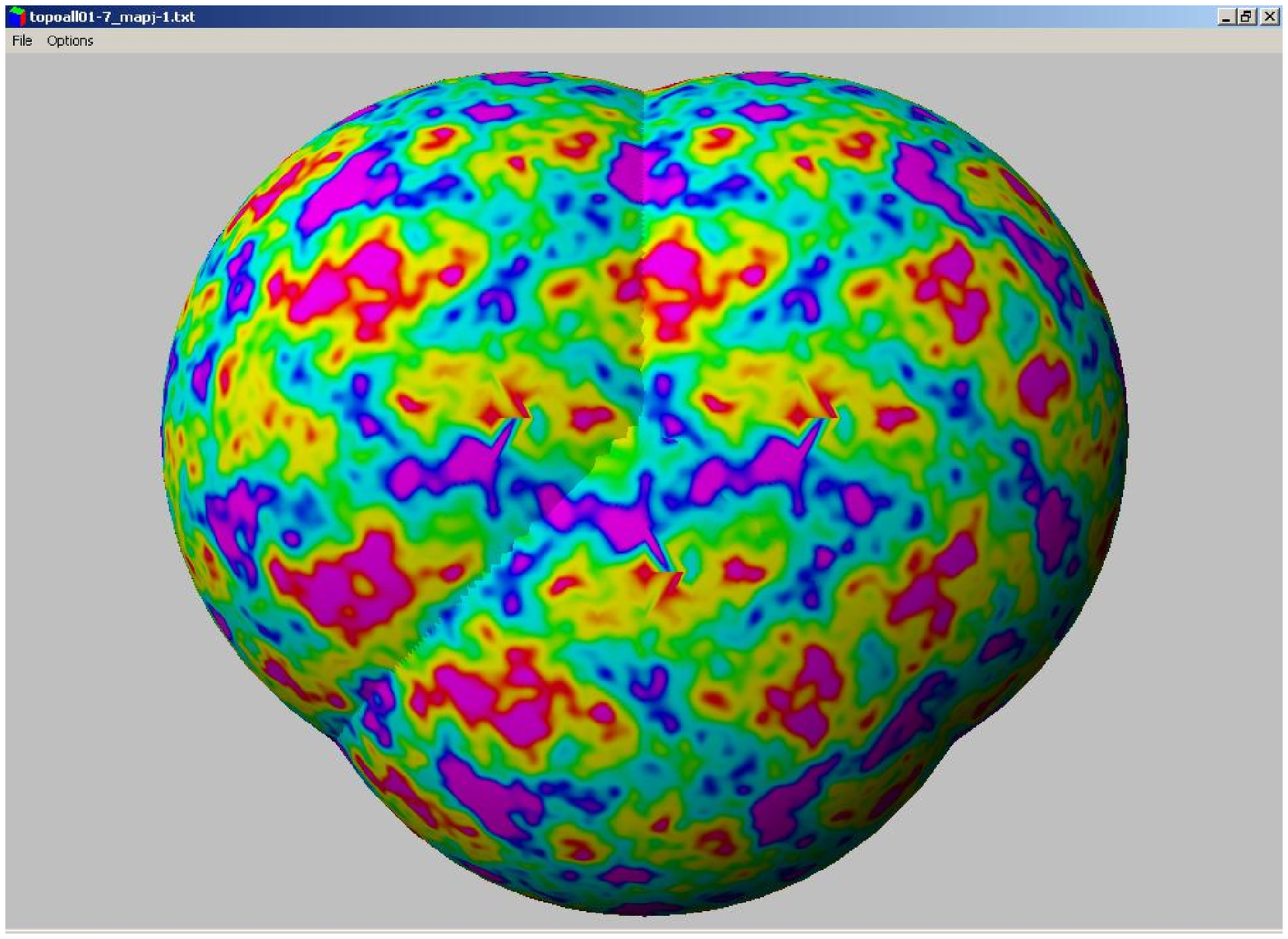,width=3.5in}
            \psfig{file=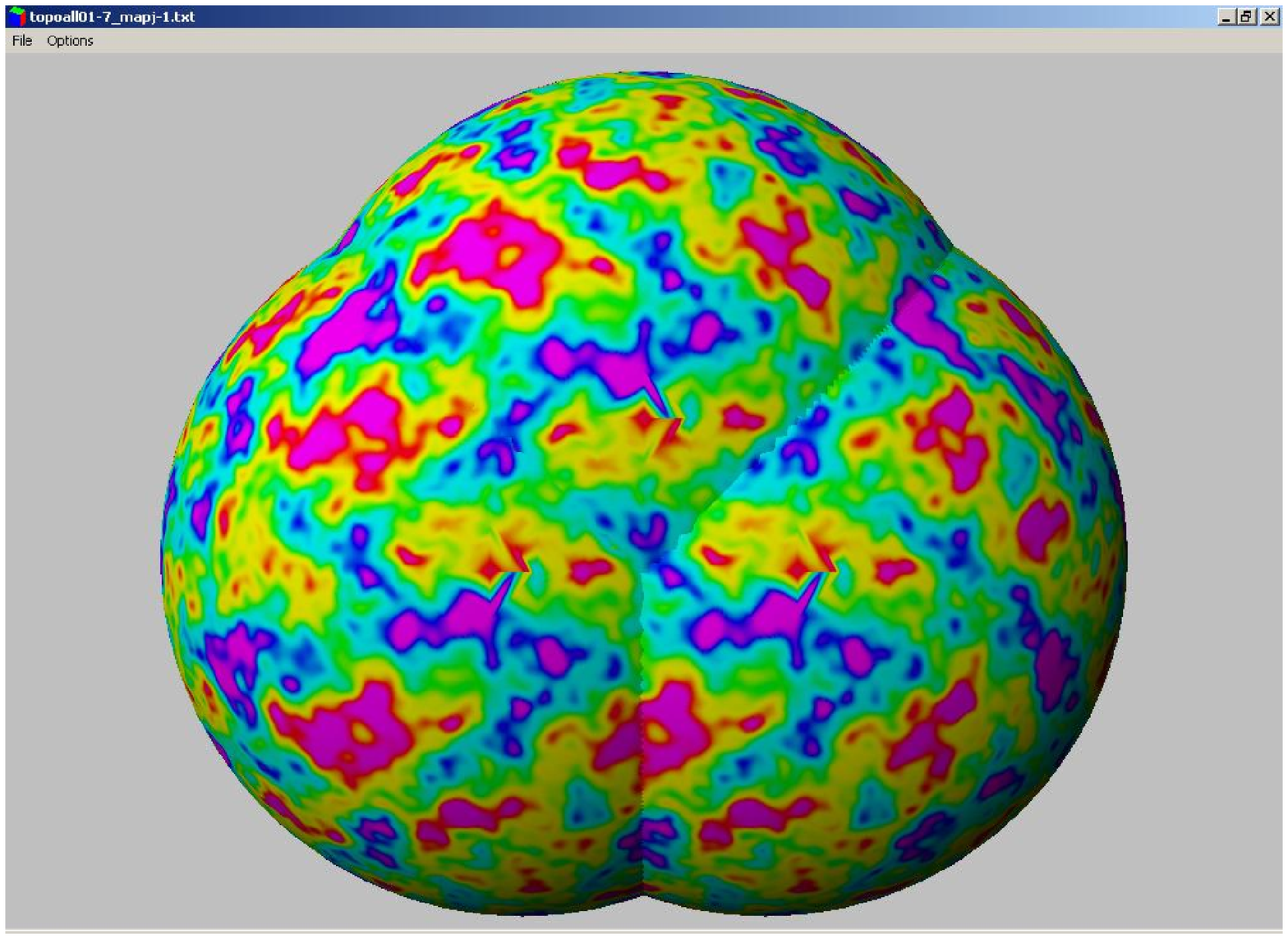,width=3.5in}}
\centerline{\psfig{file=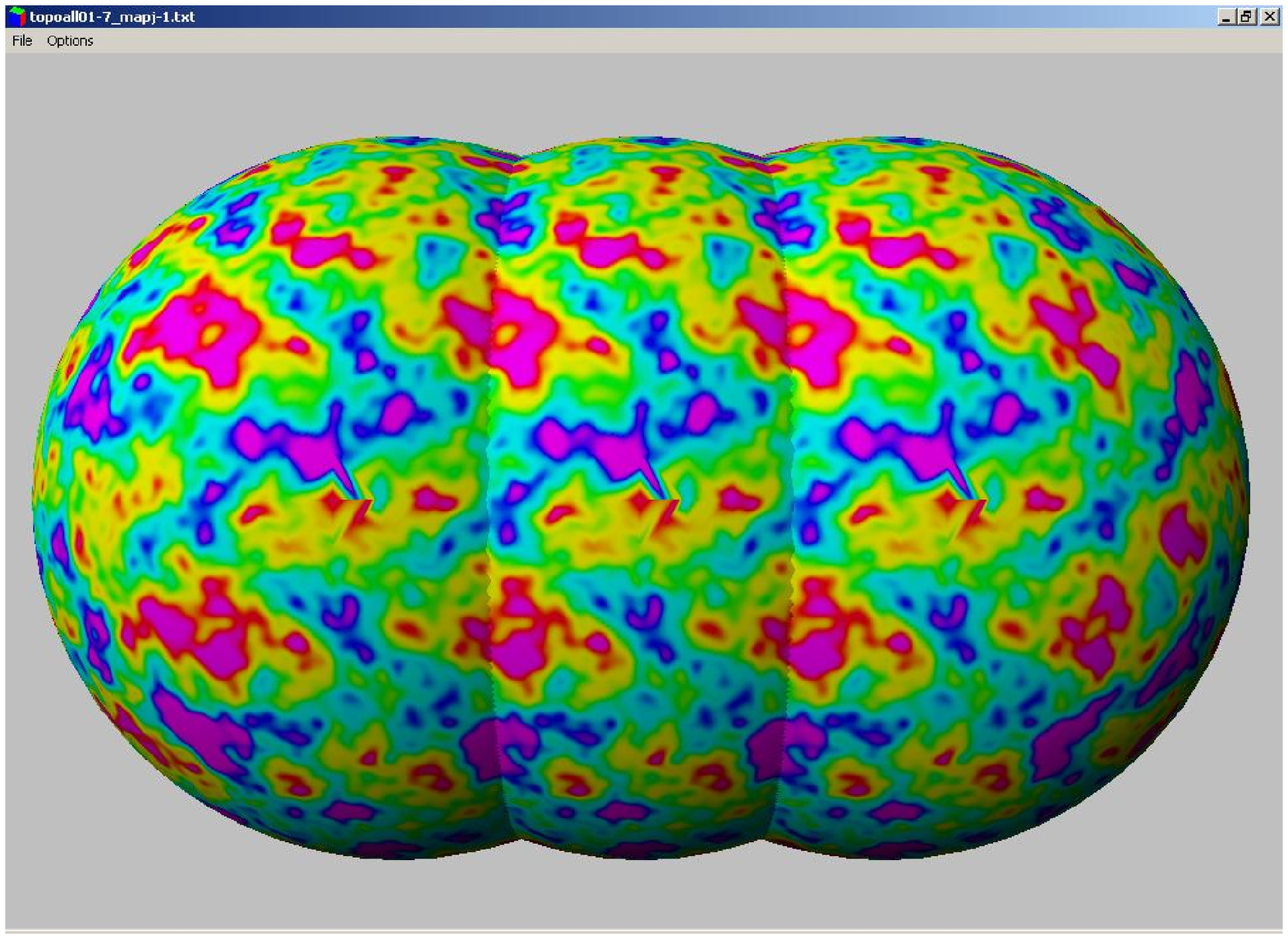,width=3.5in}
            \psfig{file=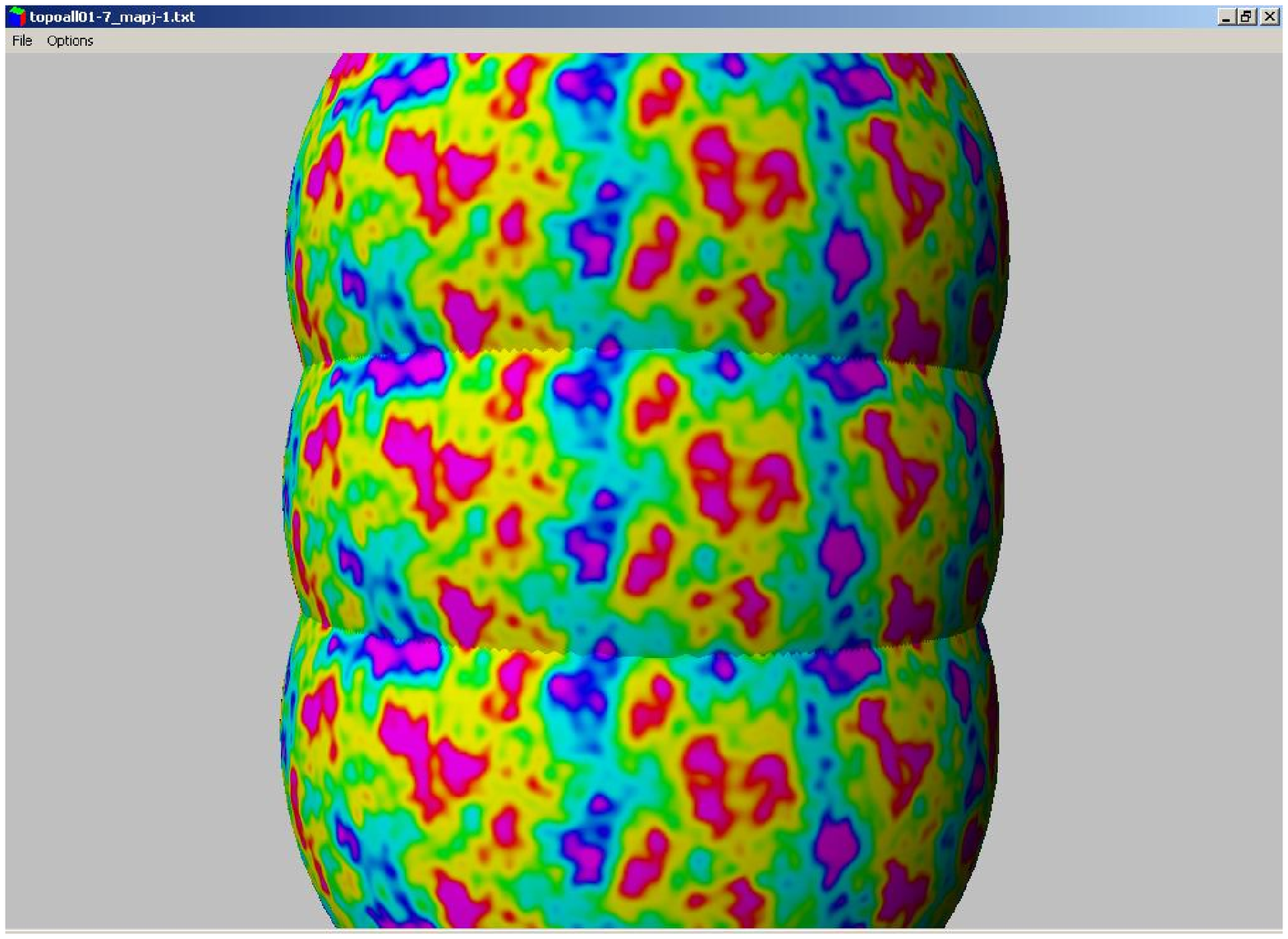,width=3.5in}}
\centerline{\psfig{file=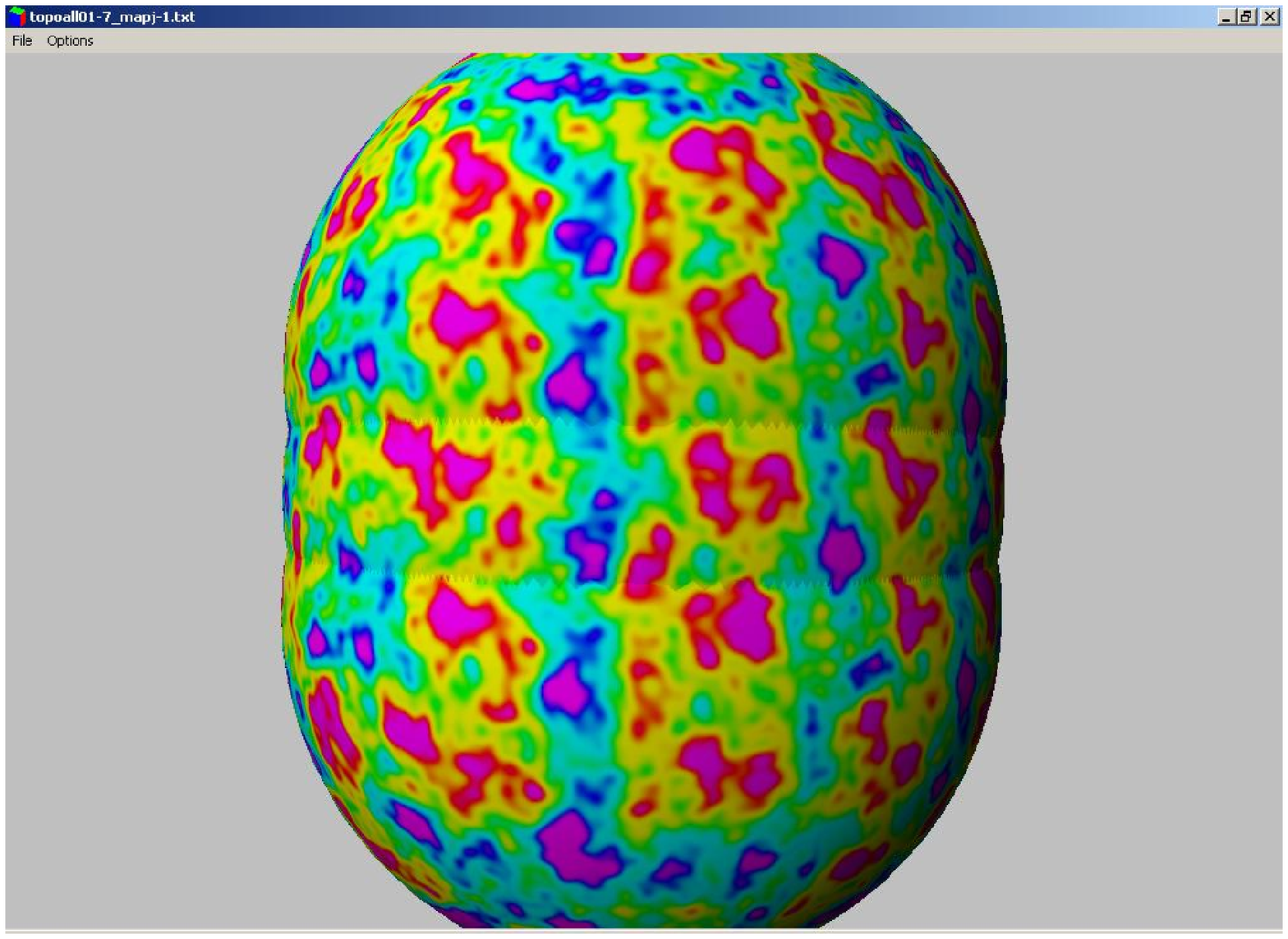,width=3.5in}
            \psfig{file=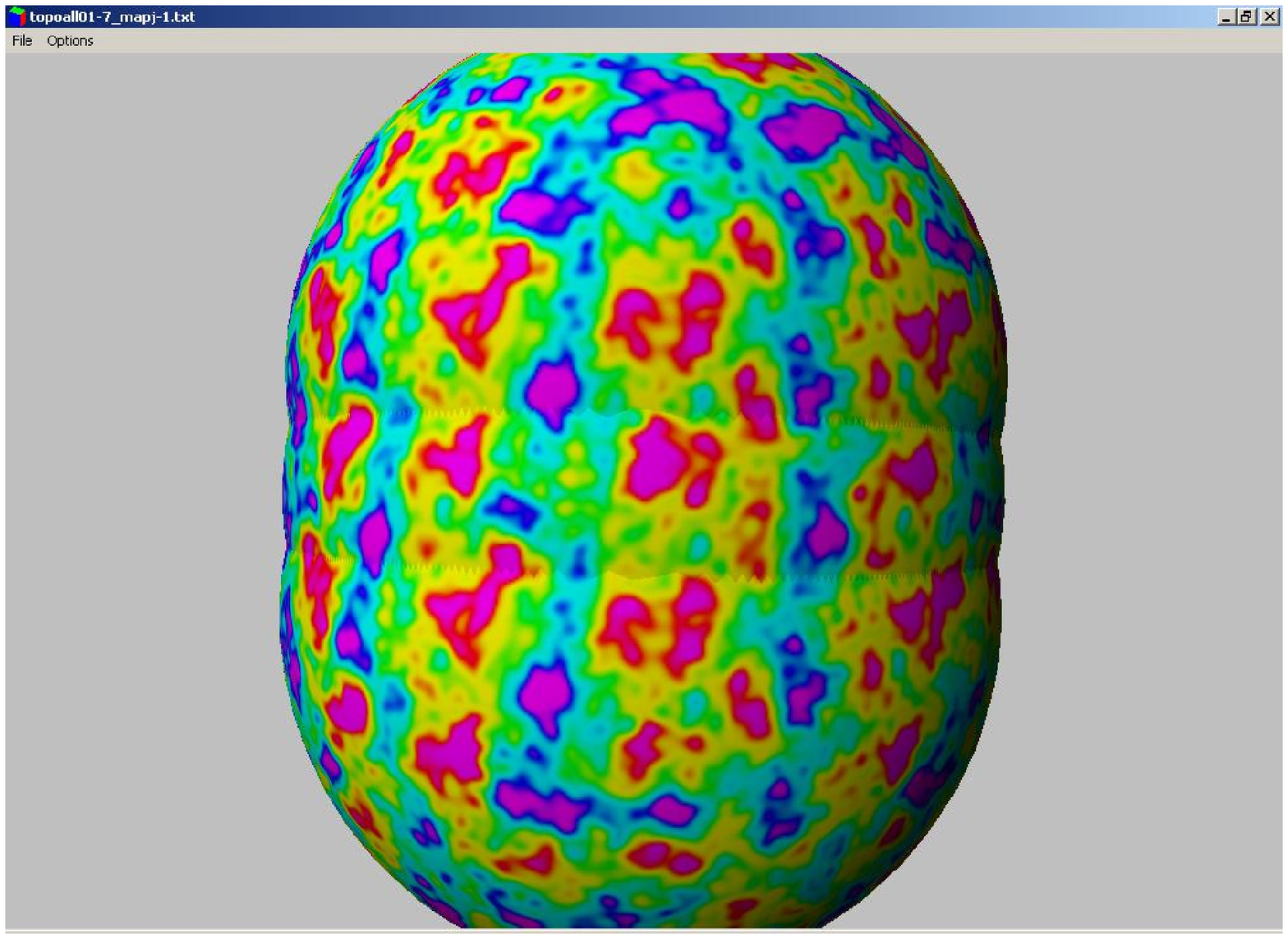,width=3.5in}}
\caption{The last scattering surface seen from outside for a Klein
space with vertical flip $E_9$ with $L_x = L_y = L_z = 0.64$. We
present the six pairs of matched circles.  The first two rows
correspond to the same holonomies as for the Klein space, while the
last row corresponds to the translation plus flip specific to this
space. Two different views of the same holonomy are
shown.}
\label{plot8}
\end{figure}

\subsubsection{Klein space with half turn}

The Klein space with half turn replaces
(\ref{KleinHorizontalFlip}) or (\ref{KleinVerticalFlip}) with
\begin{equation}
\VECTROISD{x}{y}{z}
\; \mapsto \;
\MATTROISD{- 1}{0}{0}
          {0}{- 1}{0}
          {0}{0}{\;\;\; 1} \VECTROISD{x}{y}{z} + \VECTROISD{0}{0}{L_z / 2} ,
\label{KleinHalfTurn}
\end{equation}
The orthonormal eigenbasis, which differs only slightly from that
of the Klein space with horizontal flip
(\ref{KleinHorizontalFlipModes}), is
\begin{equation}
   \begin{array}{lcl}
     \frac{1}{2}
       \left[                       \Upsilon_{2\pi( \frac{n_x}{L_x}, \frac{n_y}{L_y}, \frac{n_z}{L_z})}
        + (-1)^{n_x + n_y}       \Upsilon_{2\pi( \frac{n_x}{L_x},-\frac{n_y}{L_y}, \frac{n_z}{L_z})}\right. & & \\
 \quad \left. + (-1)^{n_z}             \Upsilon_{2\pi(-\frac{n_x}{L_x},-\frac{n_y}{L_y}, \frac{n_z}{L_z})}
        + (-1)^{n_x + n_y + n_z} \Upsilon_{2\pi(-\frac{n_x}{L_x}, \frac{n_y}{L_y}, \frac{n_z}{L_z})}
       \right]
       & \quad \mbox{for} \quad
       & n_x, n_y \in Z^+,\;n_z \in Z, \\
     \frac{1}{\sqrt{2}}
       \left[ \Upsilon_{2\pi(0,\frac{n_y}{L_y},\frac{n_z}{L_z})}
        + (-1)^{n_y} \Upsilon_{2\pi(0,-\frac{n_y}{L_y},\frac{n_z}{L_z})}
       \right]
       & \quad \mbox{for} \quad
       & n_y \in Z^+,\; n_z \in Z,\; n_y \equiv n_z \MOD{2}, \\
     \frac{1}{\sqrt{2}}
       \left[ \Upsilon_{2\pi(\frac{n_x}{L_x},0,\frac{n_z}{L_z})}
        + (-1)^{n_z} \Upsilon_{2\pi(-\frac{n_x}{L_x},0,\frac{n_z}{L_z})}
       \right]
       & \quad \mbox{for} \quad
       & n_x \in 2Z^+,\; n_z \in Z, \\
     \Upsilon_{2\pi(0,0,\frac{n_z}{L_z})}
       & \quad \mbox{for} \quad
       & n_z \in 2Z. \\
   \end{array}
\end{equation}

Following the same procedure as before, we obtain that the
analog of Eq.~(\ref{random_relation}) is given by
\begin{enumerate}
\item when $n_x, n_y \in Z^+$ and $n_z \in Z$,
$\hat e_\bk$ must satisfy
\begin{equation}
\hat e_{k_x, k_y, k_z}^* = (- 1)^{n_z} \hat e_{k_x, k_y, - k_z} ,
\end{equation}
so that it is a real random variable when $k_z = 0$,
\item when $n_y \in Z^+$, $n_z \in Z$ and $n_y \equiv n_z \MOD{2}$,
$\hat e_\bk$ must satisfy
\begin{equation}
\hat e_{0, k_y, k_z}^* = (- 1)^{n_y} \hat e_{0, k_y, - k_z} ,
\end{equation}
so that it is a real random variable when $k_z = 0$ and $n_y \in 2 Z$
and a purely imaginary random variable when $k_z = 0$ and $n_y \notin
2 Z$.
\item when $n_x \in 2 Z^+$ and $n_z \in Z$, $\hat e_\bk$ must satisfy
\begin{equation}
\hat e_{k_x, 0, k_z}^* = (- 1)^{n_z} \hat e_{k_x, 0, - k_z} ,
\end{equation}
so that it is a real random variable when $k_z = 0$
\item when $n_z \in 2 Z$, it has to be such that
\begin{equation}
\hat e_{0, 0, k_z}^* = \hat e_{0, 0, - k_z} .
\end{equation}
\end{enumerate}
\begin{figure}
\centerline{\psfig{file=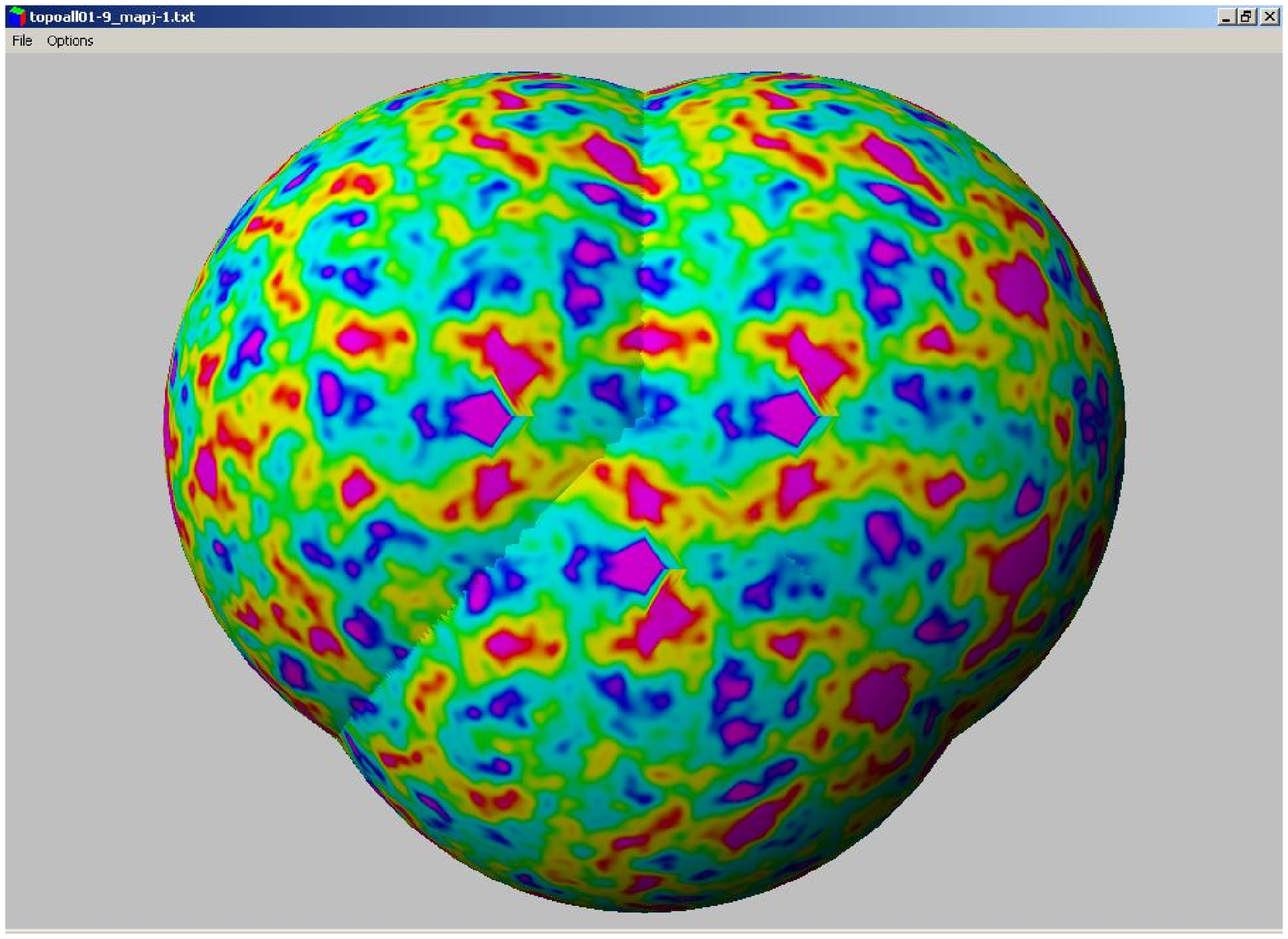,width=3.5in}
            \psfig{file=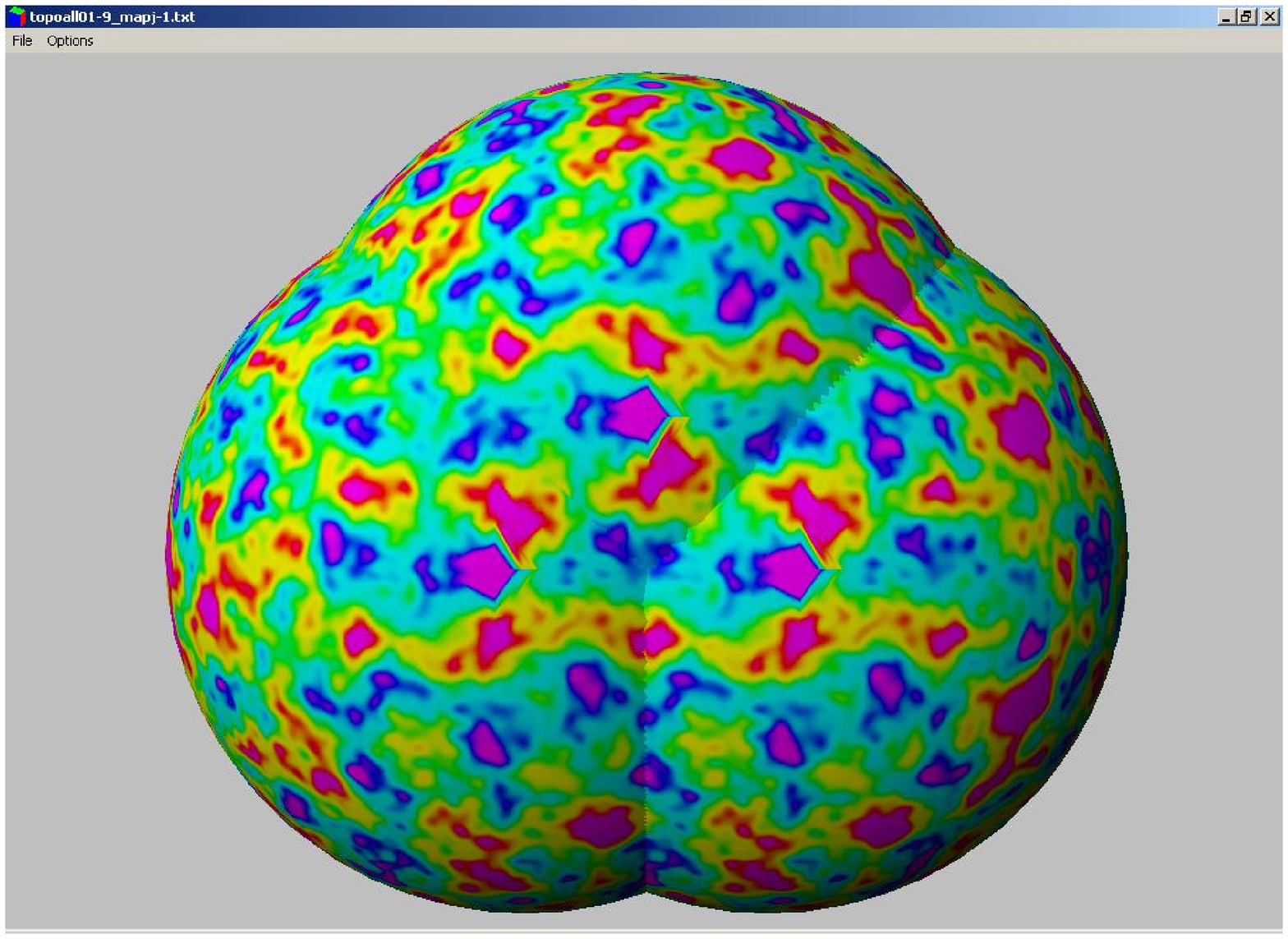,width=3.5in}}
\centerline{\psfig{file=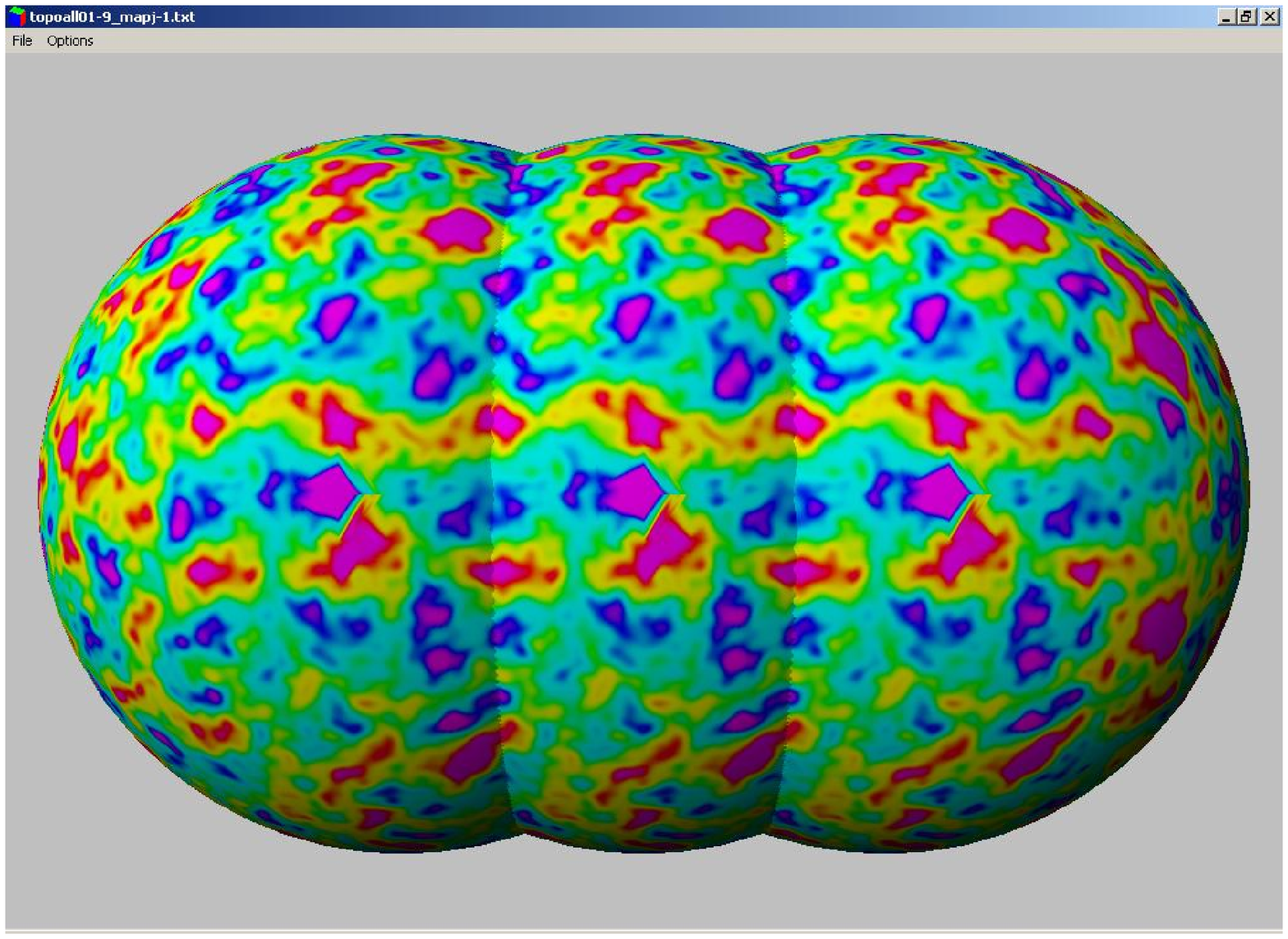,width=3.5in}
            \psfig{file=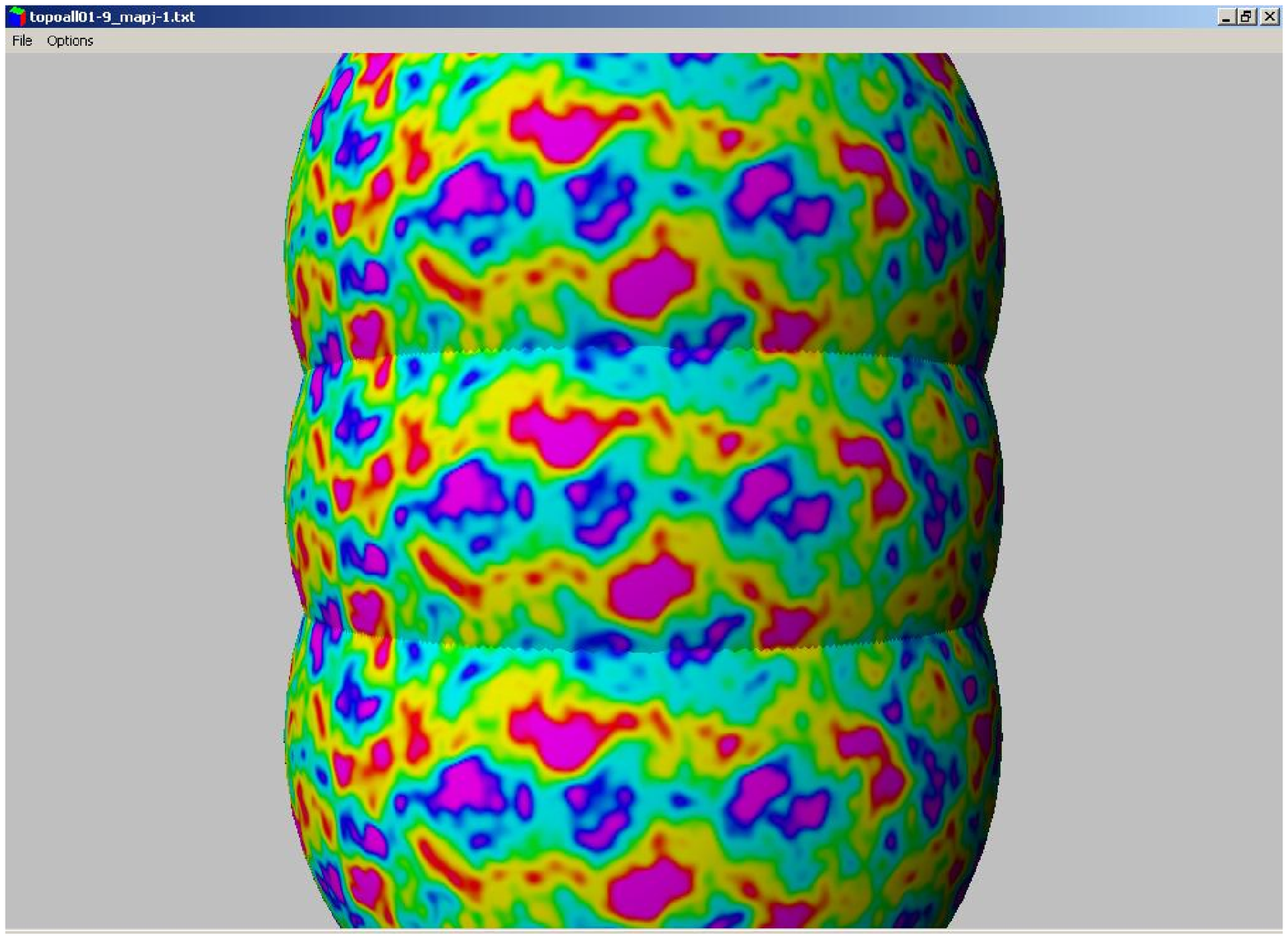,width=3.5in}}
\centerline{\psfig{file=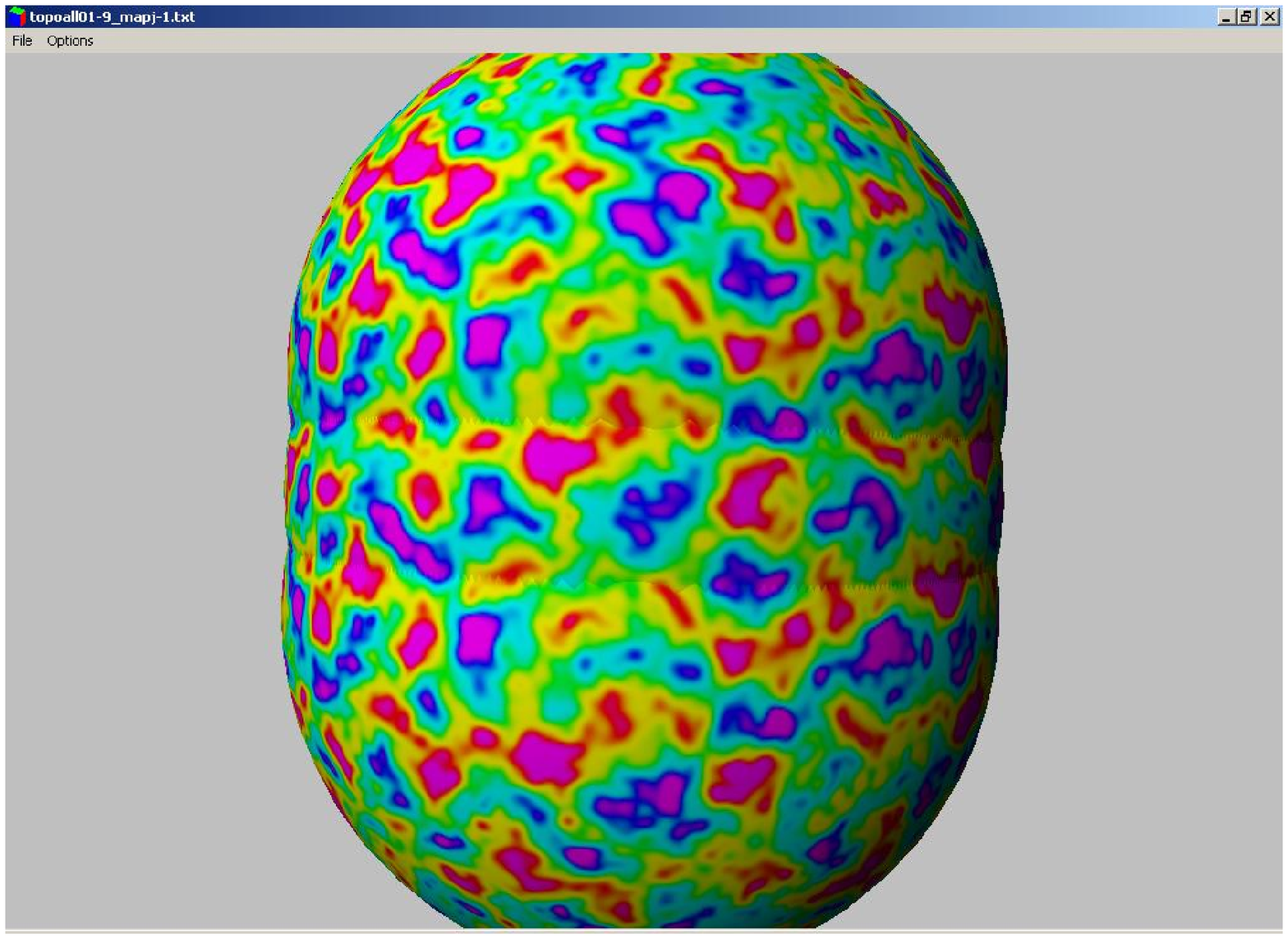,width=3.5in}
            \psfig{file=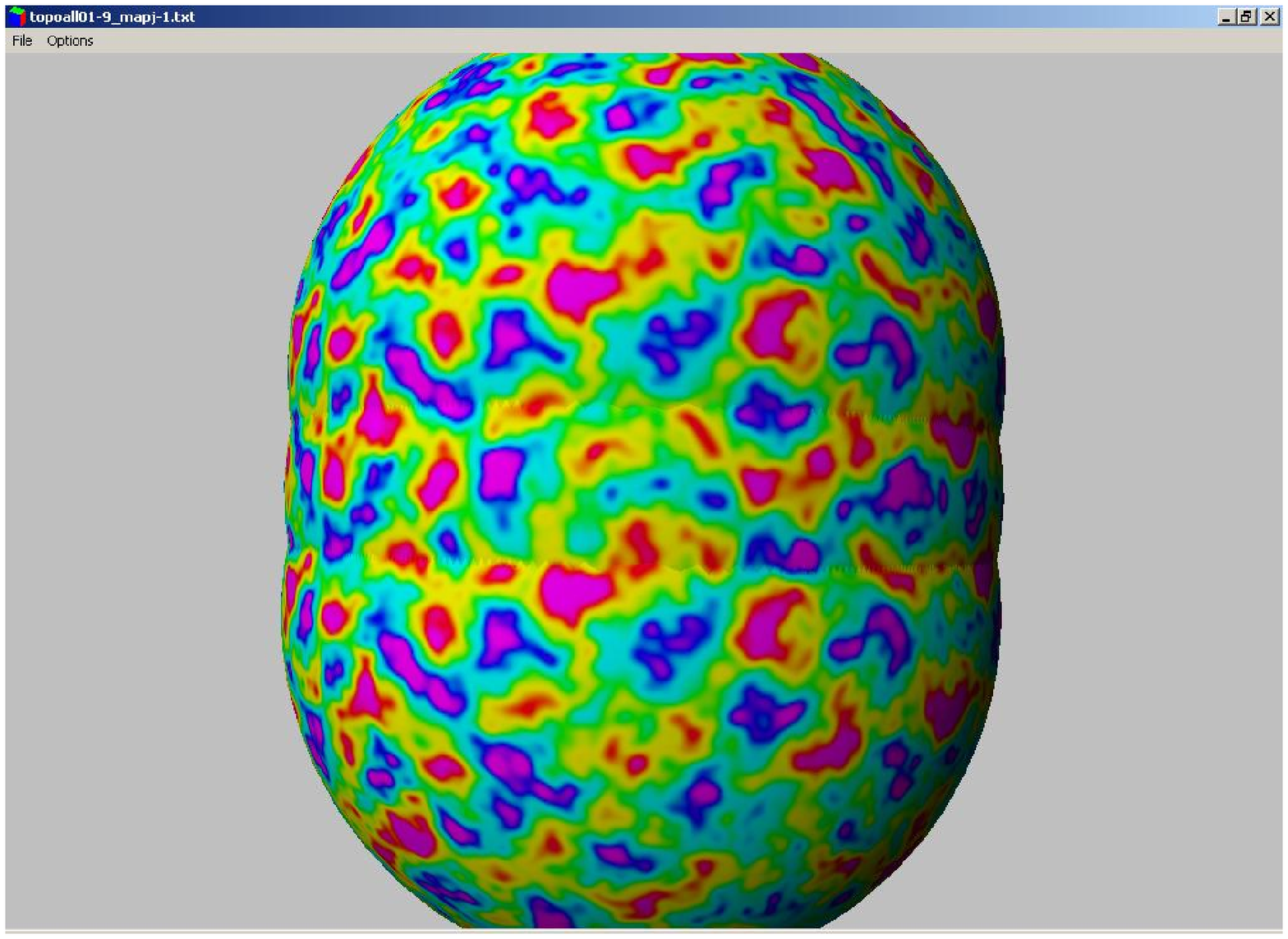,width=3.5in}}
\caption{The last scattering surface seen from outside for a Klein
space with half turn $E_{10}$ with $L_x = L_y = L_z = 0.64$. We
present the six pairs of matched circles.  The two upper left rows
correspond to the same holonomies as for the Klein space, while the
last row corresponds to the translation plus half turn holonomy
specific to this space. Two different views of the same holonomy are
shown.}
\label{plot9}
\end{figure}

\section{Doubly Periodic Spaces}
\label{SectionDoublyPeriodic}

We will first find the eigenmodes of the chimney space, and then
use them to find the eigenmodes of its quotients.

\subsection{Chimney Space}
\label{SubsectionChimneySpace}

Just as a 3-torus is the quotient of Euclidean space $\ETR$ under the
action of three linearly independent translations $\bT_1$, $\bT_2$ and
$\bT_3$, a {\it chimney space} is the quotient of $\ETR$ by only two
linearly independent translations $\bT_1$ and $\bT_2$. Its fundamental
domain is an infinitely tall chimney whose cross section is a
parallelogram (Figure~\ref{FigureChimneySpaces}). And just as the
allowable wave vectors $\bk$ for an eigenmode $\Upsilon_\bk$ of a
3-torus were defined by the intersection of three families of parallel
planes (Section~\ref{SubsectionThreeTorus}), the allowable wave
vectors $\bk$ for an eigenmode $\Upsilon_\bk$ of a chimney space are
defined by the intersection of two families of parallel planes.  Thus
the allowable wave vectors form a latticework of parallel lines.

The most important special case is the {\it rectangular chimney
space} generated by two orthogonal translations
\begin{eqnarray}
\bT_1 & = (L_x, 0, 0) , \nonumber \\
\bT_2 & = (0, L_y, 0) ,
\label{ChimneySpaceTranslations}
\end{eqnarray}
in which case the allowed wave vectors $\bk$ take the form
\begin{equation}
\label{RectangularChimneyBasis}
  \bk = 2\pi\,\left(\frac{n_x}{L_x},\, \frac{n_y}{L_y},\, r_z\right)
\end{equation}
for integer values of $n_x$ and $n_y$ and real values of $r_z$.
The corresponding orthonormal basis
\begin{equation}
   \begin{array}{lcl}
       \Upsilon_{2\pi(\frac{n_x}{L_x},\frac{n_y}{L_y},r_z)}
       & \quad \mbox{for} \quad
       & n_x, n_y \in Z,\;r_z \in R
   \end{array}
\end{equation}
is continuous, not discrete as for the 3-torus.  Nevertheless,
restricting to a fixed modulus $k = |\bk|$ recovers a
finite-dimensional basis.

The next four spaces are quotients of the chimney space, so their
eigenmodes will form subspaces of those of the chimney space
itself.

As in the case of the torus, the random variable $\hat e_\bk$ is a
complex random variable satisfying
\begin{equation}
\hat e_\bk^* = \hat e_{- \bk} .
\end{equation}

\subsection{Quotients of the Chimney Space}
\label{SubsectionChimneySpaceQuotients}

\subsubsection{Chimney space with half turn}

The chimney space with half turn
(Figure~\ref{FigureChimneySpaces}) is generated by the rectangular
chimney space's $x$-translation $\bT_1 = (L_x, 0, 0)$ along with
\begin{equation}
\VECTROISD{x}{y}{z}  
\; \mapsto \;
\MATTROISD{- 1}{0}{0}
          {0}{\;\;\; 1}{0}
          {0}{0}{- 1} \VECTROISD{x}{y}{z} + \VECTROISD{0}{0}{L_y / 2} .
\label{ChimneyHalfTurn}
\end{equation}
Even though the eigenmodes are not discrete, the Invariance Lemma
applies exactly as in the compact case, giving the eigenbasis
\begin{equation}
   \begin{array}{lcl}
     \frac{1}{\sqrt{2}}
       \left[ \Upsilon_{2\pi(\frac{n_x}{L_x},\frac{n_y}{L_y},r_z)}
        + (-1)^{n_y} \Upsilon_{2\pi(-\frac{n_x}{L_x},\frac{n_y}{L_y},-r_z)}
       \right]
       & \quad \mbox{for} \quad
       & (n_x \in Z^+, r_z \in R) \mbox{ or } (n_x = 0, r_z \in R^+), \\
     \Upsilon_{2\pi(0,\frac{n_y}{L_y},0)}
       & \quad \mbox{for} \quad
       & n_y \in 2Z. \\
   \end{array}
\end{equation}

This case is analogous to the case of the half turn space so the
analog of Eq.~(\ref{random_relation}) is given by
\begin{enumerate}
\item when $(n_x \in Z^+, r_z \in R)$ or $(n_x = 0, r_z \in R^+)$,
$\hat e_\bk$ satisfies
\begin{equation}
\hat e_{k_x, k_y, k_z}^* = (- 1)^{n_y} \hat e_{k_x, - k_y, k_z} .
\end{equation}
It is thus a real random variable when $n_y = 0$ and complex
otherwise,
\item when $n_y \in 2 Z$, $\hat e_\bk$ satisfies
\begin{equation}
\hat e_{0, k_y, 0}^* = \hat e_{0, - k_y, 0} .
\end{equation}
\end{enumerate}

\subsubsection{Chimney space with vertical flip}

The chimney space with vertical flip
(Figure~\ref{FigureChimneySpaces}) is generated by the translation
$\bT_1 = (L_x, 0, 0)$ along with
\begin{equation}
\VECTROISD{x}{y}{z}  
\; \mapsto \;
\MATTROISD{1}{0}{0}
          {0}{\;\;\; 1}{0}
          {0}{0}{- 1} \VECTROISD{x}{y}{z} + \VECTROISD{0}{L_y / 2}{0} .
\label{ChimneyVerticalFlip}
\end{equation}
The Invariance Lemma shows the orthonormal eigenbasis to be
\begin{equation}
   \begin{array}{lcl}
     \frac{1}{2}
       \left[ \Upsilon_{2\pi(\frac{n_x}{L_x},\frac{n_y}{L_y},r_z)}
        + (-1)^{n_y} \Upsilon_{2\pi(\frac{n_x}{L_x},\frac{n_y}{L_y},-r_z)}
       \right]
       & \quad \mbox{for} \quad
       & n_x, n_y \in Z,\; r_z \in R^+, \\
     \Upsilon_{2\pi(\frac{n_x}{L_x},\frac{n_y}{L_y},0)}
       & \quad \mbox{for} \quad
       & n_x \in Z,\; n_y \in 2Z. \\
   \end{array}
\end{equation}

One might also consider a chimney space with a vertical flip in
the $x$-direction as well as the $y$-direction.  Surprisingly,
such a space turns out to be equivalent to a chimney space with a
single flip, but with cross section a parallelogram rather than a
rectangle.  In other words, the chimney space with two flips has
the same topology as the chimney space with one flip, even though
they may differ geometrically.

Concerning the properties of the random variable, the analog of
Eq.~(\ref{random_relation}) is given by
\begin{enumerate}
\item when $n_x, n_y \in Z$ and $r_z \in R^+$, $\hat e_\bk$ satisfies
\begin{equation}
\hat e_{k_x, k_y, k_z}^* = (- 1)^{n_y} \hat e_{- k_x, - k_y, k_z} ,
\end{equation}
it is thus a real random variable when $k_x = k_y = 0$,
\item when $n_x \in Z$ and $n_y \in 2 Z$, $\hat e_\bk$ satisfies
\begin{equation}
\hat e_{k_x, k_y, 0}^* = \hat e_{- k_x, - k_y, 0} .
\end{equation}
\end{enumerate}

\subsubsection{Chimney space with horizontal flip}

The chimney space with horizontal flip
(Figure~\ref{FigureChimneySpaces}) is generated by the translation
$\bT_1 = (L_x, 0, 0)$ along with
\begin{equation}
\VECTROISD{x}{y}{z}
\; \mapsto \;
\MATTROISD{- 1}{0}{0}
          {0}{\;\;\; 1}{0}
          {0}{0}{\;\;\; 1} \VECTROISD{x}{y}{z} + \VECTROISD{0}{L_y / 2}{0} .
\label{ChimneyHorizontalFlip}
\end{equation}
The Invariance Lemma shows the orthonormal eigenbasis to be
\begin{equation}
   \begin{array}{lcl}
     \frac{1}{\sqrt{2}}
       \left[ \Upsilon_{2\pi(\frac{n_x}{L_x},\frac{n_y}{L_y},r_z)}
        + (-1)^{n_y} \Upsilon_{2\pi(-\frac{n_x}{L_x},\frac{n_y}{L_y},r_z)}
       \right]
       & \quad \mbox{for} \quad
       & n_x \in Z^+,\; n_y \in Z,\; r_z \in R, \\
     \Upsilon_{2\pi(0,\frac{n_y}{L_y},r_z)}
       & \quad \mbox{for} \quad
       & n_y \in 2Z,\; r_z \in R. \\
   \end{array}
\end{equation}

Concerning the properties of the random variable, the analog of
Eq.~(\ref{random_relation}) is given by
\begin{enumerate}
\item when $n_x \in Z$, $n_y \in Z$ and $r_z \in R$, $\hat e_\bk$ satisfies
\begin{equation}
\hat e_{k_x, k_y, k_z}^* = (- 1)^{n_y} \hat e_{k_x, - k_y, - k_z} ,
\end{equation}
it is thus a real random variable when $k_y = k_z = 0$,
\item when $n_y \in 2 Z$ and $r_z \in R$, $\hat e_\bk$ satisfies
\begin{equation}
\hat e_{0, k_y, 0}^* = \hat e_{0, - k_y, 0} .
\end{equation}
\end{enumerate}

\subsubsection{Chimney space with half turn and flip}

The chimney space with half turn and flip
(Figure~\ref{FigureChimneySpaces}) is generated by
\begin{equation}
\VECTROISD{x}{y}{z}
\; \mapsto \;
\MATTROISD{- 1}{0}{0}
          {0}{\;\;\; 1}{0}
          {0}{0}{- 1} \VECTROISD{x}{y}{z} + \VECTROISD{0}{L_y / 2}{0} , 
\label{ChimneyHalfTurnFlip1}
\end{equation}
and
\begin{equation}
\VECTROISD{x}{y}{z}
\; \mapsto \; 
\MATTROISD{1}{0}{0}
          {0}{\;\;\; 1}{0}
          {0}{0}{- 1} \VECTROISD{x}{y}{z} + \VECTROISD{L_x / 2}{0}{0} .
\label{ChimneyHalfTurnFlip2}
\end{equation}
It is a four-fold quotient of the plain chimney space, unlike the
preceding examples, which were two-fold quotients. The Invariance
Lemma gives an orthonormal basis for its eigenmodes
\begin{equation}
   \begin{array}{lcl}
     \frac{1}{2}
       \left[                 \Upsilon_{2\pi( \frac{n_x}{L_x}, \frac{n_y}{L_y}, r_z)}
        + (-1)^{n_y}       \Upsilon_{2\pi(-\frac{n_x}{L_x}, \frac{n_y}{L_y},-r_z)}
        \right.& & \\
 \quad \left. + (-1)^{n_x}       \Upsilon_{2\pi( \frac{n_x}{L_x}, \frac{n_y}{L_y},-r_z)}
        + (-1)^{n_x + n_y} \Upsilon_{2\pi(-\frac{n_x}{L_x}, \frac{n_y}{L_y}, r_z)}
       \right]
       & \quad \mbox{for} \quad
       & n_x \in Z^+,\; n_y \in Z,\; r_z \in R^+, \\
     \frac{1}{\sqrt{2}}
       \left[\Upsilon_{2\pi( 0, \frac{n_y}{L_y}, r_z)}
        + \Upsilon_{2\pi( 0, \frac{n_y}{L_y}, -r_z)}
       \right]
       & \quad \mbox{for} \quad
       & n_y \in 2Z,\; r_z \in R^+, \\
     \frac{1}{\sqrt{2}}
       \left[ \Upsilon_{2\pi( \frac{n_x}{L_x}, \frac{n_y}{L_y}, 0)}
        + (-1)^{n_y} \Upsilon_{2\pi( -\frac{n_x}{L_x}, \frac{n_y}{L_y}, 0)}
       \right]
       & \quad \mbox{for} \quad
       & n_x \in 2Z^+,\; n_y \in Z, \\
     \Upsilon_{2\pi( 0, \frac{n_y}{L_y}, 0)}
       & \quad \mbox{for} \quad
       & n_y \in 2Z. \\
   \end{array}
\end{equation}

Following the same procedure as before, we obtain that the
analog of Eq.~(\ref{random_relation}) is given by
\begin{enumerate}
\item when $n_x \in Z^+$, $n_y \in Z$ and $r_z \in R^+$, $\hat e_\bk$
must satisfy
\begin{equation}
\hat e_{k_x, k_y, k_z}^* = (- 1)^{n_y} \hat e_{k_x, - k_y, k_z} ,
\end{equation}
so that it is a real random variable when $k_y = 0$,
\item when $n_y \in 2 Z$ and $r_z \in R^+$, $\hat e_\bk$ must satisfy
\begin{equation}
\hat e_{0, k_y, k_z}^* = \hat e_{0, - k_y, k_z} ,
\end{equation}
so that it is a real random variable when $k_y = 0$,
\item when $n_x \in 2 Z^+$ and $n_y \in Z$, $\hat e_\bk$ must satisfy
\begin{equation}
\hat e_{k_x, 0, k_z}^* = (- 1)^{n_y} \hat e_{k_x , - k_y, 0} ,
\end{equation}
so that it is a real random variable when $k_y = 0$,
\item when $n_y \in 2 Z$, it has to be such that
\begin{equation}
\hat e_{0, k_y, 0}^* = \hat e_{0, - k_y, 0} .
\end{equation}
\end{enumerate}

\section{Singly Periodic Spaces}
\label{SectionSinglyPeriodic}

We will first find the eigenmodes of the slab space, and then use
them to find the eigenmodes of the slab space with flip.

\subsection{Slab Space}
\label{SubsectionSlabSpace}

Just as a 3-torus is the quotient of Euclidean space under the action
of three linearly independent translations and a chimney space is the
quotient by two translations, a {\it slab space} is the quotient of
$\ETR$ by a single translation. Its fundamental domain is an
infinitely tall and wide slab (Figure~\ref{FigureSlabSpaces}), with
opposite faces identified straight across.  The allowable wave vectors
$\bk$ for an eigenmode $\Upsilon_\bk$ of a slab space define a family
of parallel planes.

If we choose coordinates so that the translation takes the form
\begin{equation}
\bT = (0, 0, L_z) ,
\label{SlabSpaceTranslation}
\end{equation}
then the allowed wave vectors $\bk$ are
\begin{equation}
\label{SlabSpaceBasis}
\bk = 2 \pi \, \left(r_x, \, r_y, \, \frac{n_z}{L_z} \right) ,
\end{equation}
for real values of $r_x$ and $r_y$ and integer values of $n_z$.
The corresponding orthonormal basis is
\begin{equation}
   \begin{array}{lcl}
       \Upsilon_{2\pi\,(r_x,\, r_y,\, \frac{n_z}{L_z})}
       & \quad \mbox{for} \quad
       & r_x, r_y \in R,\;n_z \in Z.
   \end{array}
\end{equation}
Even if we restrict to a fixed modulus $k = |\bk|$, the eigenmodes
of slab space remain continuous, not discrete.

If desired, one could construct a more general slab space by
identifying opposite faces with a rotation.  Such a space would be
topologically the same as a standard slab space, but geometrically
different.

As in the case of the torus and of the chimney space, the random
variable $\hat e_\bk$ is a complex random variable satisfying
\begin{equation}
\hat e_\bk^* = \hat e_{- \bk} .
\end{equation}

\subsection{Slab Space with Flip}
\label{SubsectionSlabSpaceWithFlip}

A {\it slab space with flip} is generated by
\begin{equation}
\VECTROISD{x}{y}{z}
\; \mapsto \;
\MATTROISD{- 1}{0}{0}
          {0}{\;\;\; 1}{0}
          {0}{0}{\;\;\; 1} \VECTROISD{x}{y}{z} + \VECTROISD{0}{0}{L_z / 2} .
\label{SlabSpaceWithFlip}
\end{equation}
The Invariance Lemma provides the orthonormal basis for its
eigenmodes
\begin{equation}
   \begin{array}{lcl}
       \left[
                      \Upsilon_{2\pi\,( r_x,\, r_y,\, \frac{n_z}{L_z})}
         + (-1)^{n_z} \Upsilon_{2\pi\,(-r_x,\, r_y,\, \frac{n_z}{L_z})}
       \right]
         & \quad \mbox{for} \quad
         & r_x \in R^+,\; r_y \in R,\; n_z \in Z, \\
       \Upsilon_{2\pi\,( 0,\, r_y,\, \frac{n_z}{L_z})}
         & \quad \mbox{for} \quad
         & r_y \in R,\;n_z \in 2Z.
   \end{array}
\end{equation}

Concerning the properties of the random variable, the analog of
Eq.~(\ref{random_relation}) is given by
\begin{enumerate}
\item when $r_x \in R^+$, $r_y\in R$ and $n_z \in Z$, $\hat e_\bk$ satisfies
\begin{equation}
\hat e_{k_x, k_y, k_z}^* = (- 1)^{n_z} \hat e_{k_x, - k_y, - k_z} .
\end{equation}
It is thus a real random variable when $k_y = k_z = 0$,
\item when $r_y \in R$ and $n_z \in 2 Z$, $\hat e_\bk$ satisfies
\begin{equation}
\hat e_{0, k_y, k_z}^* = \hat e_{0, - k_y, - k_z} .
\end{equation}
\end{enumerate}

\section{Numerical simulations}
\label{SectionCMBMaps}

We can now compute the correlation matrix and simulate CMB maps
for the 17 multi-connected spaces described in the previous
sections.

In all the simulations, we have considered a flat $\Lambda$CDM model
with $\Omega_\Lambda = 0.7$, a Hubble parameter $H_0 \equiv 100 h
\UUNIT{km}{} \UUNIT{s}{-1} \UUNIT{Mpc}{-1}$ with $h = 0.62$, a baryon
density $\omega_\BAR \equiv \Omega_\BAR h^2 = 0.019$ and a spectral
index $n_\SCAL = 1$. With this the radius of the last scattering
surface is $R_\LSS = 15.0 \UUNIT{Gpc}{}$.

We present a series of CMB maps with a resolution of $\ell = 120$ for
the different spaces. These maps are represented on a sphere
portraying the last scattering surface seen from outside in the
universal cover. Images of the last scattering surface under the
action of a holonomy and its inverse are shown and their intersection
gives a pair of matched circles.  All the plots presented here contain
only the Sachs-Wolfe contribution and omit both the Doppler and
integrated Sachs-Wolfe contributions.

For the compact spaces, the characteristics of the fundamental
polyhedra are
\begin{itemize}

\item $L_x = L_y = 0.64$ for all spaces,

\item $L_z = 1.28$ for the half- and quarter-turn spaces
 (Figs.~\ref{plot1},~\ref{plot2})

\item $L_z = 1.92$ for the third- and sixth-turn spaces
(Figs.~\ref{plot3} and~\ref{plot4})

\item $L_z = 0.64$ for the Hantzsche-Wendt and Klein spaces
(Figs.~\ref{plot5} through \ref{plot9}).

\end{itemize}

Turning to the $C_\ell$, let us examine the effects
of the topology and of the volume of the fundamental domain.

To understand the properties of the angular power spectrum on large
scales, let us develop a simple geometrical argument based on the
properties of the eigenmodes of the Laplacian operator (see
Ref.~\cite{inoue2} for an analogous discussion and Ref.~\cite{prl} for
the spherical case). In the simply connected Euclidean space $\ETR$,
the number $N_\SC$ of modes between $k$ and $k + \Delta k$ is simply
given by $N_\SC (k) = 4 \pi k^2 \Delta k$, whatever the scale. Now,
due to the topology, most modes will disappear from the spectrum and
we are left with wavenumbers of modulus
\begin{equation}
k =   2 \pi \sqrt{\left(  \frac{n_x}{L_x}\right)^2
                        + \left(\frac{n_y}{L_y}\right)^2
                        + \left(\frac{n_z}{L_z}\right)^2 } .
\end{equation}
On very small scales (large $k$), the Weyl formula~\cite{lwugl} allows
us to determine the number $N_\MC$ of modes remaining in the spectrum:
asymptotically, $N_\MC (< k) \sim V k^3 / 6 \pi^2$ (see, e.g., Fig.~2
of Ref.~\cite{rulw02}). It follows that the number of modes between
$k$ and $k + \Delta k$ is now given by $N_{\MC, \infty} (k) \sim V k^2
\Delta k / 2 \pi^2 \sim V N_\SC / (2 \pi)^3$.  Thus we may set the
overall normalization on small scales where the effect of the topology
reduces to an overall rescaling.  But this has implications concerning
the large scales.

Consider a rectangular torus with a square cross section
of size $L_x = L_y$ and with height $L_z$, and let the relative
proportions of $L_x$ and $L_z$ vary.

When $L_x \gg L_z $, the space looks like a slab space and the modes
on large scales (i.e., such that $2 \pi / L_x \ll k \ll 2 \pi / L_z$)
have a modulus $k \sim \frac{2 \pi}{L_x} \sqrt{n_x^2 + n_y^2}$, so
that they approach a two-dimensional distribution.  Since the number
of modes with $k < k_0$ is given by $N (< k_0) = \sum_{n = 0}^{k_0^2}
r_2(n) = \pi k_0^2 + {\cal O} (k_0)$, where $r_n (p)$ is the number of
representations of $p$ by $n$ squares, allowing zeros and
distinguishing signs and order (e.g., $r_2 (5) = 8$ and $r_3 (4) =
6$), we obtain that the number of modes between $k$ and $k + \Delta k$
is now given by $N_{\MC, 0} (k) \sim L_x^2 k \Delta k / 2
\pi$. Defining the relative weight as
\begin{equation}
w (k) \equiv \frac{(2 \pi)^3}{V} \frac{N_{\MC, 0} (k)}{N_\SC (k)} ,
\end{equation}
we obtain that $w \sim (\pi / k L_z) \gg 1$ so that the large scale
modes are boosted compared with the mode distribution of the simply
connected space exactly as if the spectral index $n_\SCAL$ were
lowered by 1. In the hypothesis of a scale invariant spectrum $n_\SCAL
= 1$, one therefore expects that the $\ell (\ell + 1) C_\ell$ spectrum
will behave as $\ell^{- 1}$ for the relevant scales

When $L_x \ll L_z$, the space looks like a chimney space and the modes
on large scales (i.e., such that $2 \pi / L_z \ll k \ll 2 \pi / L_x$)
have a modulus $k \sim 2 \pi n_z / L_z$ so that they approach a
one-dimensional distribution. It follows that the number of modes
between $k$ and $k + \Delta k$ is now given by $N_{\MC, 0} (k)
\sim L_z \Delta k / \pi$, so that $w(k) \sim (2 \pi / k^2 L_x^2) \gg
1$. Again, this will imply a relative boost of the spectrum on large
scales as if the spectral index were lowered by 2.

When $L_x \sim L_z$, as long as we are above the mode cut-off, one has
a three-dimensional distribution of modes so that the relative weight of
large scale modes is $w \sim 1$, as in a simply connected space. The
signature of the topology in the $C_\ell$ exists at sufficiently large
scales in the form of small spikes around the expected value in a
simply connected space due to the discrete nature of the $k$-spectrum.

\begin{figure}
\centerline{\psfig{file=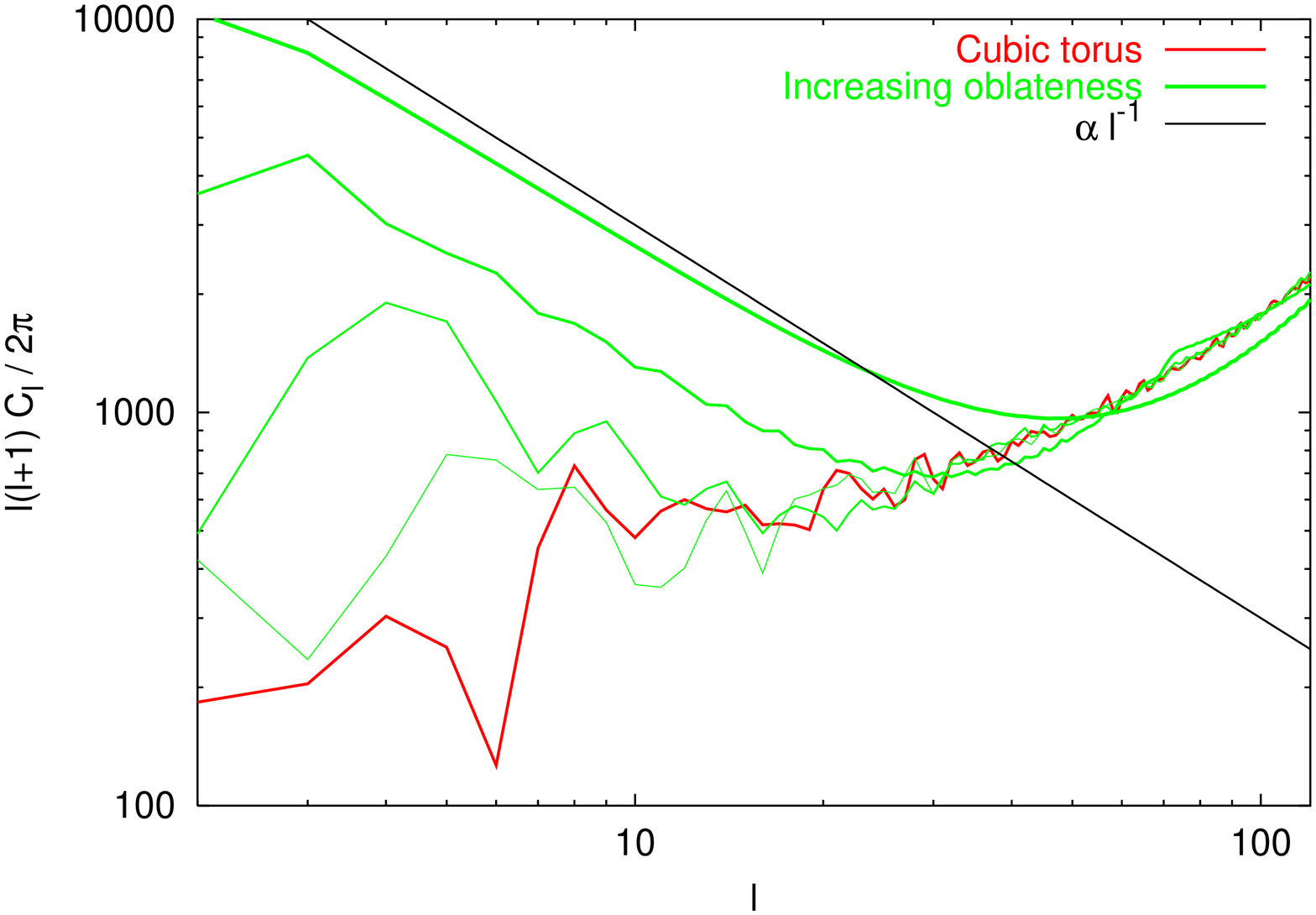,width=3.5in,angle=270}
            \psfig{file=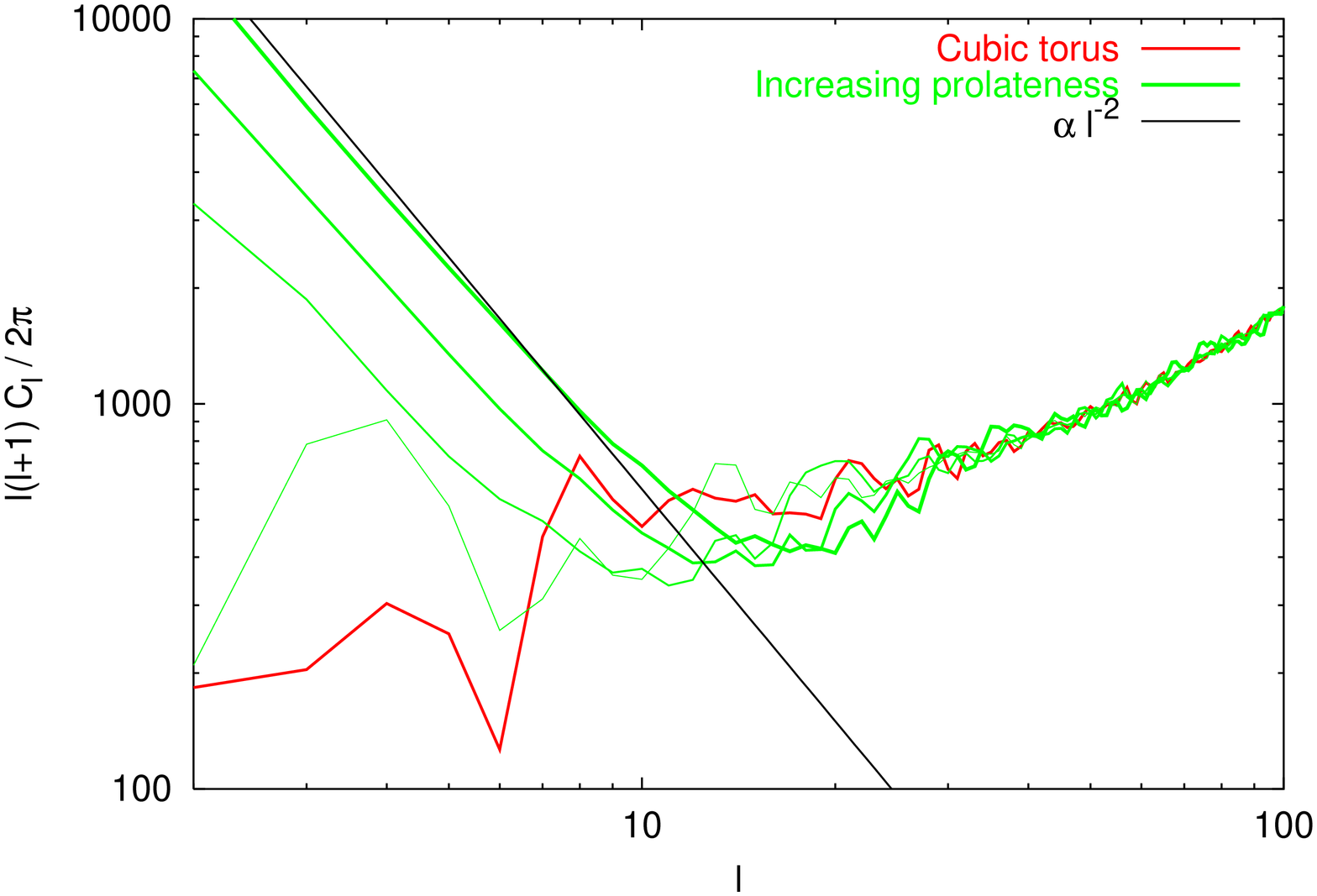,width=3.5in,angle=270}}
\caption{CMB anisotropies in a rectangular torus with $L_x = L_y$ and
$L_z < L_x$ (left) of $L_z > L_x$ (right). The volume of the
fundamental domain is always the same ($0.262 R_\LSS^3$). The spectra
start from a cubic torus and $L_z$ varies by a factor $2$ from each
spectrum to the next (so that the ratio $L_z / L_x$ reaches $64^{\pm
1}$ for the most anisotropic configurations showed here.  The spectrum
is boosted and behaves as $\ell^{- 1}$ of $\ell^{- 2}$ depending on
whether the fundamental domain is flattened or elongated along the $z$
direction. Note that the scale at which the cutoff in the spectrum
occurs increases with $\max (L_x, L_z)$ and that the spikes in the
spectrum are less present for large $L_z$. This is because the mode
spacing decays as $\min (L_x^{- 1}, L_z^{- 1})$ and hence the
discreteness of the spectrum is less obvious. This is of course all
the more true when there are two large directions where the mode
density is higher (left panel).}
\label{oprol}
\end{figure}

These results are summarized in Fig.~\ref{oprol}.

\section{Location of the Observer}
\label{SectionLocationOfObserver}

\begin{figure}
\centerline{\psfig{file=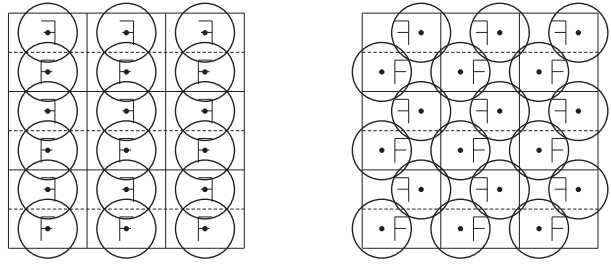,width=8cm}}
\caption{The repeating images of the last scattering circle in a
two-dimensional Klein bottle align in rows if the observer happens to
sit on an axis of glide symmetry (left), but form a different pattern
if the observer sits elsewhere (right).}
\label{KleinBottleBasepoints}
\end{figure}

The 3-torus, chimney space, and slab space are exceptional because
they are globally homogeneous.  A globally homogeneous space looks
the same to all observers within it;  that is, a global isometry
will take any point to any other point.  The remaining
multi-connected flat spaces, by contrast, are not globally
homogeneous and may look different to different observers.  For
ease of illustration, consider the two-dimensional Klein bottle: the
self-intersections of the ``last scattering circle'' are different
for an observer sitting on an axis of glide symmetry
(Figure~\ref{KleinBottleBasepoints} left) than for an observer
sitting elsewhere (Figure~\ref{KleinBottleBasepoints} right).
Analogously in three dimensions, the lattice of images of the last
scattering surface may differ tremendously for observers sitting
at different locations within the same space.  The power spectrum,
the statistical anisotropies, and the matching circles may all
differ.

Moving the observer to a new basepoint would needlessly complicate
existing computer software for simulating CMB maps.  It is much easier
to move the whole universe, leaving the observer fixed!  In technical
terms, we want to replace an eigenmode $\Upsilon_\bk (\bx)$ with the
translated mode $\Upsilon_\bk (\bx + \bx_\OBS)$, where $\bx_\OBS$ is
the desired location for the observer.  The translated mode is quite
easy to compute:
\begin{eqnarray}
\label{TranslatedObserver}
\Upsilon_\bk (\bx)
 & \mapsto & \Upsilon_\bk(\bx + \bx_\OBS) \nonumber \\
 & = &       \ee{i \bk \cdot (\bx + \bx_\OBS)} \nonumber \\
 & = &       \ee{i \bk \cdot \bx_\OBS} \; \ee{i \bk \bx} \nonumber \\
 & = &       \ee{i \bk \cdot \bx_\OBS} \; \Upsilon_{\bk} (\bx) .
\end{eqnarray}
For a simple mode $\Upsilon_\bk (\bx)$, the translation produces a
phase shift (by a factor of $\ee{i \bk \cdot \bx_\OBS}$) and
nothing more.  The full effect is seen when one considers linear
combinations of simple modes:
\begin{eqnarray}
\label{psh}
   a_1 \Upsilon_{\bk_1} (\bx)
 + a_2 \Upsilon_{\bk_2} (\bx)
\quad\mapsto\quad 
   a_1 \ee{i \bk_1 \cdot \bx_\OBS} \Upsilon_{\bk_1} (\bx)
 + a_2 \ee{i \bk_2 \cdot \bx_\OBS} \Upsilon_{\bk_2} (\bx) .
\end{eqnarray}
Each term undergoes a different phase shift, so the final sum may
be qualitatively different from the original.

\begin{figure}
\centerline{\psfig{file=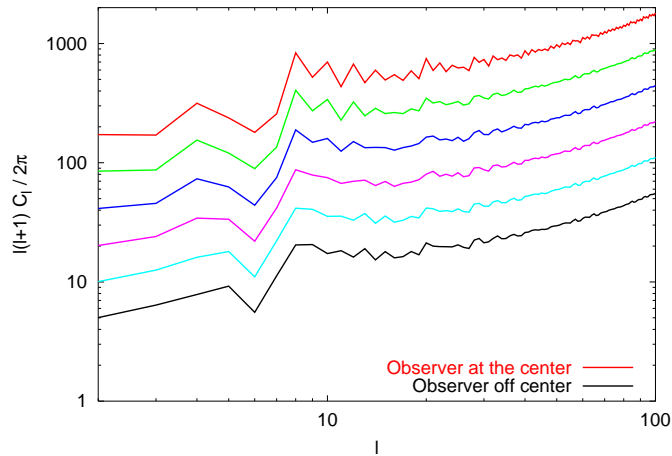,width=3.5in,angle=270}}
\caption{CMB anisotropies in a half-turn space with $L_x = L_y =
0.64$, $L_z = 1.28$ for various positions of the observer. The
observer starts from the center of the fundamental domain and moves
along the $x$ axis. As the position and size of some of the matching
circles vary, the isotropic part of the angular spectrum also
varies. The global structure of the spectrum remains unchanged, but
the local ``spikes'' in the spectrum which originate from the discrete
nature of the $k$-spectrum are more or less smoothed depending on the
position of the observer. For a better visibility, each spectrum has
been offset by a factor of 2 relative to the preceding one (vertical
units are arbitrary).}
\label{figsw}
\end{figure}

\begin{figure}
\centerline{\psfig{file=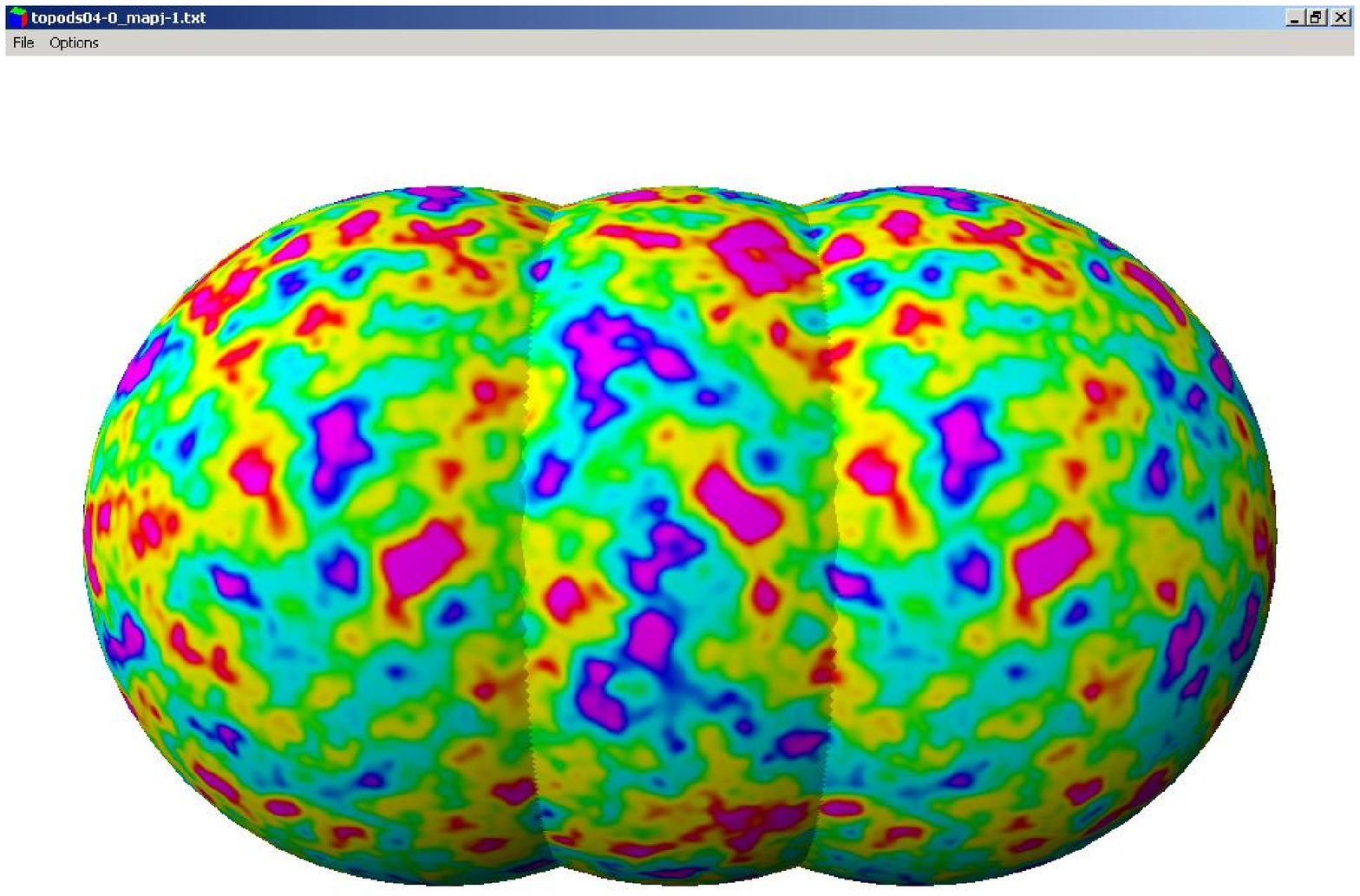,width=3.5in}
            \psfig{file=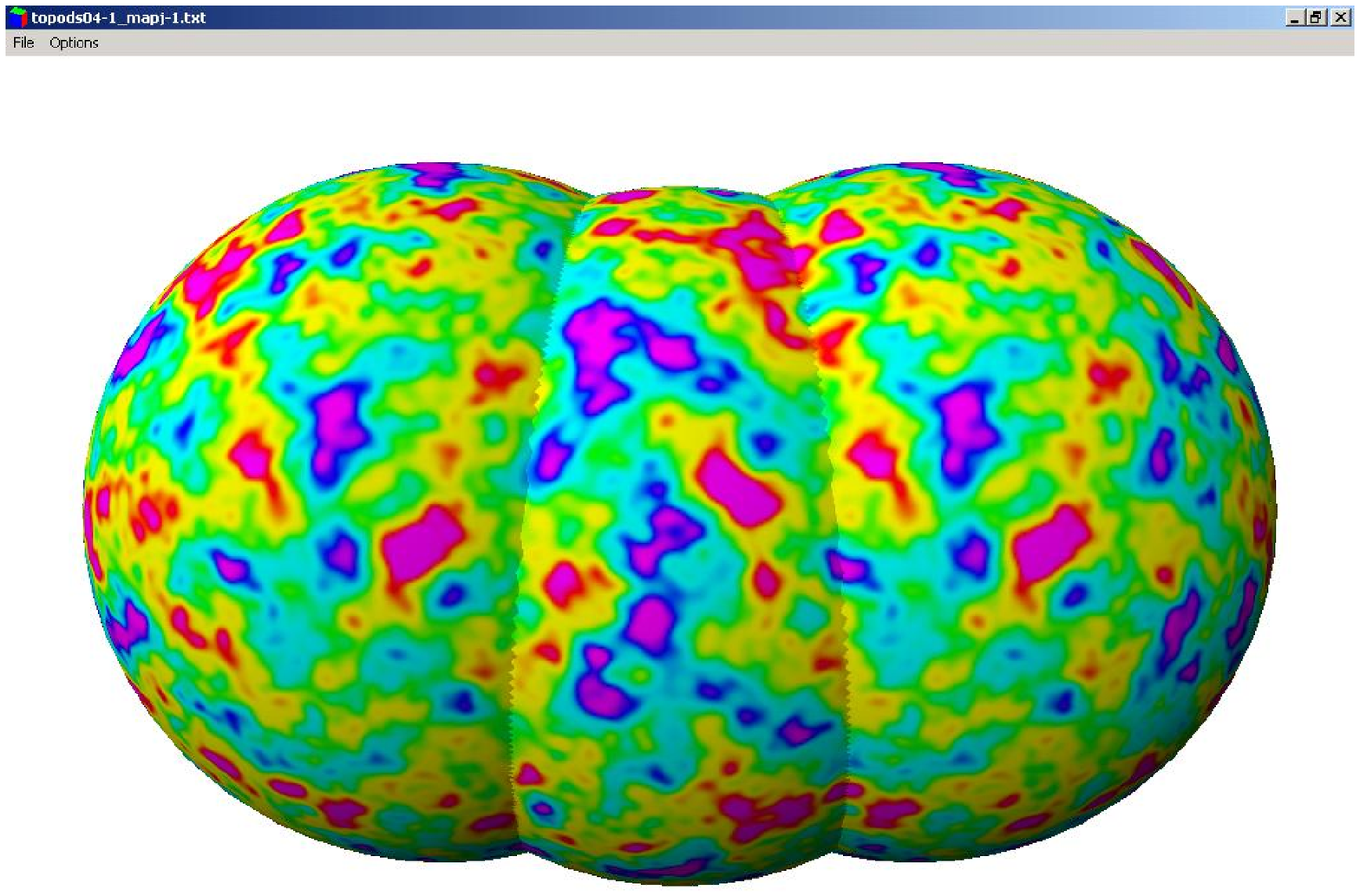,width=3.5in}}
\centerline{\psfig{file=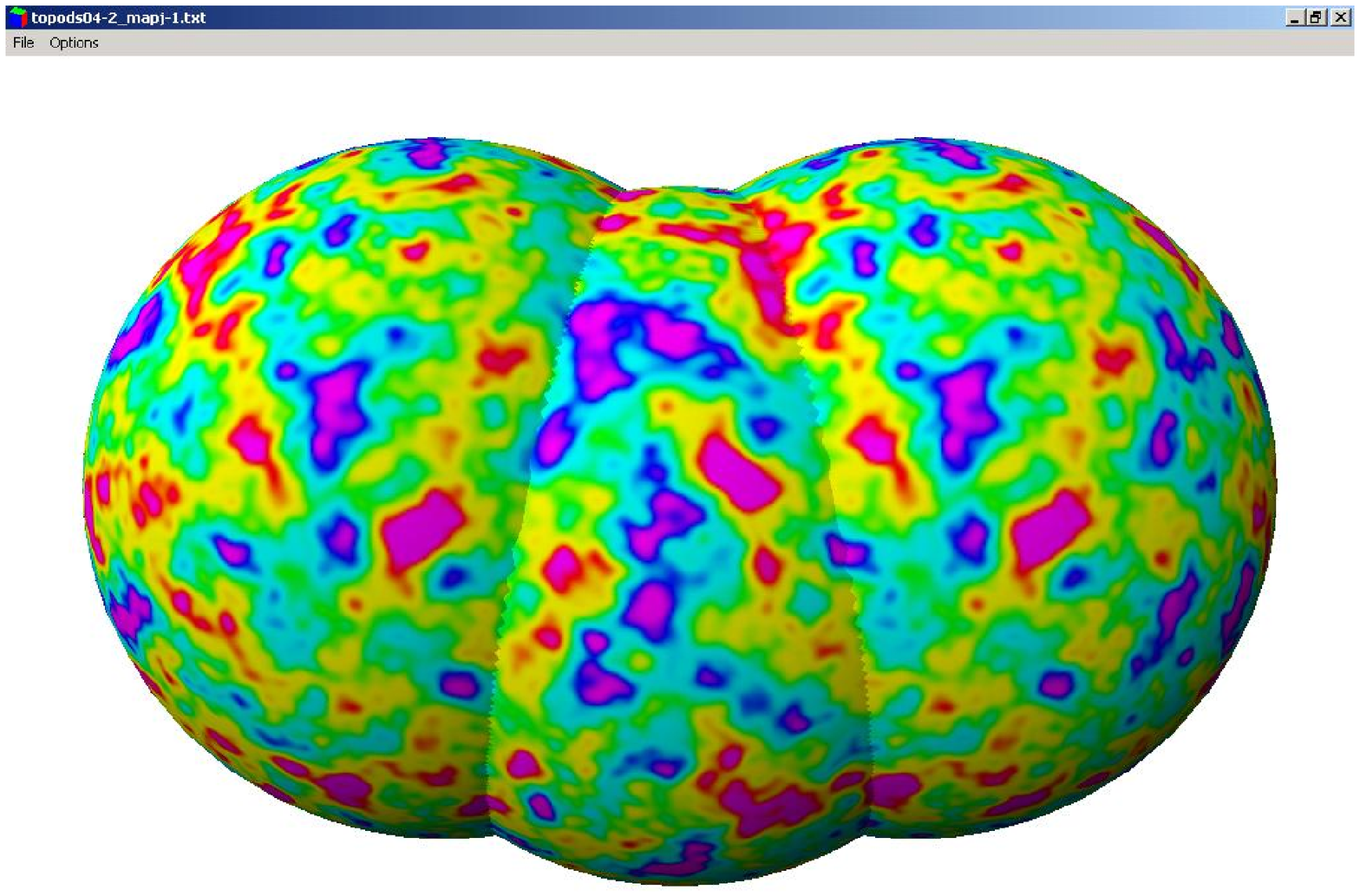,width=3.5in}
            \psfig{file=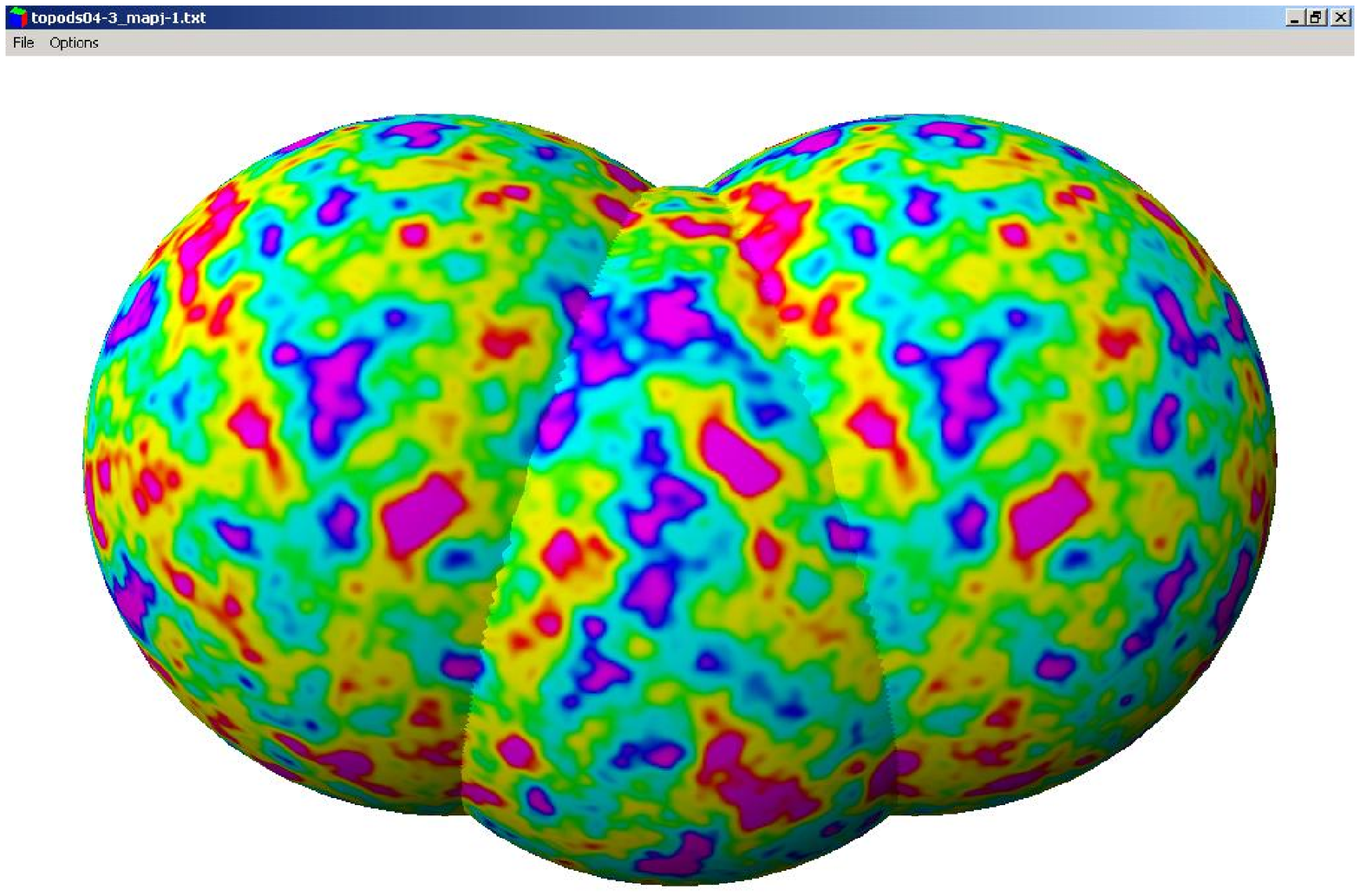,width=3.5in}}
\caption{The last scattering surface seen from outside for a
half-turn space $E_2$ with $L_x = L_y = 0.64$, $L_z = 1.28$ as in
Fig.~\ref{plot1}. The observer starts from position $(0, 0, 0)$, and
slowly moves in the $x$ direction. Due to the non-homogeneity of the
space, the CMB maps look different.}
\label{genobs1}
\end{figure}

Note that the phase shift (\ref{psh}) induced by the change of the
position of the observer does not influence the properties of the
statistical variable $\hat e_\bk$, but does influence the way a given
mode contributes to a given angular scale. This is depicted in
Figure~\ref{figsw}, where the angular power spectrum is shown in a
half-turn space for various positions of the observer. Corresponding
examples of maps are shown in Fig.~\ref{genobs1}.


\section{Conclusions}\label{SectionConclusions}

This article has presented the tools required to compute CMB maps
for all multi-connected flat spaces. We gave for each space
\begin{itemize}
\item the polyhedron and holonomy group,
\item the eigenmodes of the Laplacian.
\end{itemize}
We then presented simulated maps for all of the nine compact non
homogeneous spaces. On the basis of the angular power spectra we
compared the effect of different topologies and different
configurations for a given topology. We also implemented the effect of
an arbitrary position of the observer which yields significant effects
for non-homogeneous spaces. We investigated this effect both on
simulated maps and angular power spectra. In particular, it shows that
generically matched circles are not back-to-back and that their
relative position depends on the position of the observer.

All these tools and simulations will be of great help for extending the
conclusions reached on the torus and to investigate their genericity
as well as for providing test maps for any method wishing to detect
(an interpret) the breakdown of global isotropy.

\section*{Acknowledgements}

We thank Adam Weeks Marano for drawing the figures of the fundamental
polyhedra. We also thank Fran\c{c}ois Bouchet and Simon Prunet for
discussions, and Neil Cornish, David Spergel, Glenn Starkman and Max
Tegmark for fruitful exchanges. J.W.\ thanks the MacArthur Foundation
for its support.


\end{document}